\documentclass[iop]{emulateapj}
\usepackage{epsfig}
\usepackage{natbib}                
\usepackage{lscape}
\usepackage{verbatim}              
\usepackage{graphicx}              
\usepackage{amsmath,amsthm,amsfonts,amsopn,amssymb} 
\usepackage{longtable,pdflscape}
\usepackage{afterpage}
\usepackage{rotating}					
\usepackage{ulem}

\shorttitle{Stellar Diameters and Temperatures III} 
\shortauthors{Boyajian \& von Braun et al.}
\begin{document}


\title{Stellar Diameters and Temperatures \\
III. Main Sequence A, F, G, \& K Stars: Additional high-precision measurements and empirical relations}

\author{Tabetha S. Boyajian\altaffilmark{1,2}, 
Kaspar von Braun\altaffilmark{3,4},
Gerard van Belle\altaffilmark{5},  \\ 
Chris Farrington\altaffilmark{6}, 
Gail Schaefer\altaffilmark{6}, 
Jeremy Jones\altaffilmark{1},
Russel White\altaffilmark{1},
Harold A. McAlister\altaffilmark{1}, 
Theo A. ten Brummelaar\altaffilmark{6}, 
Stephen Ridgway\altaffilmark{7}, 
Douglas Gies\altaffilmark{1},
Laszlo Sturmann\altaffilmark{6}, 
Judit Sturmann\altaffilmark{6},  
Nils H. Turner\altaffilmark{6},
P. J. Goldfinger\altaffilmark{6}, 
Norm Vargas\altaffilmark{6} 
}

\altaffiltext{1}{Center for High Angular Resolution Astronomy and Department of Physics and Astronomy, Georgia State University, P. O. Box 4106, Atlanta, GA 30302-4106} 
\altaffiltext{2}{Department of Astronomy, Yale University, New Haven, CT 06511} 
\altaffiltext{3}{NASA Exoplanet Science Institute, California Institute of Technology, MC 100-22, Pasadena, CA 91125} 
\altaffiltext{4}{Max Planck Institute for Astronomy, K\"{o}nigstuhl 17, 69117 Heidelberg } 
\altaffiltext{5}{Lowell Observatory, Flagstaff, AZ 86001}
\altaffiltext{6}{The CHARA Array, Mount Wilson Observatory, Mount Wilson, CA 91023}
\altaffiltext{7}{National Optical Astronomy Observatory, P.O. Box 26732, Tucson, AZ 85726-6732}



\begin{abstract}

Based on CHARA Array measurements, we present the angular diameters of 23 nearby, main-sequence stars, ranging from spectral type A7 to K0, five of which are exoplanet host stars.  We derive linear radii, effective temperatures, and absolute luminosities of the stars using {\it HIPPARCOS} parallaxes and measured bolometric fluxes.  The new data are combined with previously published values to create an {\it Angular Diameter Anthology} of measured angular diameters to main-sequence stars (luminosity class V and IV). This compilation consists of 125 stars with diameter uncertainties of less than 5\%, ranging in spectral types from A to M.  The large quantity of empirical data are used to derive color-temperature relations to an assortment of color indices in the Johnson $(BVR_{\rm J}I_{\rm J}JHK)$, Cousins $(R_{\rm C}I_{\rm C})$, Kron $(R_{\rm K}I_{\rm K})$, Sloan $(griz)$, and WISE $(W_{\rm 3}W_{\rm 4})$ photometric systems. These relations have an average standard deviation of $\sim 3$\% and are valid for stars with spectral types ~A0 to M4. To derive even more accurate relations for Sun-like stars, we also determined these temperature relations omitting early-type stars ($T_{\rm eff} > 6750$~K) that may have biased luminosity estimates because of rapid rotation; for this subset the dispersion is only $\sim 2.5$\%.
We find effective temperatures in agreement within a couple percent for the interferometrically characterized sample of main sequence stars compared to those derived via the infrared-flux method and spectroscopic analysis.

\end{abstract}




\section{Introduction}                              
\label{sec:introduction}


Aside from our Sun, stars are often-times considered unresolved point sources, with no readily measurable 2-D structure obtainable. However, current technology enables measurements of angular diameters of stars with somewhat large angular sizes ($\theta >$ a few $\times \sim 0.1$~mas, where $\theta$ is the angular diameter) to be spatially resolved via long-baseline optical/infrared interferometry (LBOI) (refer to references in Table~\ref{tab:diameters_literature} for examples).  The two general flavors of observable stellar diameters include evolved stars (giants/supergiants), whose extended linear diameter compensate for their relatively large distance from the observer, and main-sequence stars, whose linear size remains non-inflated from stellar evolution and therefore must reside in the observer's close vicinity.  It is these nearby stars with known parallaxes and interferometrically measured angular sizes that enable us to empirically determine the absolute properties of the star, namely the linear radius and effective temperature (e.g. \citealt{boy12a}).

Stellar properties can also be indirectly estimated from comparisons of spectral lines and predictions from atmospheric models.  Strengths in such stellar atmosphere and evolutionary models as well as less-direct methods of characterizing stellar properties do not just rely upon the input physics; very often it is necessary to calibrate the zero points from direct measurements. The ability to characterize empirically the fundamental properties of stars through interferometry provides us with the critical information needed to constrain and allow improvements for stellar atmosphere and evolutionary models \citep{and91, tor10}.  

General characterization of stars using spectroscopic analysis in combination with evolutionary models (e.g., \citealt{val05,tak07}) are dependent on the accuracies of models and uniqueness of solutions obtainable.  Such model atmosphere codes used to analyze the stellar spectrum are dependent on many variables such as metallicity, temperature, gravity, and microturbulent velocity, and existing degeneracies between parameters make for difficult analysis given such a large and correlated parameter space. Spectroscopic solutions for effective temperature, surface gravity, and atmospheric abundances are the leading constraints to subsequent analysis using evolutionary models, where the stellar mass and radius may be determined. 

The work in \citet{boy12a} compares interferometrically determined properties to those using model-dependent methods. They find that the use of spectroscopically or photometrically defined properties tend to overestimate the effective temperatures compared to directly measured values.  This discrepancy in temperature is strongly correlated to an offset in spectroscopically measured surface gravities - consequently yielding higher masses and younger ages for the stars studied (see their figures 22 and 23).  Offsets in spectroscopic surface gravities have also been noted to be present through spectroscopic analysis alone, as discussed in section~7.4 of \citet{val05}. However as they note, the lack of data available to calibrate these properties limits the accuracies of their solutions.  Iterative techniques using interferometrically constrained parameters in combination with spectroscopic analysis have proven to yield robust results, such as the one used in \citet{cre12}.  Unfortunately, however, such targets are scarce, given the observability requirements (brightness and proximity) of LBOI.

Stellar temperatures from the infrared flux method (IRFM), a technique first developed by \citet{bla77}, is a popular substitute for defining stellar properties, and the least model dependent behind interferometric measurements.  The photometrically based IRFM is advantageous in approach because it may be applied to a large number of stars, spanning a large range in metallicities.  Tremendous work has blossomed in the field over the past few decades, however its true validity is somewhat plagued by systematic differences between the temperature scales used in the literature, which can be as large as $\sim 100$~K (see \citealt{gon09, cas10}, and references therein).  As many argue, the zero-point calibration of the IRFM lacks the empirical data as a good foundation - always referring to the paucity of interferometric measurements available.

A few years past, we embarked on a interferometric survey of main-sequence stars, as previously reported in the works of \citet{boy12a, boy12b} (these are papers entitled Stellar {\bf D}iameters and {\bf T}emperatures {\bf I} and {\bf II} we hereafter abbreviate as DT1 and DT2, respectively). This work is the third installment of stellar diameters pertaining to this survey (abbreviated DT3), where we continue to populate the literature with accurate stellar parameters of these nearby stars measured with interferometry.  In this paper, we report new angular diameters of 18 stars and improved precision on 5 additional stars, with average uncertainty in the angular diameter of 2\% (Section~\ref{sec:observations_chara}).  In Section~\ref{sec:observations_literature_averages}, we present an overview of angular diameters, listing all main-sequence stars that have interferometrically measured angular diameters with better than 5\% precision.  Sections~\ref{sec:stellar_params} and \ref{sec:evolution} describe the radii, temperatures, bolometric fluxes, luminosities, masses, and ages for the entirely interferometrically characterized sample. Finally, in Section~\ref{sec:discussion} we present the results to color - temperature relations calibrated using our empirical data set, and we present our conclusions in Section~\ref{sec:conclusion}.


\section{Targets and Angular Diameters}
\label{sec:observations}

A census of angular diameter measurements of lower-mass K- and M-dwarfs recently enumerated in DT2 yield a total of 33 stars.  In this work, we expand on the DT2 sample to describe fully the current state of measured angular diameters including all A-, F-, and G-type main-sequence stars.  We follow the same method and criteria as in DT2, admitting only stars where the angular diameter was measured to better than 5\%.

In addition to the collection of literature measurements, in this work we present new angular diameters for 23 stars (Section~\ref{sec:observations_chara}).  Stars with multiple measurements are also examined in Section~\ref{sec:observations_literature_averages}, where we determine mean values for use in the determination of their fundamental properties (Section~\ref{sec:stellar_params}) and the analysis of the data (Section~\ref{sec:discussion}).

\subsection{Observations with the CHARA Array}
\label{sec:observations_chara}

Akin to the observing outlined in DT1 and DT2, observations for this project were made with the CHARA Array, a long-baseline optical/infrared interferometer located on Mount Wilson Observatory in southern California (see \citealt{ten05} for details).  The target stars were selected based on their approximate angular size (a function of their intrinsic linear size and distance to the observer).  We limit the selection to stars with angular sizes $>0.45$~mas, in order to adequately resolve their sizes to a few percent precision with the selected instrument set-up. Note that all stars that meet this requirement are brighter than the instrumental limits of our detector by several magnitudes. The stars also have no known stellar companion within three arcseconds to avoid contamination of incoherent light in the interferometers field of view.  From 2008 to 2012, we used the CHARA Classic beam combiner operating in $H$-band ($\lambda_H=1.67 \mu$m) and $K^{\prime}$-band ($\lambda_{K^{\prime}}=2.14 \mu$m) to collect observations of 23 stars using CHARA's longest baseline combinations.  A log of the observations can be found in Table~\ref{tab:observations}.
 
As is customary, all science targets were observed in bracketed sequences along with calibrator stars.  To choose an appropriate calibrator star in the vicinity of the science target, we used the SearchCal tool developed by the JMMC Working Group \citep{bon06, bon11}.  These calibrator stars are listed in Table~\ref{tab:observations}, and the value of the estimated angular diameters $\theta_{\rm EST}$ is taken from the SearchCal catalog value for the estimated limb-darkened angular diameter.  In order to ensure carefully calibrated observations and to minimize systematics, we employ the same observing directive we initiated and followed in DT1 and DT2: each star must be observed 1) on more than one night 2) using more than one baseline, and 3) with more than one calibrator.  Of the 23 stars in Table~\ref{tab:observations}, only HD~136202 did not meet the second requirement of revisiting it on another baseline, but the data give us no reason to reject it only based on this shortcoming, since a sufficient number of observations were collected over time on the nights we did observe this star.  All other directive requirements were met by all stars.

In addition to the observing directives mentioned above, we also follow the guidelines described in \citet{van05} for choosing unresolved calibrators in order to alleviate any bias in the measurements introduced with the assumed calibrator diameter.  At the CHARA Array, this limit on the calibrator's estimated angular diameter is $\theta_{EST} < 0.45$~mas, and this criteria is met for 30 of the observed calibrators in this paper.  In practice however, we find that some science stars do not have more than one suitably unresolved calibrators available nearby to observe.  As such, we must extend this calibrator size limit to slightly larger, $\theta_{EST} < 0.5$~mas sizes, adding an additional 14 additional calibrator stars to our program\footnote{We mark stars with $\theta_{EST} >0.45$ in Table~\ref{tab:observations}}.  While this is less than ideal, it is important to note that any star observed with a slightly larger calibrator star is also observed with a more unresolved calibrator, and calibration tests show no variance in the calibrated visibilities from these objects compared to each other\footnote{These calibrator size limits are also maintained when pertaining to the calibrators observed in DT1 and DT2, as described within the observations section of each respective paper.}.  As a whole, the calibrators observed have average magnitudes of $V=6.0$, $H=5.0$, $K=4.9$ and an average angular diameter of $0.41 \pm 0.03$~mas.

The calibrated visibilities for each object are fit to the uniform disk $\theta_{\rm UD}$ and limb-darkened $\theta_{\rm LD}$ angular diameter functions, as defined in \citet{han74}.  We use a non-linear least-squares fitting routine written in IDL to solve for each value of $\theta_{\rm UD}$ and $\theta_{\rm LD}$ as well as the errors, assuming a reduced $\chi^2 = 1$.  In order to correct for limb-darkening, we use the linear limb-darkening coefficients from \citet{cla00}, calculated from ATLAS models.  We employ an iterative procedure to identify the correct limb-darkening coefficients to use since those coefficients are dependent on the assumed atmospheric properties of the source.  For main-sequence stars in the range of this sample, we find that only the assumed temperature contributes to a marked change in the limb-darkened value, whereas both surface gravity and metallicity do not provide additional constraints. As such, initial guesses of the object's temperature are used for the preliminary fit to determine $\theta_{\rm LD}$. This value for $\theta_{\rm LD}$ is used with the measured bolometric flux to derive a temperature, as described in Section~\ref{sec:stellar_params}. This new temperature, often not so different from the initial guess, is then used to search for a tweaked limb-darkening coefficient, if needed. This procedure is typically repeated only once, for changes within the grid increments are 250~K, and the average correction needed is on the order of only a few percent\footnote{Although the implementation of this iterative procedure was practiced within DT2, it was not within the analysis of DT1 diameters. Therefore, we performed a complete re-evaluation of the limb-darkened angular diameter fits for all the stars in DT1.  We found that in response to the iterative procedure, the limb-darkening coefficient did not change for 19 of the 44 stars, even though the assumed initial temperature stayed the same for only 6 of these 19. Using the modified limb-darkening coefficients changed the diameters of only 11 of the 44 stars by $<0.1$ sigma and the other 14 of the 44 stars by 0.1 to 0.3 sigma.  This change of much less than 1-sigma using modified coefficients is on the order of what is quoted for the errors in the limb-darkening  coefficients themselves.}.   

\begin{figure}										
\centering
\begin{tabular}{cc}
\epsfig{file=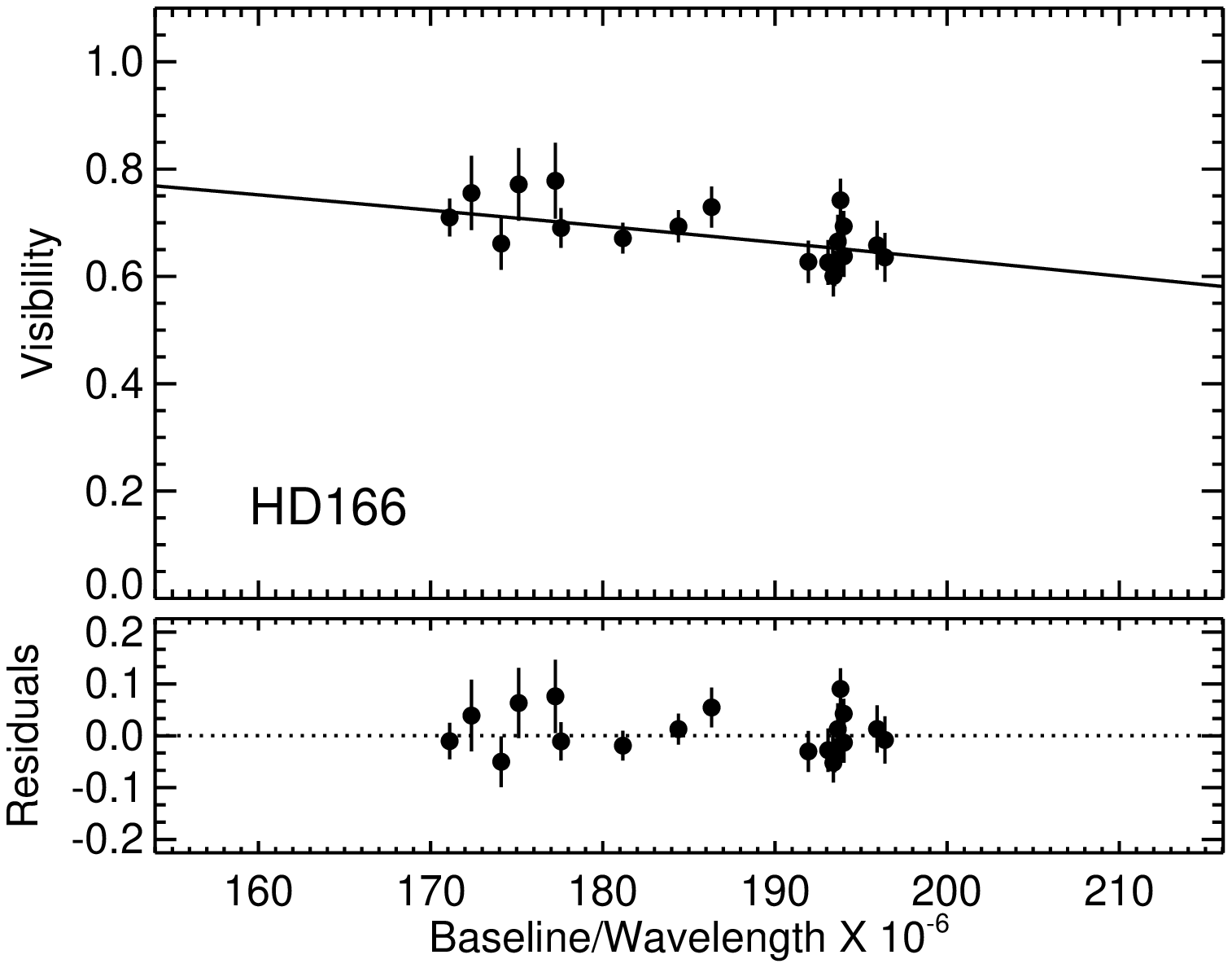, width=0.5\linewidth, clip=} &
\epsfig{file=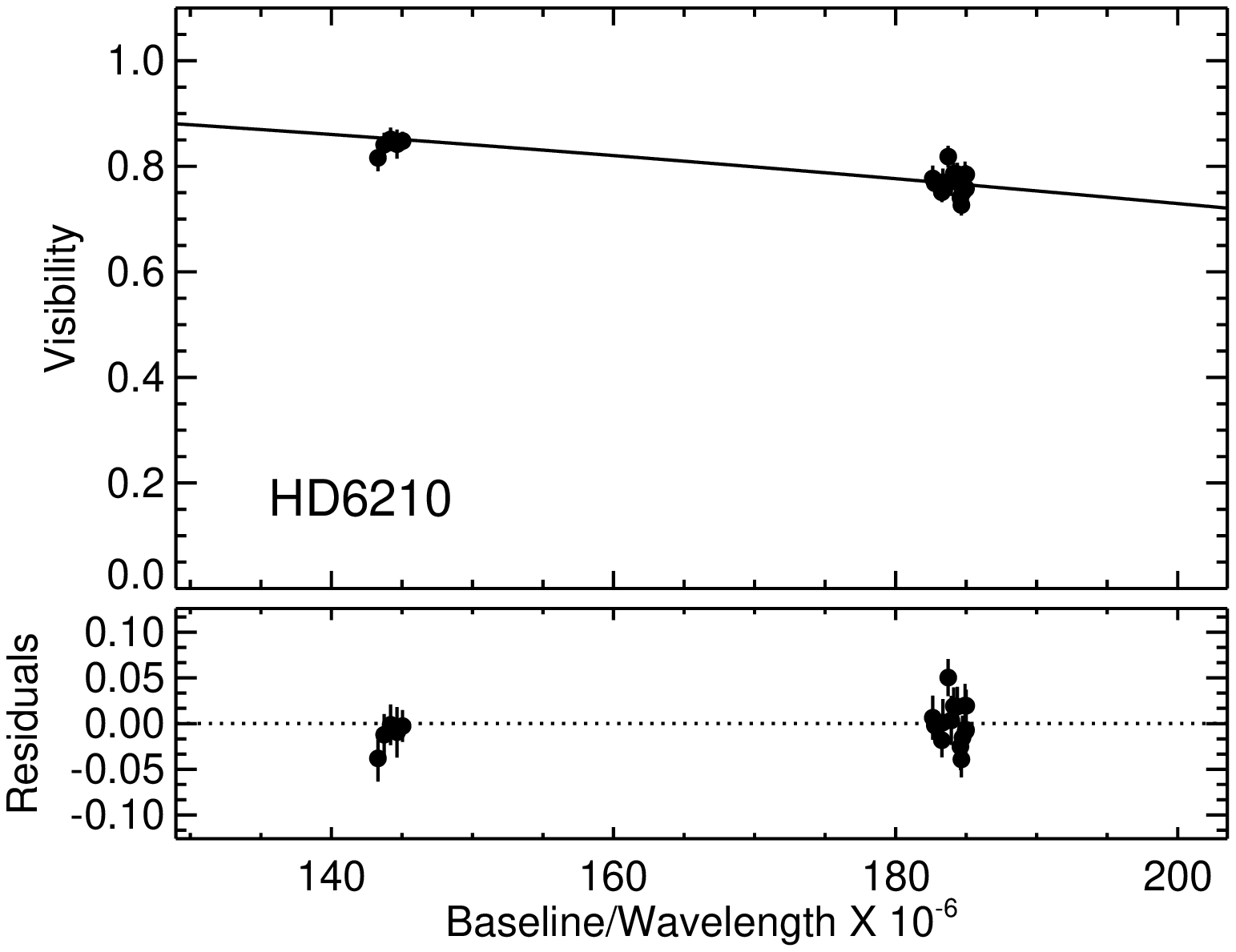, width=0.5\linewidth, clip=} \\
\epsfig{file=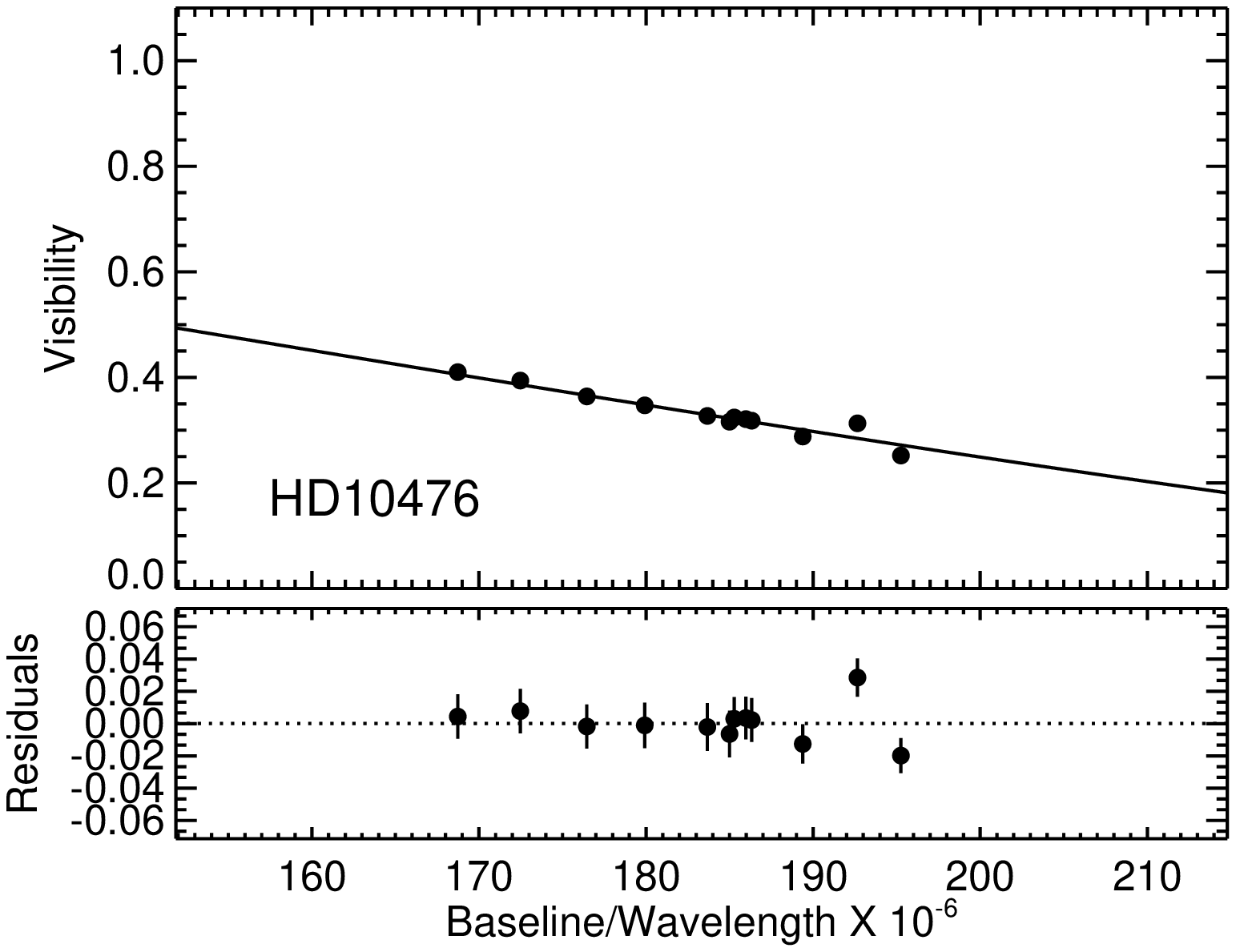, width=0.5\linewidth, clip=} &
\epsfig{file=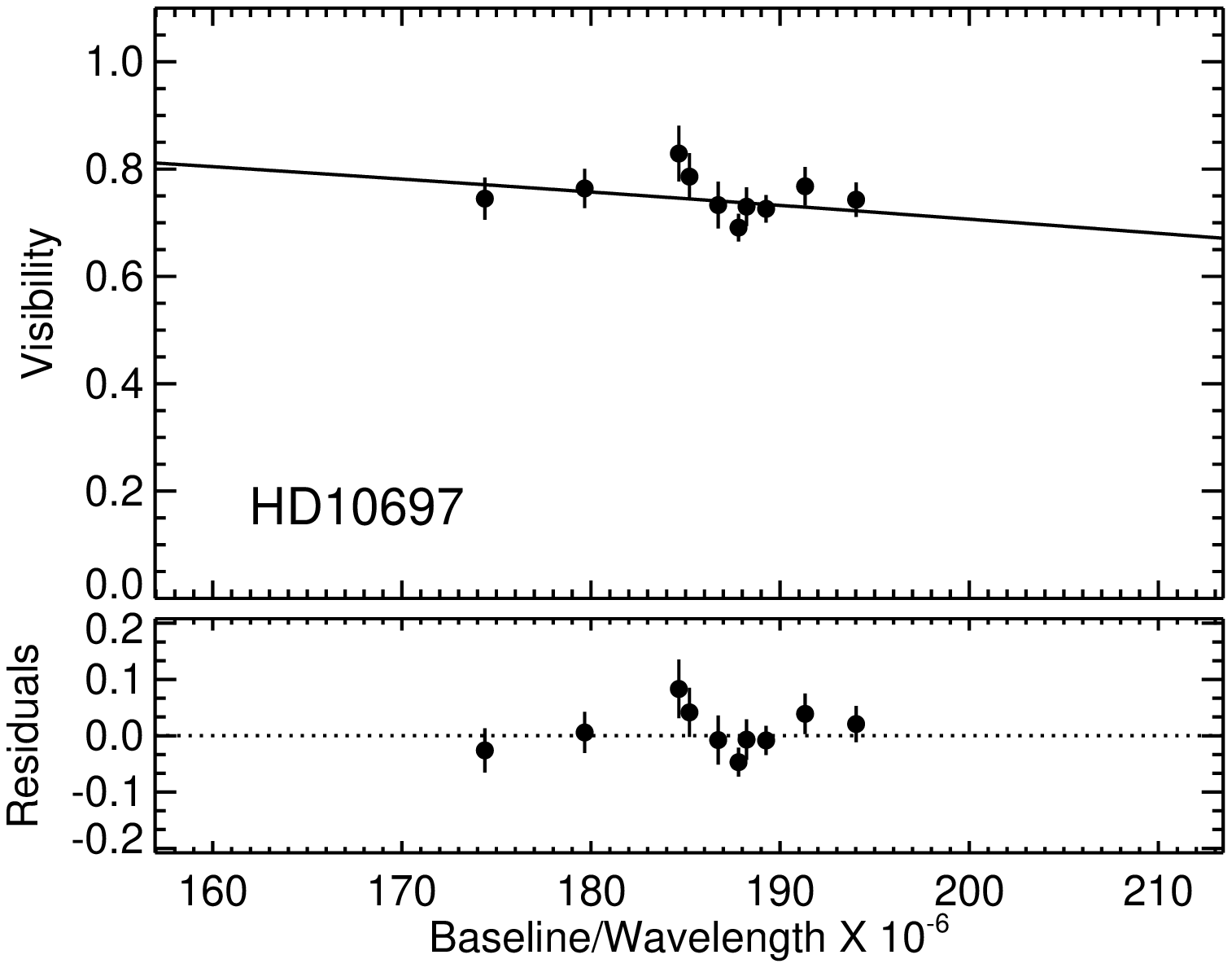, width=0.5\linewidth, clip=} \\
\epsfig{file=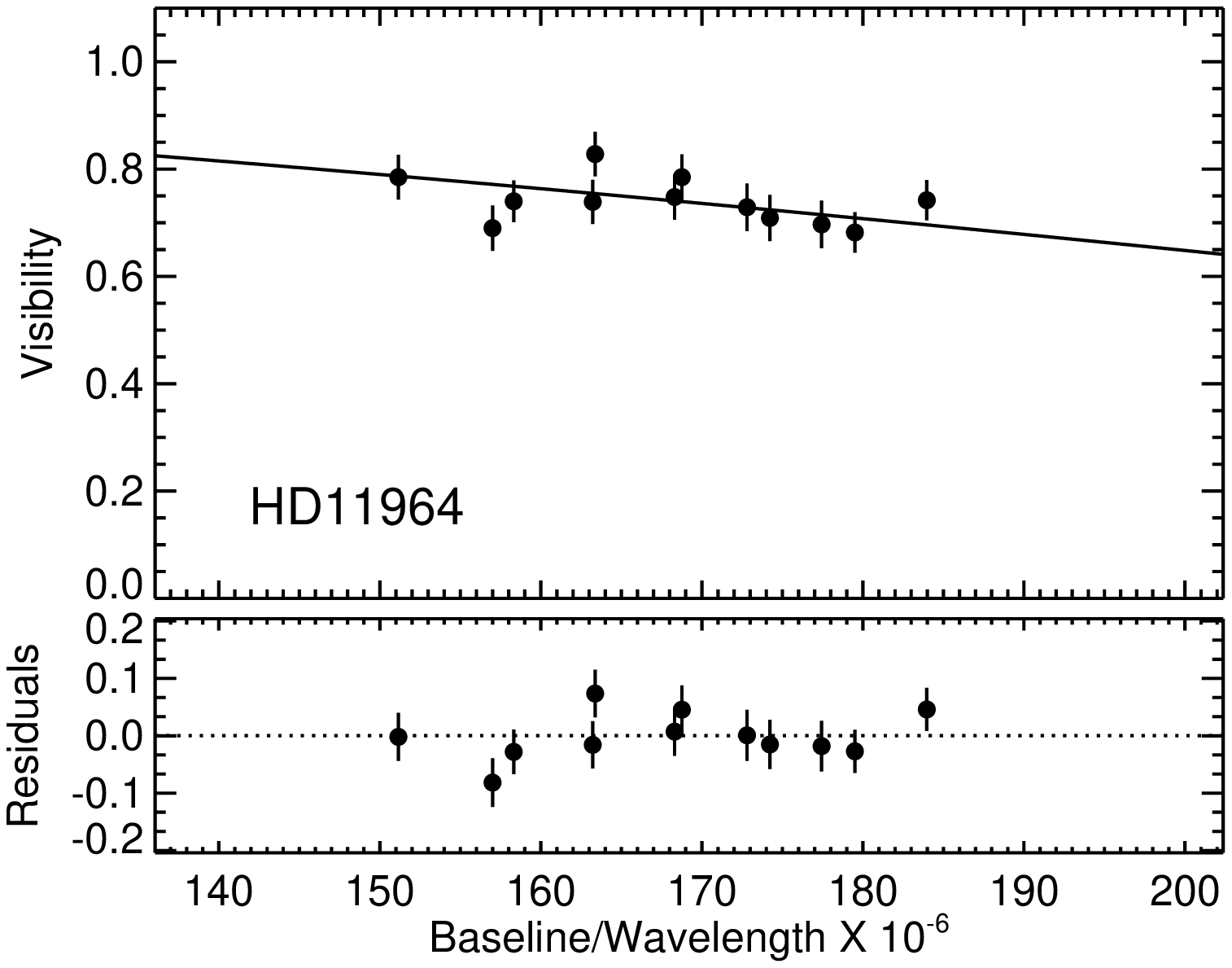, width=0.5\linewidth, clip=} &
\epsfig{file=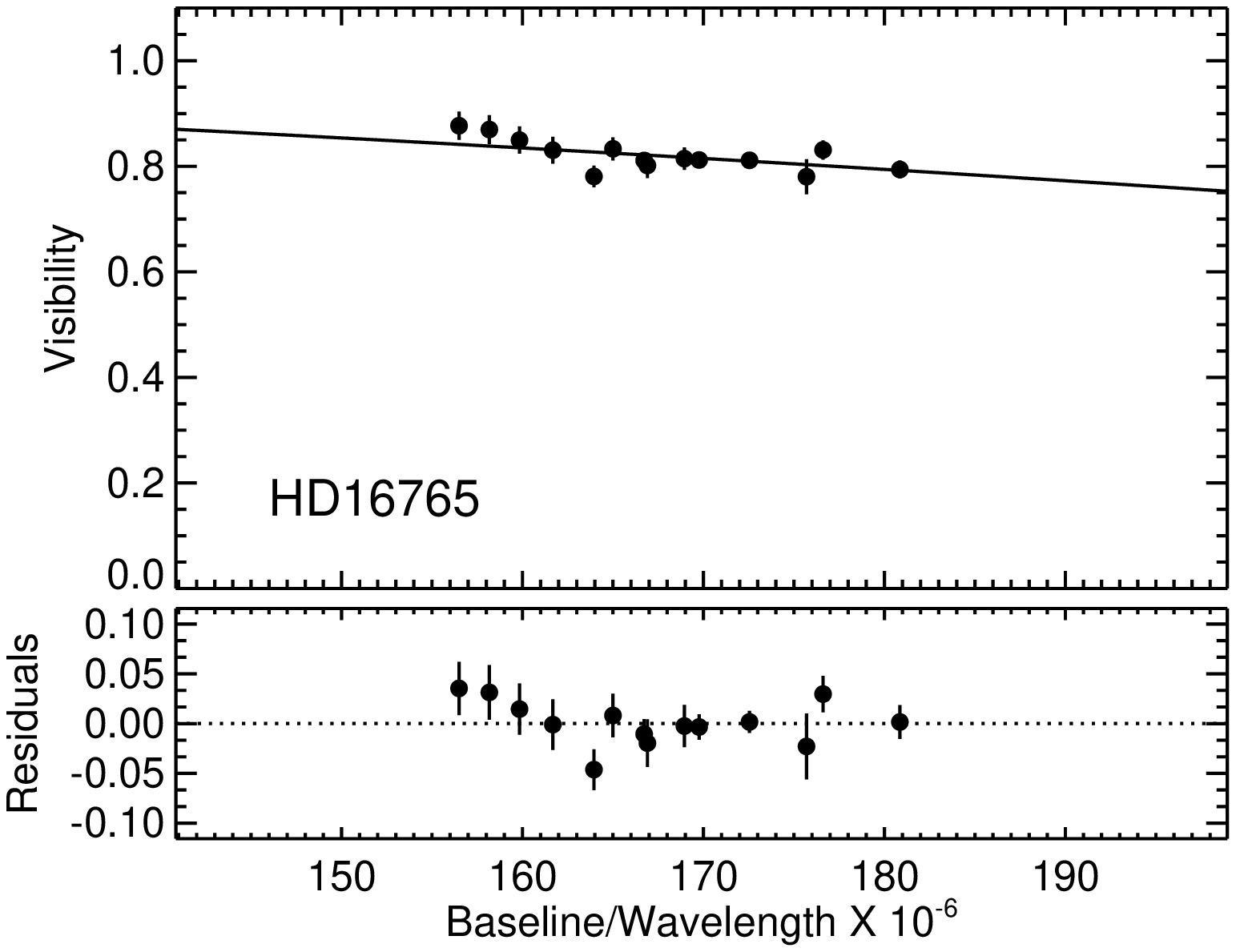, width=0.5\linewidth, clip=} 
 \end{tabular}
 \caption[Angular Diameters] {Calibrated observations plotted with the limb-darkened angular diameter fit for each star. See Section~\ref{sec:observations_chara} for details.}
 \label{fig:diameters1}
 \end{figure}

\begin{figure}										
\centering
\begin{tabular}{cc}
\epsfig{file=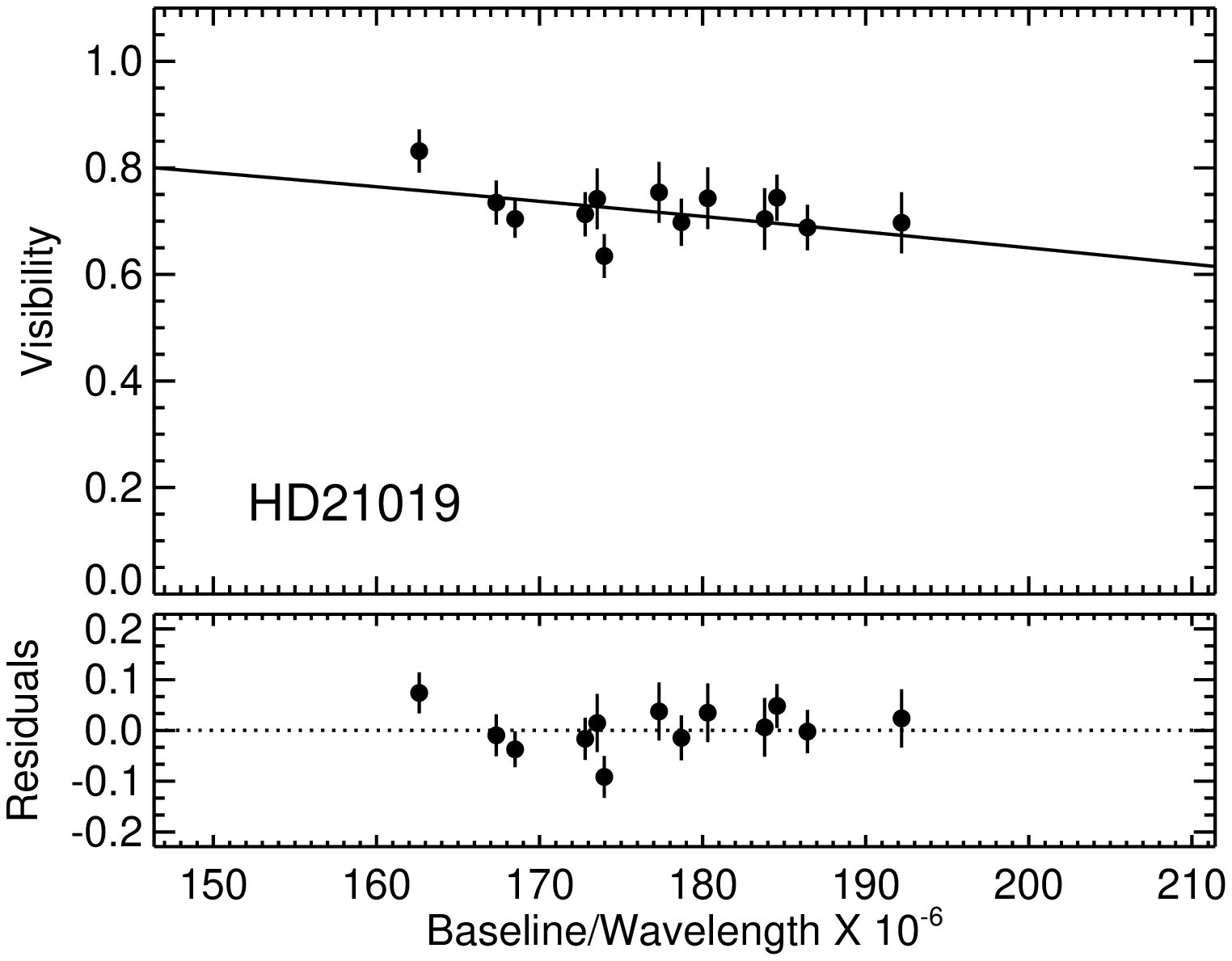, width=0.5\linewidth, clip=} &
\epsfig{file=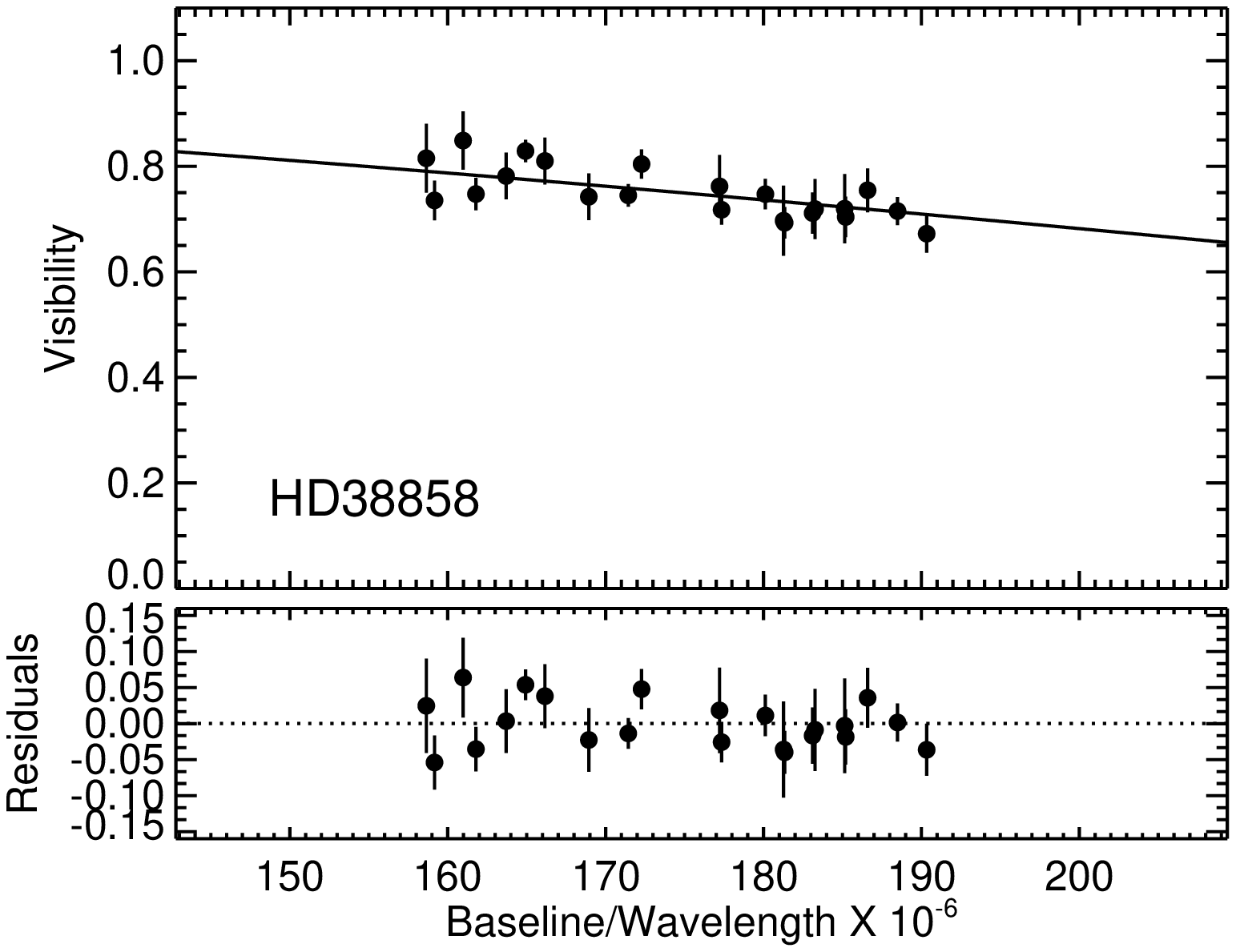, width=0.5\linewidth, clip=} \\
\epsfig{file=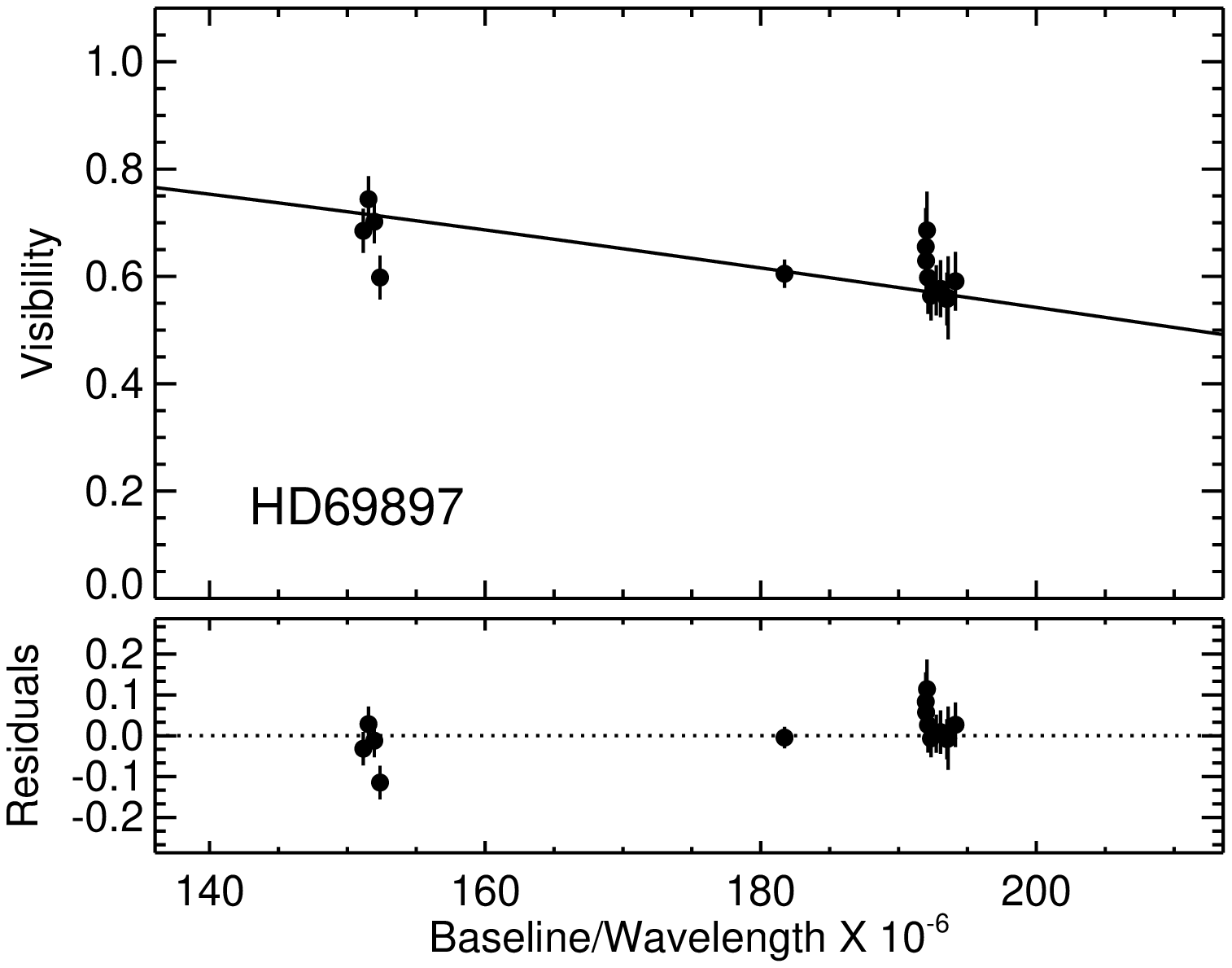, width=0.5\linewidth, clip=} &
\epsfig{file=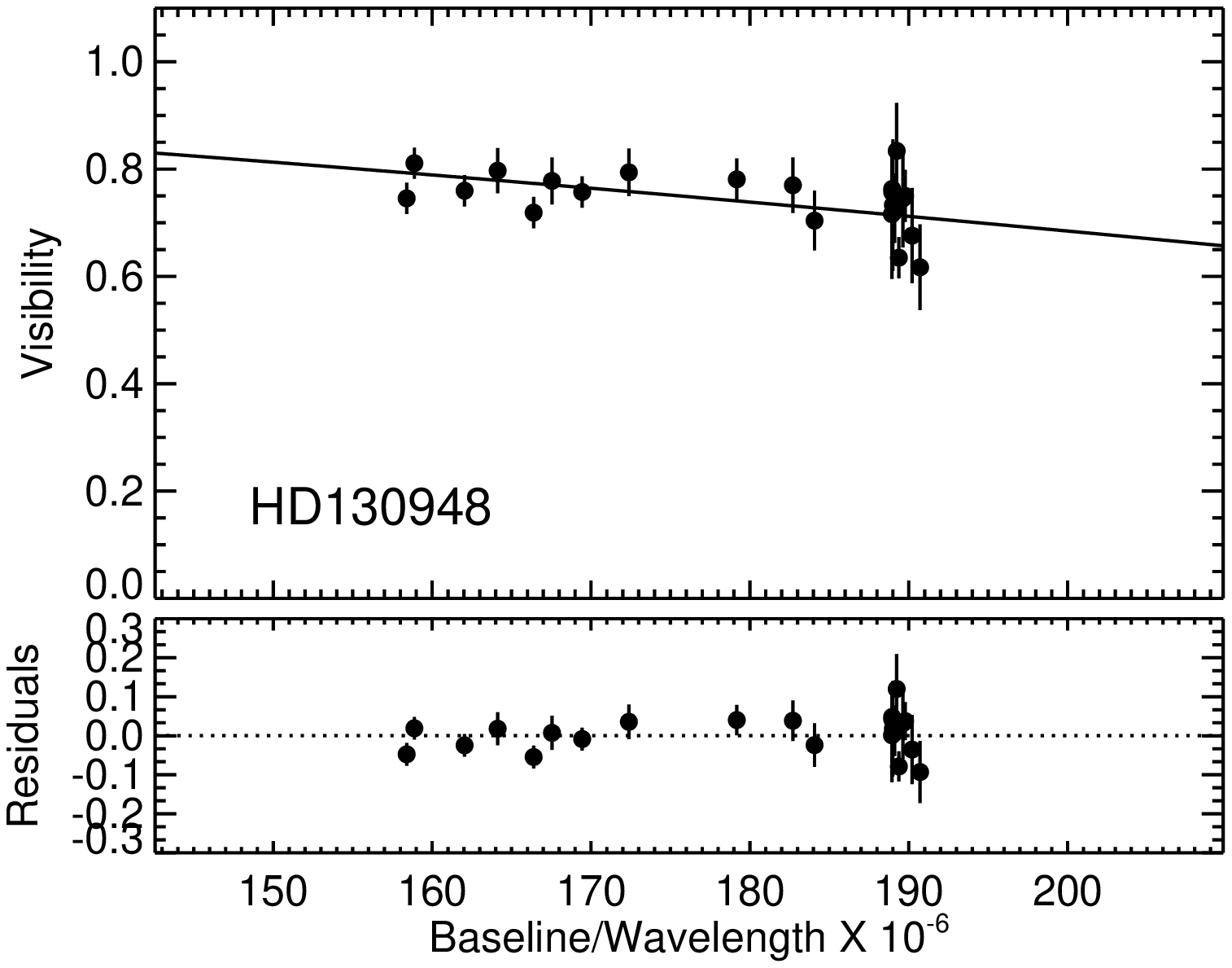, width=0.5\linewidth, clip=} \\
\epsfig{file=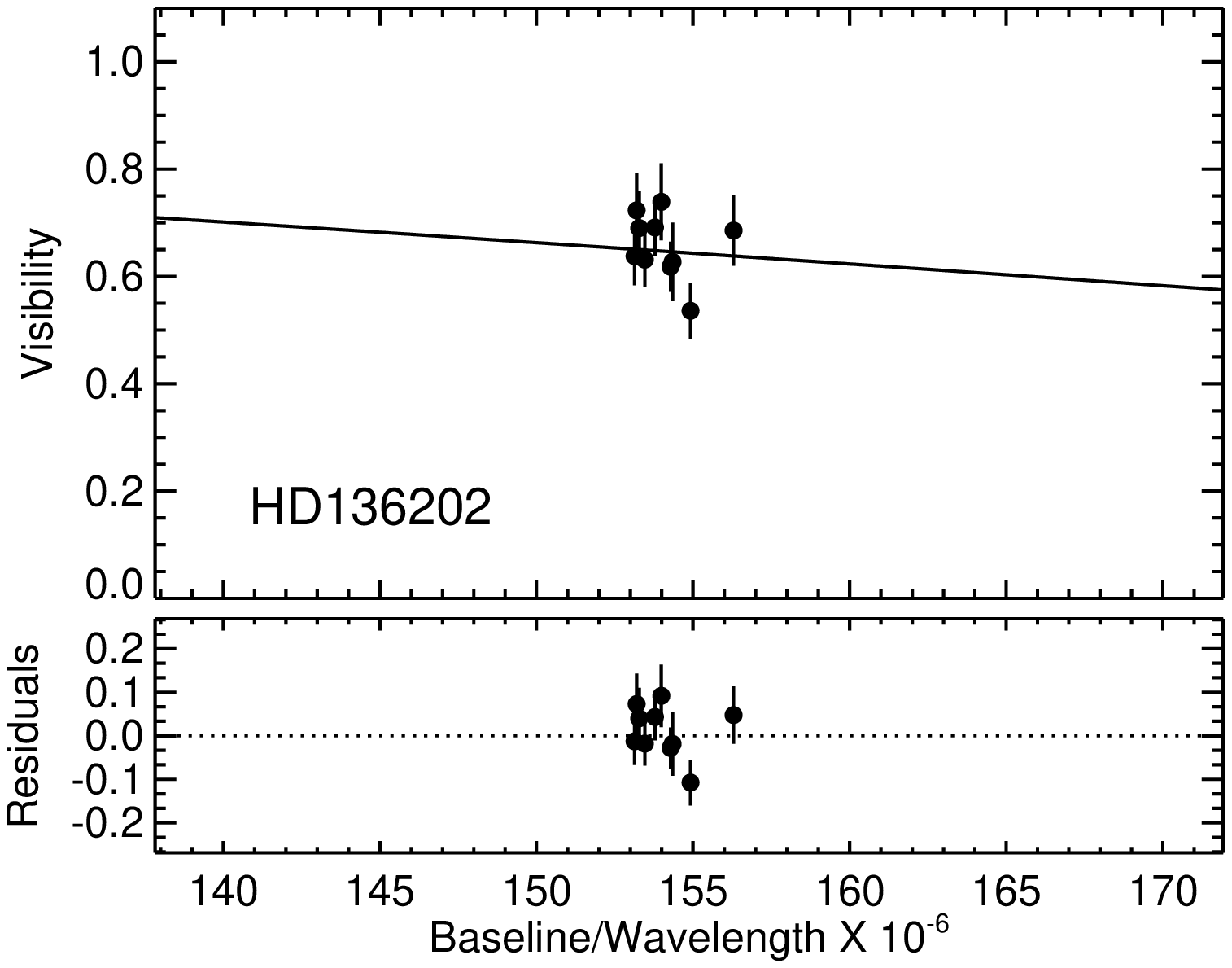, width=0.5\linewidth, clip=} &
\epsfig{file=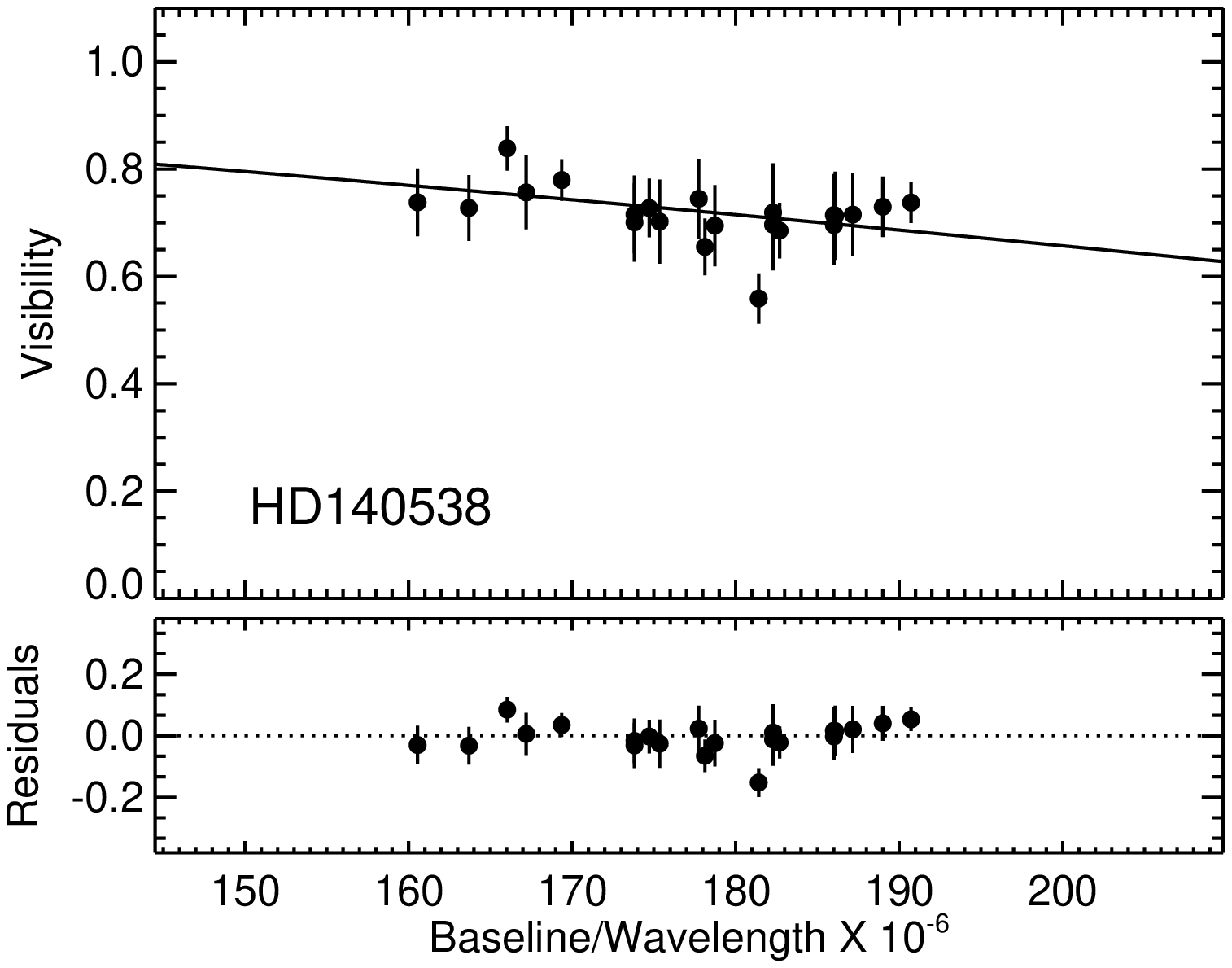, width=0.5\linewidth, clip=} 
 \end{tabular}
 \caption[Angular Diameters] {Calibrated observations plotted with the limb-darkened angular diameter fit for each star. See Section~\ref{sec:observations_chara} for details.}
 \label{fig:diameters2}
 \end{figure}

\begin{figure}										
\centering
\begin{tabular}{cc}
\epsfig{file=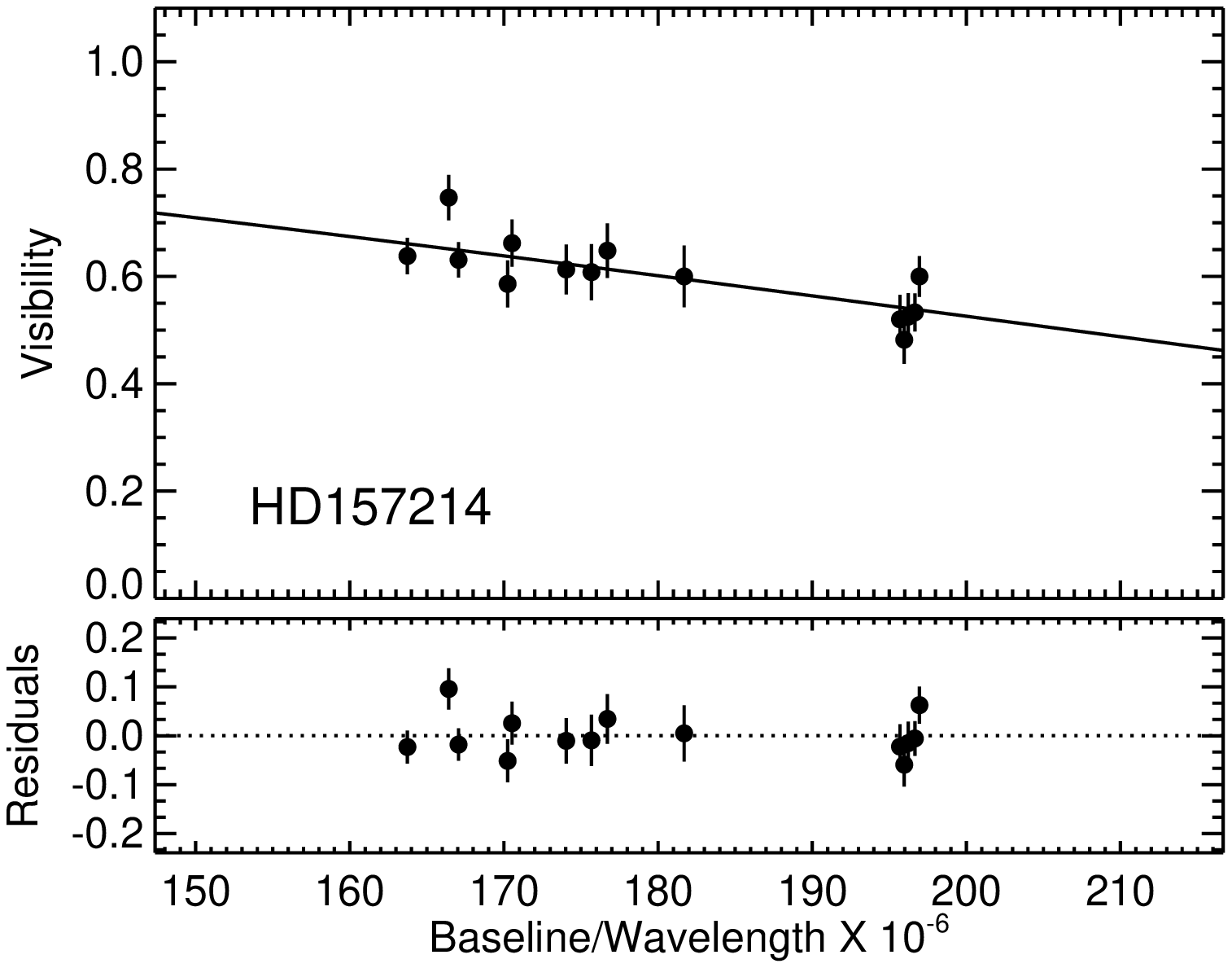, width=0.5\linewidth, clip=} &
\epsfig{file=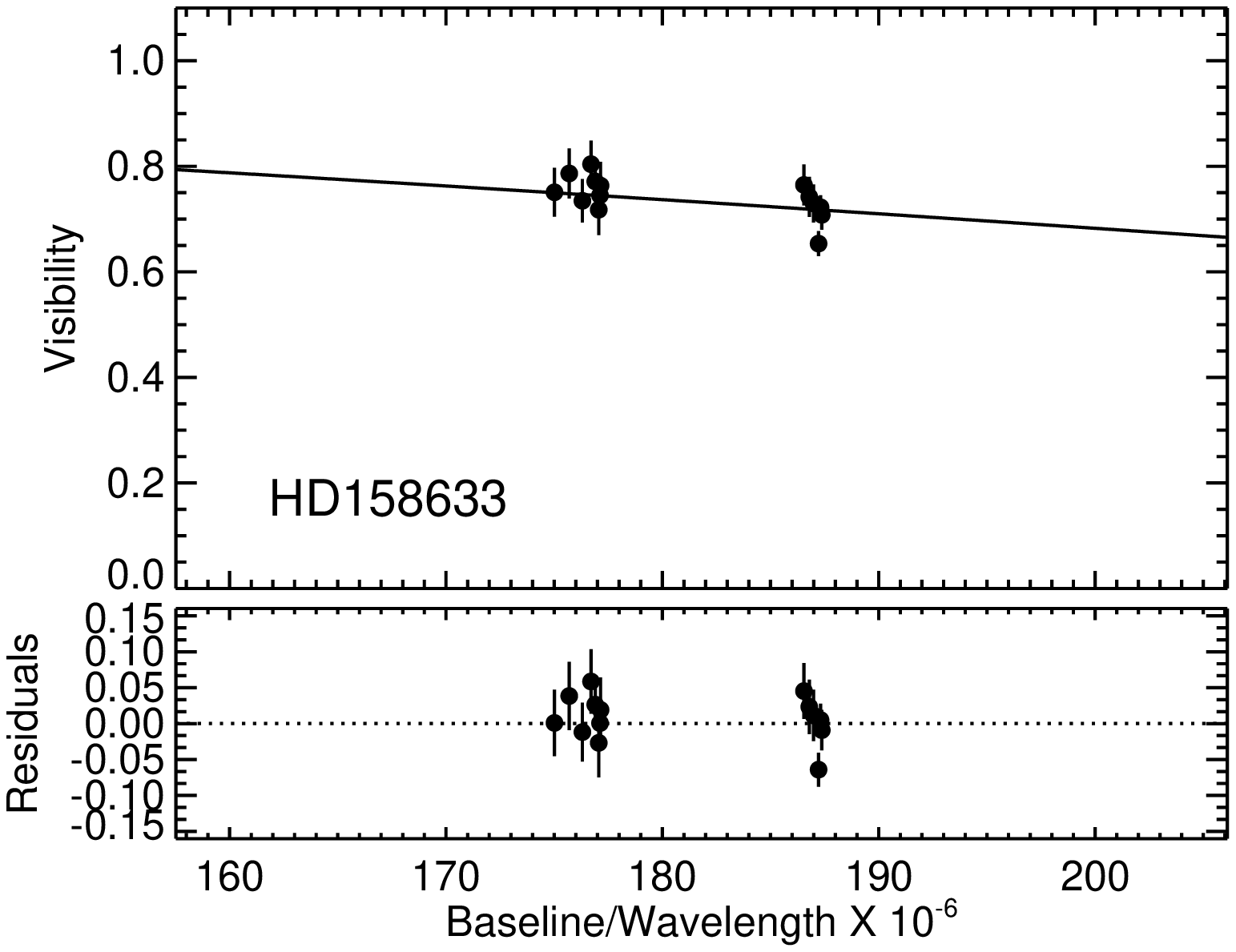, width=0.5\linewidth, clip=} \\
\epsfig{file=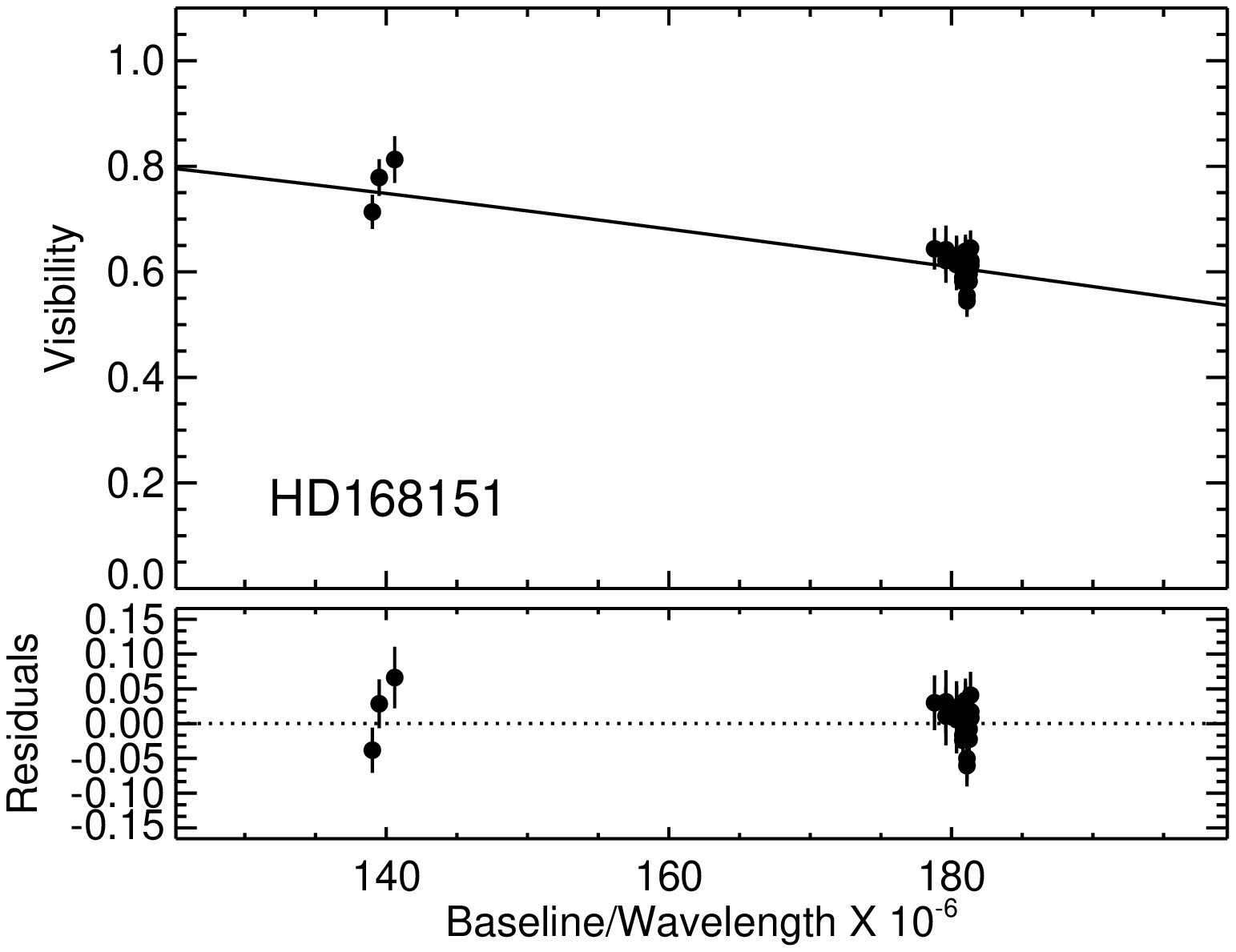, width=0.5\linewidth, clip=} &
\epsfig{file=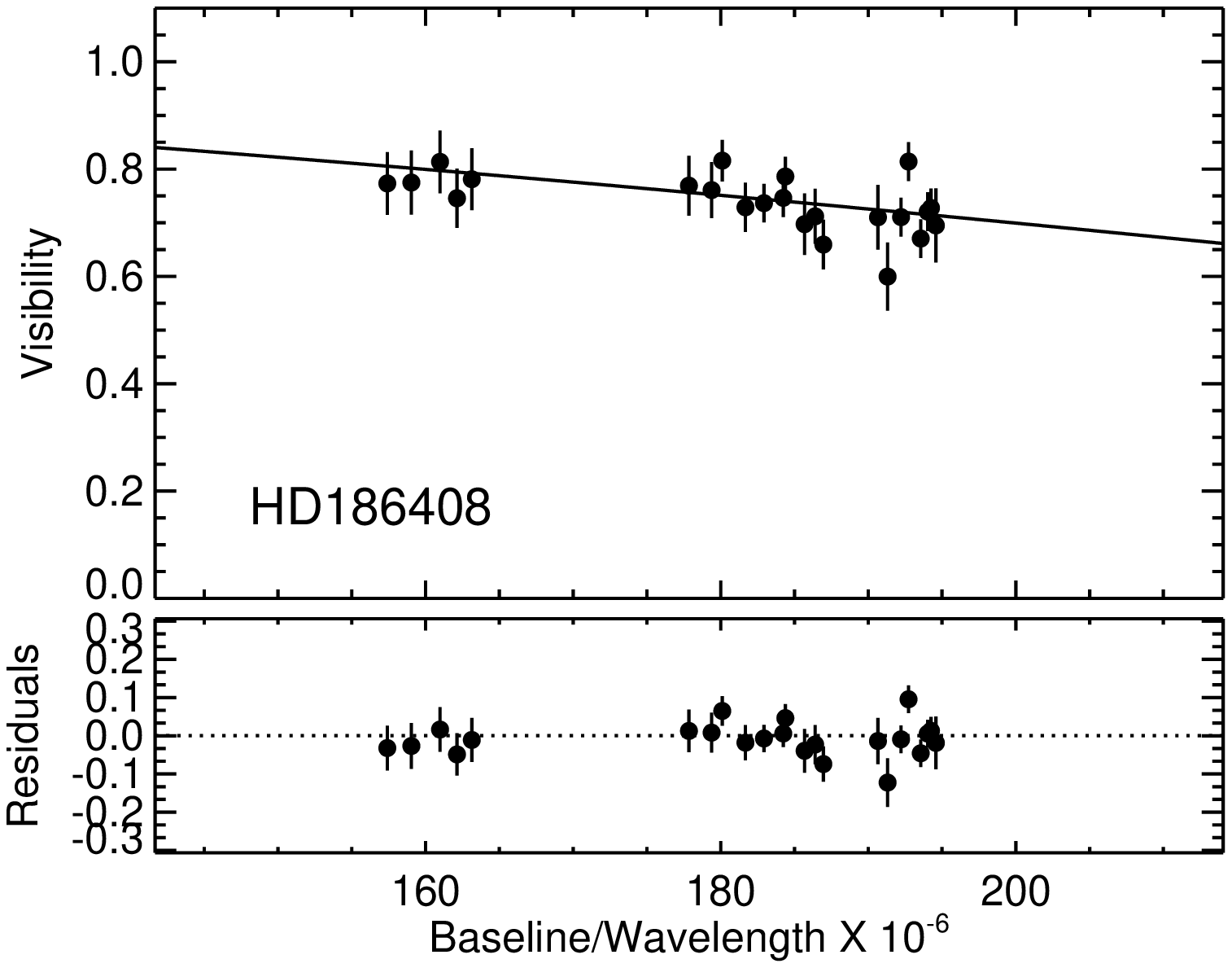, width=0.5\linewidth, clip=} \\
\epsfig{file=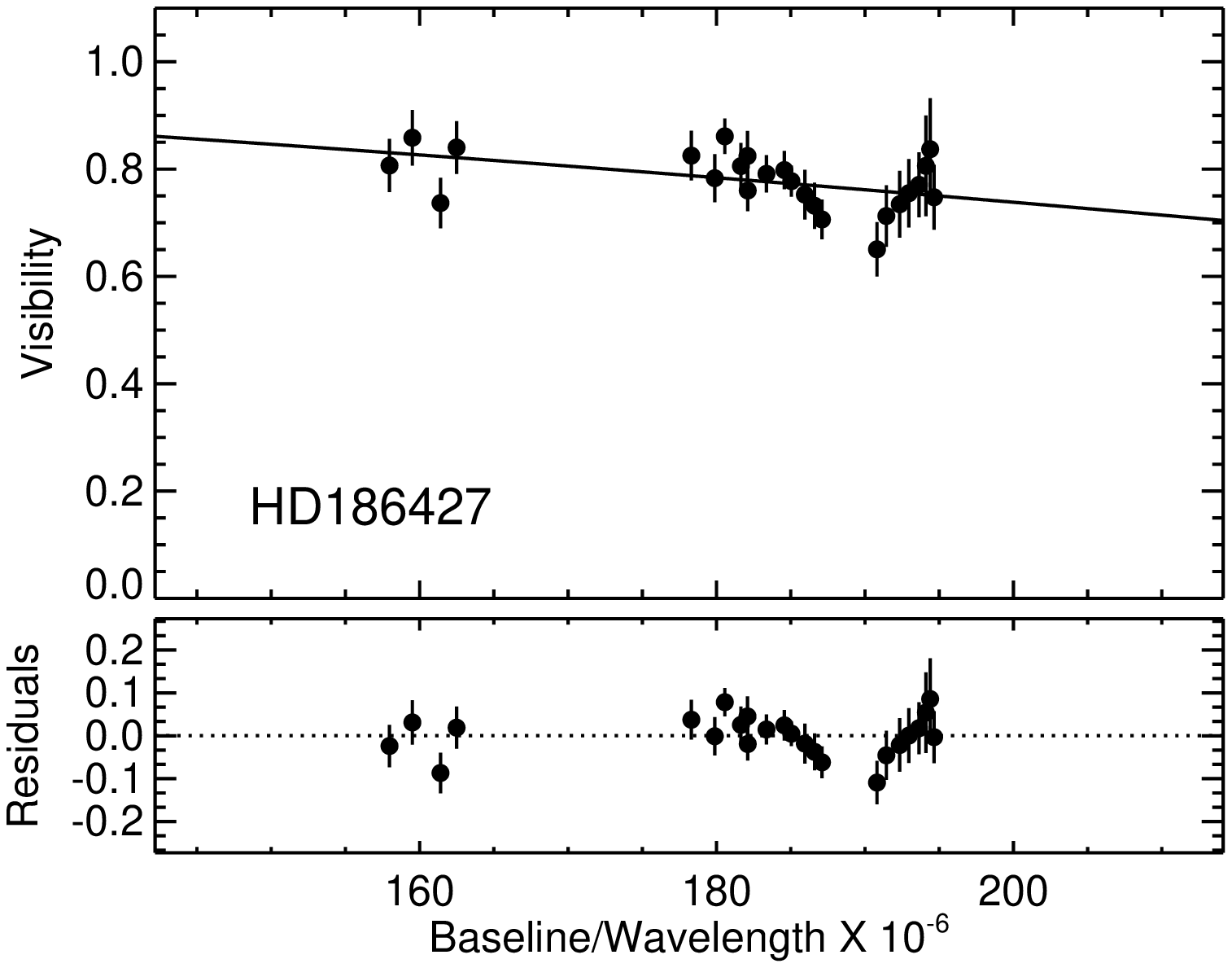, width=0.5\linewidth, clip=} &
\epsfig{file=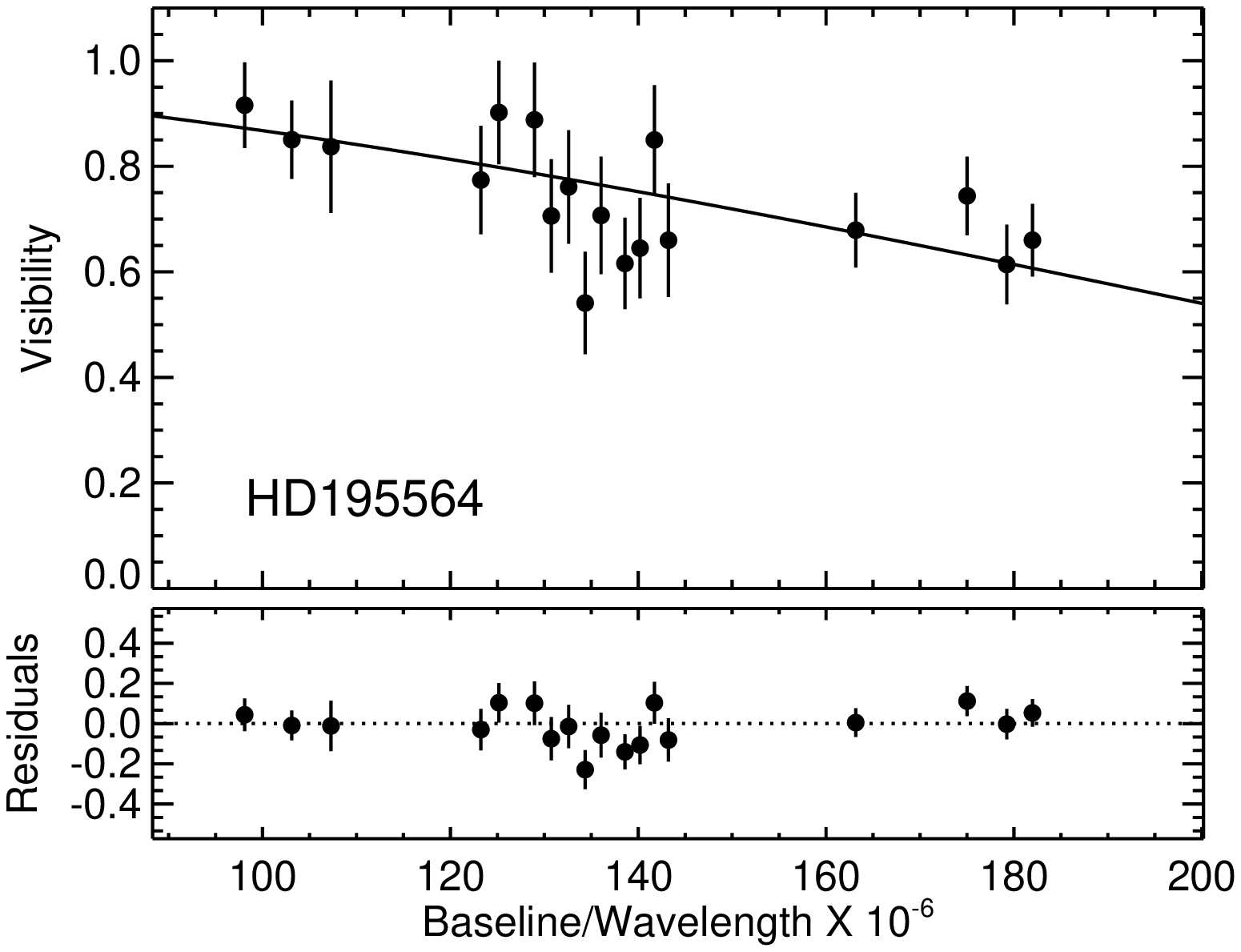, width=0.5\linewidth, clip=} 
 \end{tabular}
 \caption[Angular Diameters] {Calibrated observations plotted with the limb-darkened angular diameter fit for each star. See Section~\ref{sec:observations_chara} for details.}
 \label{fig:diameters3}
 \end{figure}

\begin{figure}										
\centering
\begin{tabular}{cc}
\epsfig{file=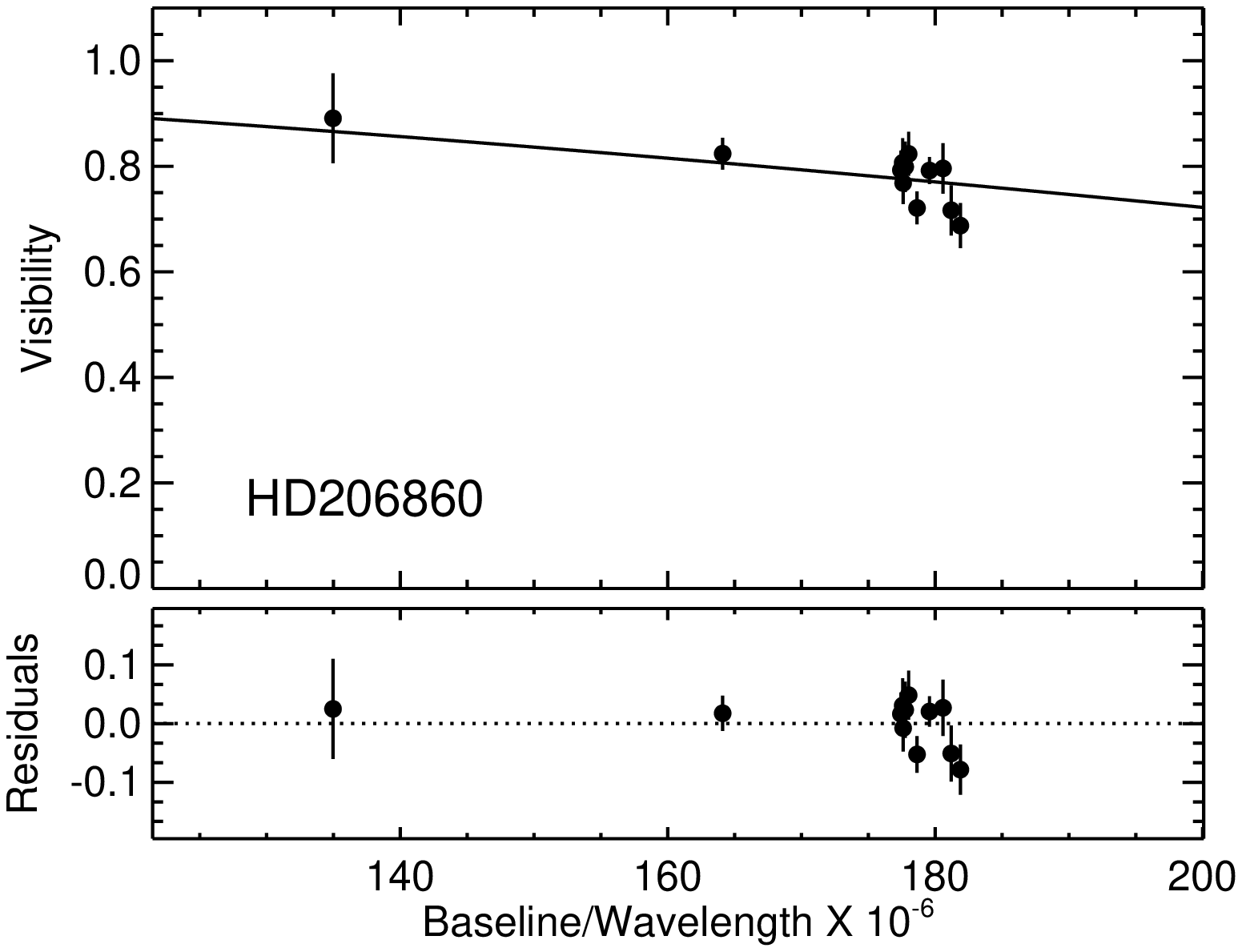, width=0.5\linewidth, clip=} &
\epsfig{file=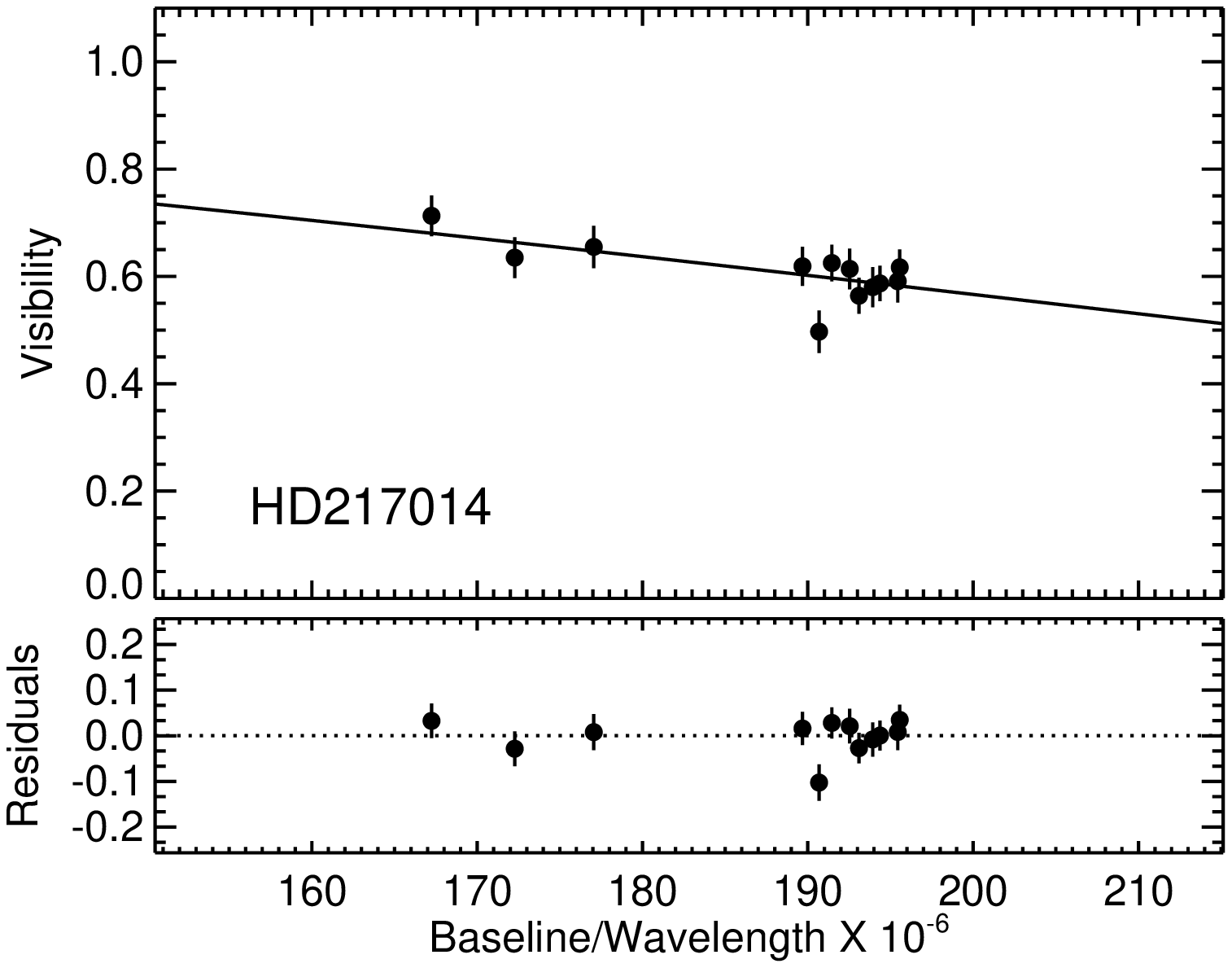, width=0.5\linewidth, clip=} \\
\epsfig{file=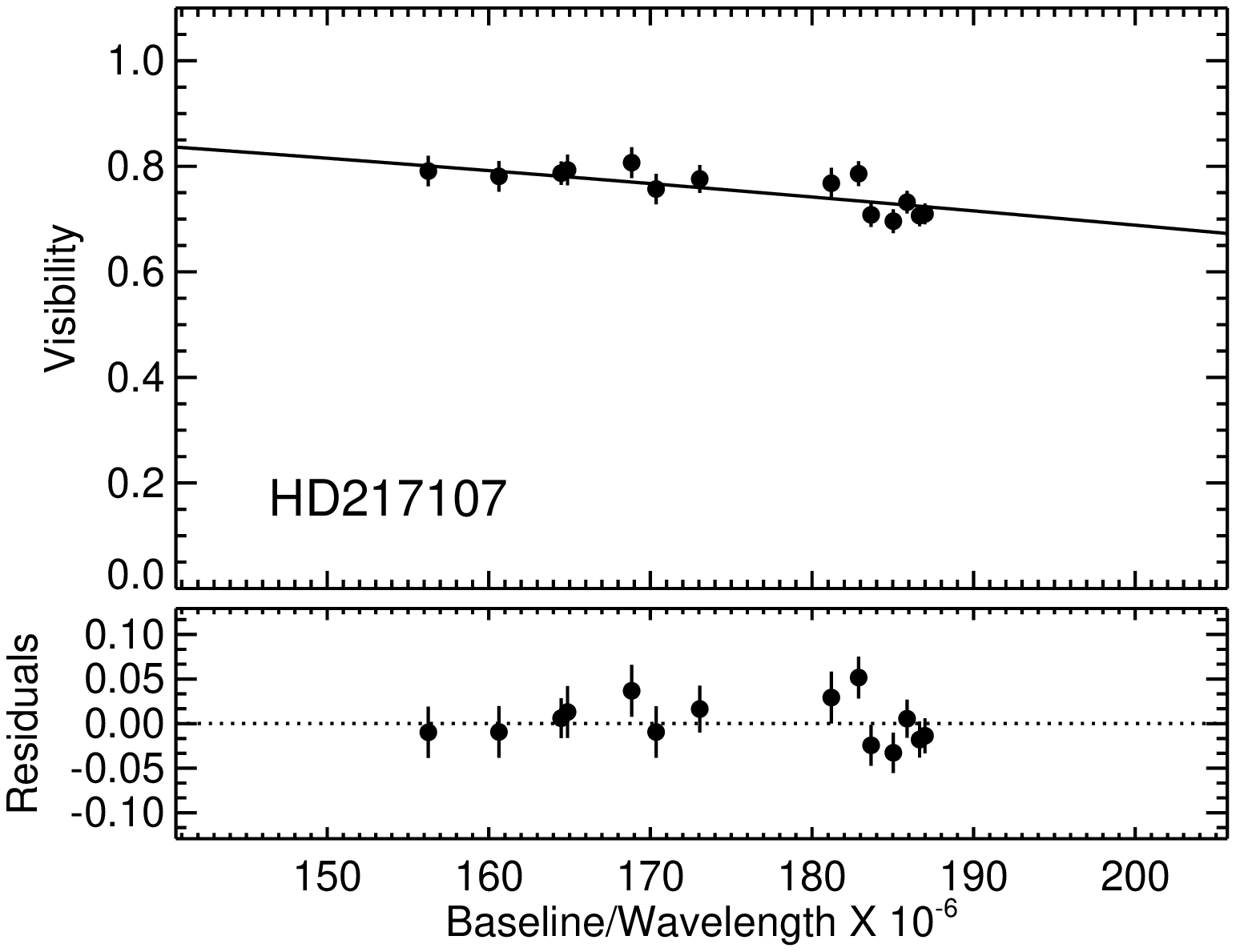, width=0.5\linewidth, clip=} &
\epsfig{file=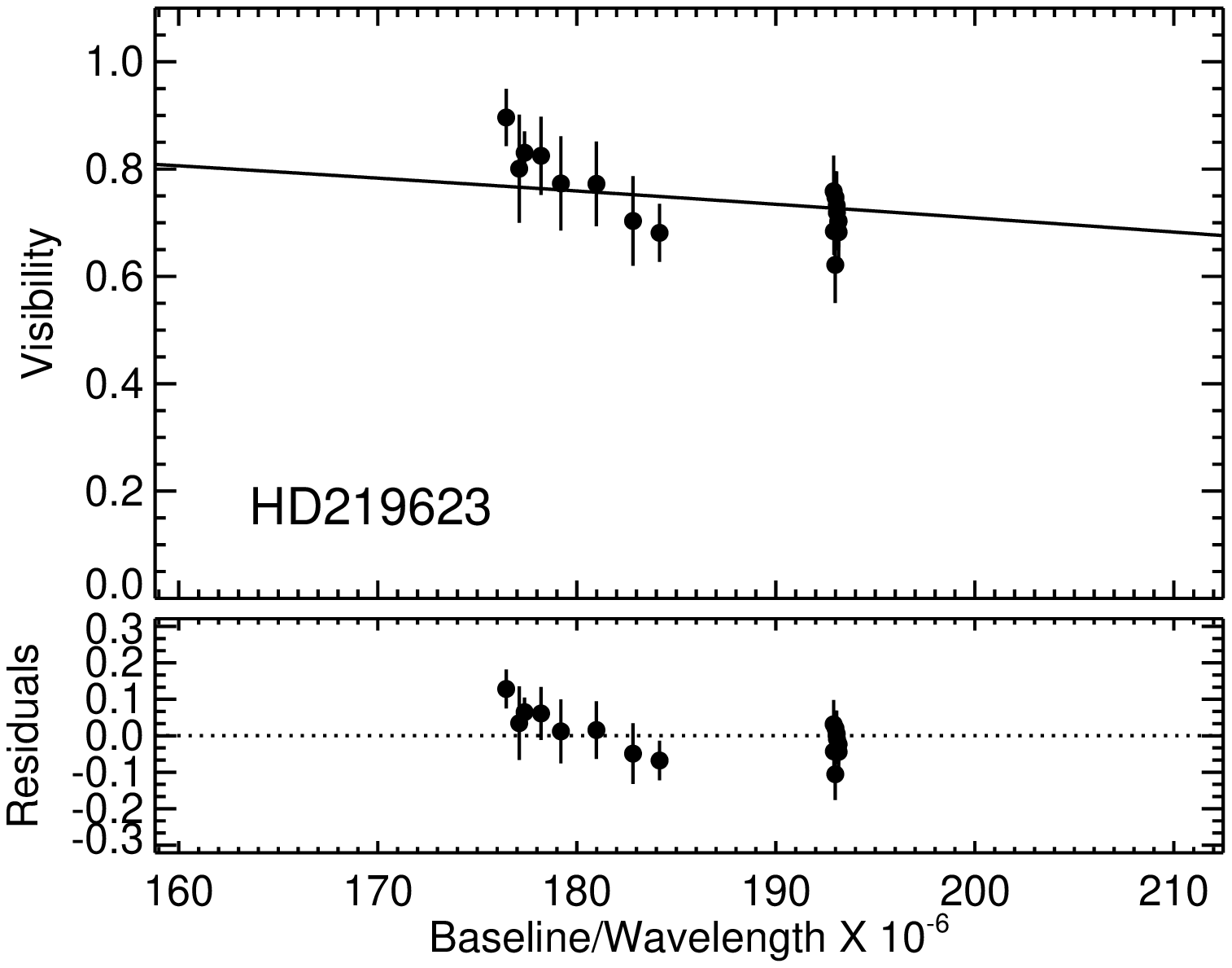, width=0.5\linewidth, clip=} \\
\epsfig{file=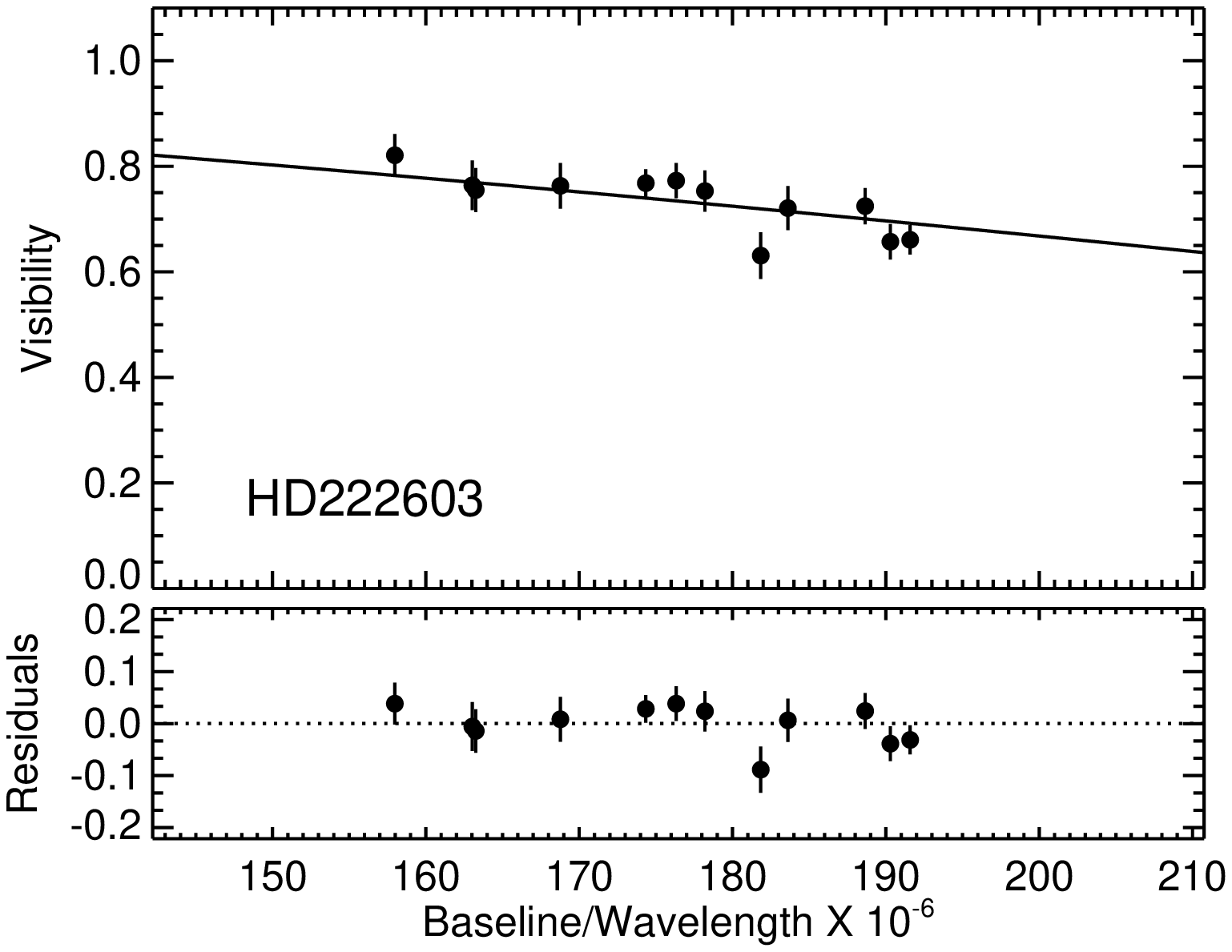, width=0.5\linewidth, clip=} 
 \end{tabular}
  \caption[Angular Diameters] {Calibrated observations plotted with the limb-darkened angular diameter fit for each star. See Section~\ref{sec:observations_chara} for details.}
  \label{fig:diameters4}
  \end{figure}

A table of the new angular diameters of the target stars can be found in Table~\ref{tab:diameters}. In Figures~\ref{fig:diameters1}, \ref{fig:diameters2}, \ref{fig:diameters3}, and \ref{fig:diameters4} we show the data and limb-darkened diameter fit for each star.  

\begin{figure}										
\centering
\epsfig{file=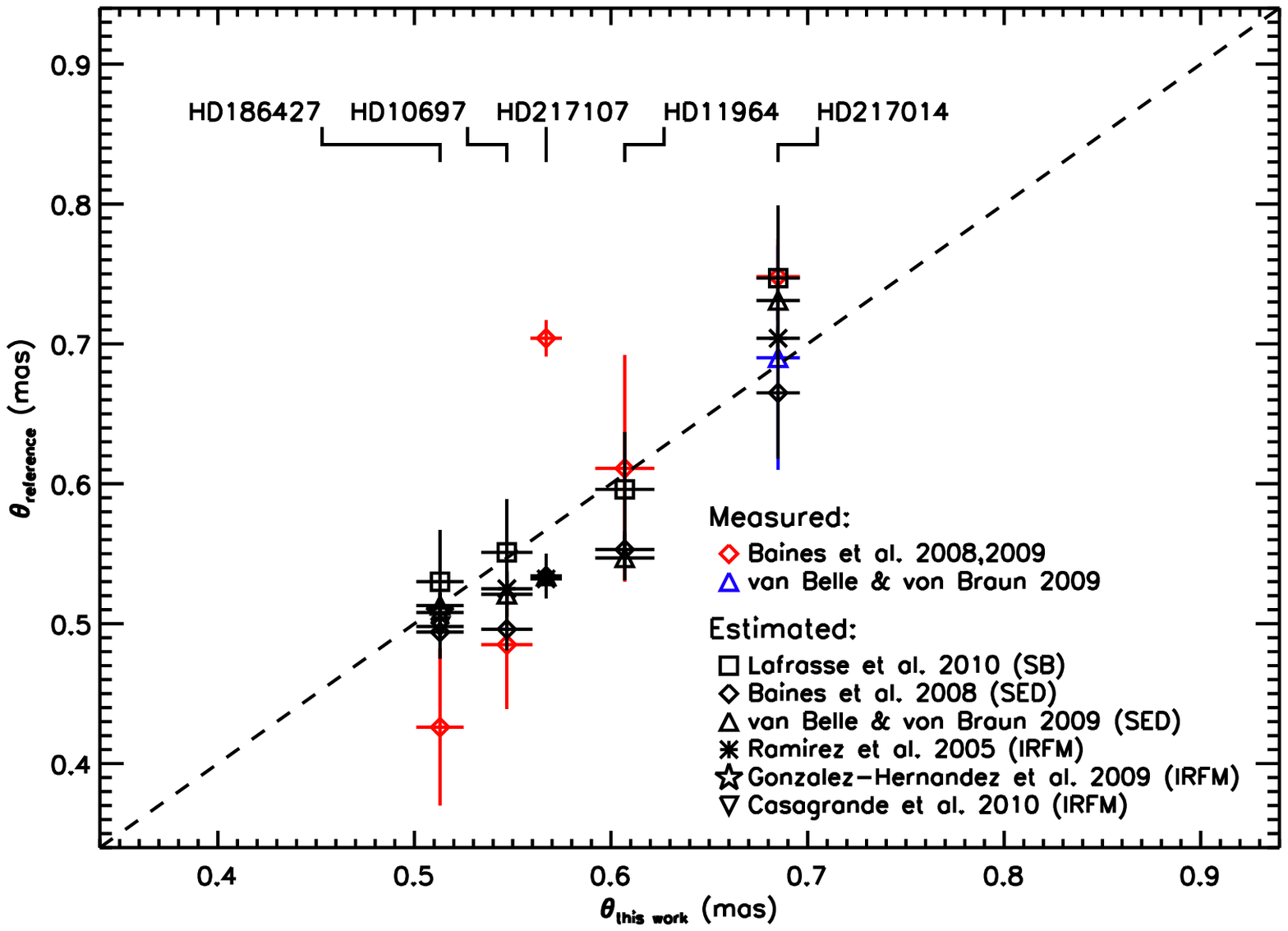, width=0.55\linewidth, clip=} 
 \caption[Comparing Diameters] {New angular diameter measurements of exoplanet host stars compared to previously published measurements from \citet{bai08}, \citet{bai09}, and \citet{van09}.  We also show the agreement with indirect diameter determinations using the surface brightness (SB) relation \citep{laf10}, spectral energy distribution fitting (SED) \citep{bai08, bai09, van09}, and the infrared flux method (IRFM) \citep{ram05, gon09, cas10}. Each of the four objects are identified with a vertical marker at the top end of the plot. The dashed line indicates a 1:1 relation.  See legend within plot and Section~\ref{sec:observations_chara} for details.}
  \label{fig:Compare_EHS_diameters_new}
  \end{figure}

Of the 23 angular diameters we present in this paper, we measured the diameters of five stars known to host exoplanets: HD~10697, HD~11964, HD~186427, HD~217014, and HD~217107.  Each of these stars have directly measured diameters in the literature from previous works, although with the exception of HD~217014, the previously published values have large errors (see \citealt{bai08, bai09, van09}).  Numerous values for indirectly derived angular diameters are cited for these stars as well, spawning from the application of the infrared flux method (IRFM), spectral energy distribution (SED) fitting, and surface-brightness (SB) relations \citep{ram05, cas10, gon09, van08, laf10}.  Our new measurements of the 5 exoplanet host star diameters are compared to the various literature values in Figure~\ref{fig:Compare_EHS_diameters_new}.  Figure~\ref{fig:Compare_EHS_diameters_new} shows that the SB technique ({\it squares}; \citealt{laf10}) provides the best agreement with our directly measured diameters, where $\theta_{\rm this~work}/\theta_{\rm SB} = 1.028 \pm 0.047$ is the average and standard deviation of the two methods. This is similar agreement of angular diameters measured in DT2 compared to values by other interferometers of $\theta_{\rm DT2}/\theta_{\rm Reference} = 1.008$. 


\subsection{Angular Diameters in the Literature and Stars with Multiple Measurements: The Anthology}

\label{sec:observations_literature_averages}

All stars with published angular diameters are listed in Table~\ref{tab:diameters_literature}, which includes 94 measurements from 24 papers.  Like DT2, this collection only admits stars with diameter errors $<5$\%.  Each star's respective state of evolution is also considered, and we filter the results to stars on or near the main-sequence stars (luminosity classes V or IV). There are several stars meeting these requirements that have multiple measurements, and we mark them as such in Table~\ref{tab:diameters_literature}, reducing the total count from 94 down to 71 unique sources.  In the bottom partition of Table~\ref{tab:diameters_literature}, we list the weighted mean of these values for each of these sources with multiple measurements. These stars with measurements from multiple sources agree by $<1$\% on average, with the exception of two cases: HD~146233 and HD~185395.  The reason for the disagreement between these two measurements can only be associated to errors in calibration, and thus these data are omitted in the remainder of the analysis. 

We do not include data for the rapidly rotating early-type stars observed by \citet{mon07,van01} (Altair; $\alpha$~Aql; HR~7557; HD~187642: A7~Vn), \citet{zha09, vanbelle06} (Alderamin; $\alpha$~Cep; HR~8162; HD~203280: A8~Vn), \citet{zha09} (Rasalhague; $\alpha$~Oph; HR~6556; HD~159561: A5~IVnn), \citet{che11} (Caph; $\beta$~Cas; HR~21, HD~432: F2~III), and \citet{che11, mca05} (Regulus; $\alpha$~Leo; HR~3982; HD~87901: B8~IVn) in Table~\ref{tab:diameters_literature}.  Due to their high rotational velocities observed at close to break-up speeds, observations of stars such as these show polar to equatorial temperature gradients on the order of several thousands of Kelvin.  These characteristics make the stars unfavorable as calibrators for the relationships we derive in this paper linking color to effective temperature.  

Finally, we note that we do not repeat the information in Table~\ref{tab:diameters_literature} for the low-mass K- and M- dwarfs studied in DT2. That selection consists of 33 stars, and their stellar properties are collected in a manner identical to the one followed here. Inclusion of the low-mass stars in DT2 leads to a total of 125 main-sequence stars studied with interferometry (33 from DT2, 69 from literature + 23 from this work that are new).


\subsection{Stellar Radii, Effective Temperatures, and Luminosities}     
\label{sec:stellar_params}

Each measurement of the stellar angular diameter is converted to a linear radius using {\it Hipparcos} distances from \citet{van07}.  Errors in distance and interferometrically measured angular diameter are propagated into the uncertainty of the linear radius, however due to the close proximity of the targets to the Sun, the error in angular diameter is the dominant source of error, not the distance.

We present new measurements of the stellar bolometric flux $F_{\rm BOL}$ for all stars with interferometric measurements listed in Table~\ref{tab:diameters_literature}.  The technique is described in detail in \citet{van08}, and is the same tool we employed in several previous works (for example, see \citealt{von11a, von11b, boy12a, boy12b}).  This approach involves collecting all broad-band photometric measurements available in the literature and fitting an observed spectral template from the \citet{pic98} spectral atlas, essentially resulting in a model-independent bolometric flux for each star.  

We have further expanded upon this technique by adding the spectrophotometric data found in the catalogs of 
\citet{Burnashev1985BCrAO..66..152B}, 
\citet{Kharitonov1988scsb.book.....K},
\citet{Alekseeva1996BaltA...5..603A,Alekseeva1997BaltA...6..481A}, and 
\citet{Glushneva1998yCat.3207....0G,Glushneva1998yCat.3208....0G}.
Once an initial $F_{\rm BOL}$ fit was derived using the established technique, spectrophotometry from these catalogs were included in a second SED fit, which typically resulted in an improvement in the formal error for $F_{\rm BOL}$ dropping from $\sim 0.56$\% to $\sim 0.14$\% for 61 of our stars present in these catalogs.  This iterative approach allowed us to screen for outlying spectrophotometric data that did not agree with the photometry; the multiple spectrophotometric data sets permitted a further check against each other for those stars present in multiple catalogs. {\it Note that only statistical uncertainties are taken into account, assuming that photometry from different sources have uncorrelated error bars.}  Although our SED fitting code has the option to fit the data for reddening, we fixed $A_{\rm V}=0$ for these stars, given their distances were all $d < 40$pc. For each star, Table~\ref{tab:monsterPhot} lists the input photometry and corresponding reference. The results and description of the iterative SED fitting routine are in Table~\ref{tab:fbols_table}. 	

The bolometric flux is then used to calculate the temperature of the star through the Stephan-Boltzmann equation:

 \begin{equation}
 T_{\rm eff} = 2341 (F_{\rm BOL}/ \theta^2)^{0.25}
 \end{equation} 
 
 \noindent where the units for $F_{\rm BOL}$ are in $10^{-8}$~erg/s/cm$^2$ and the angular diameter $\theta$ is the interferometrically measured limb-darkened angular diameter in units of milli-arcseconds.  The stellar absolute luminosity is also calculated from the bolometric flux and {\it Hipparcos} distance.  These values are tabulated in Table~\ref{tab:diameters_literature}, which includes properties for all new stars presented in this work (Section~\ref{sec:observations_chara}), as well as the collection of literature stars described in Section~\ref{sec:observations_literature_averages}.  

\begin{figure}										
\centering
\epsfig{file=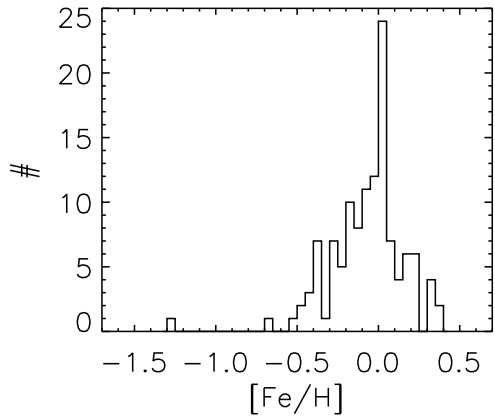, width=0.5\linewidth, clip=} 
 \caption[FeH Histogram] {Histogram of metallicities for the stars with interferometrically determined radii discussed in this work and presented in Table~\ref{tab:diameters_literature}. See Section~\ref{sec:discussion} for details.}
  \label{fig:FeH_histogram}
  \end{figure}

No single publication has metallicity estimates for all the stars in the sample, so instead we use metallicities gathered from the \citet{and11} catalog, where the values they quote are averages from numerous available references. The 4 stars that have no metallicity data are HD~56537, HD~213558, HD~218396, and HD~222603, as noted in Table~\ref{tab:diameters_literature}\footnote{Because these stars are A-type stars and are likely to have solar abundances, we assign them a [Fe/H]$=0$ when constructing the color - temperature relations (Section~\ref{sec:discussion}).}.  A histogram showing the distribution of the stellar metallicities is plotted in Figure~\ref{fig:FeH_histogram}.  Figure~\ref{fig:FeH_histogram} shows that the metallicity distribution of the stars is fairly evenly distributed around $-0.5 <$~[Fe/H]~$< 0.4$, with a strong peak for stars with solar metallicity. 

In Figure~\ref{fig:Lumin_VS_Teff_VS_Radius_color} and Figure~\ref{fig:Teff_VS_Radius_color} we show H-R diagrams on the temperature - luminosity and temperature-radius planes for all the stars in Table~\ref{tab:diameters_literature} and the stars in table~7 of DT2.  In these Figures, the color of the respective data point reflects the metallicity [Fe/H] of the star, ranging from $-1.26$ to $+0.38$~dex, and the size of the respective data point reflects the linear radius $R$ ranging from 0.1869 to 4.517~R$_{\odot}$.  Temperatures range from 3104 to 9711~K and luminosities range from 0.00338 to 58.457~L$_{\odot}$. 
A representative view of main sequence stellar properties is summarized in Table~\ref{tab:spectral_types}, showing the spectral type, number of stars $n$, mean color index, and mean effective temperature of each spectral type for the stars in Table~\ref{tab:diameters_literature}.

Figure~\ref{fig:Lumin_VS_Teff_multiplot} marks the accomplishments of our work in supplying fundamental measurements to main sequence stars over the past few years. Each panel in Figure~\ref{fig:Lumin_VS_Teff_multiplot} shows measurements plotted as black open circles, dubbed as {\it other}.  These data are published measurements from works other than those included in DT1, DT2, and DT3 (this work).  The {\it other} measurements also include stars in DT1, DT2, and DT3 that have multiple measurements, and thus are not unique contributions to the ensemble of data (i.e., stars marked with a $^{\dagger}$ in Table~\ref{tab:diameters_literature} and a $^{\dagger}$ or $^{\dagger\dagger}$ in table~7 of DT2).  The descending panels add the contributions of DT1, DT2, and DT3, indicated as red, green and blue points within the plot, respectively. A breakdown of the number of stars in each category are as follows.  The {\it other} category totals 52 stars.  With the additional measurements presented in this work ($n=23$), our contributions have more than doubled the number of existing main-sequence diameter measurements, yielding a total of 75 unique sources.


\subsection{Estimated Stellar Masses and Ages}
\label{sec:evolution}

The sample of stars with interferometric measurements represents the largest (in linear size, inversely proportional to distance) and brightest (inversely proportional to the square of the distance) population of stars in the local neighborhood.  Once we can determine to great accuracy the fundamental properties of these nearby stars, the knowledge may then be used to extend to much broader applications.  For stars more massive than $\sim 0.8$~M$_{\odot}$, their physical properties are likely to have been affected by stellar evolution as they have lived long enough to 
display observable characteristics marking their journey off the zero age main-sequence. 

We derive ages and masses for stars in Table~\ref{tab:diameters_literature} by fitting the measured radii and temperatures to the Yonsei-Yale ($Y^2$) stellar isochrones \citep{dem04, yi03,kim02,yi01}. 
Isochrones are generated in increments of 0.1~Gyr steps for each star's metallicity [Fe/H] (Table~\ref{tab:diameters_literature}), assuming an alpha-element enrichment of [$\alpha$/Fe]$=0$, acceptable for stars with iron abundances close to solar.  Errors in the age and mass are dependent on the measurement errors in radii and temperature but also in metallicity.  However, metallicities for the stars in our sample are averages from numerous available references (\citealt{and11}; see Section~\ref{sec:stellar_params}), and thus do not come with uncertainties. Simply assigning a characteristic error on this average metallicity is also not justified, because the stars cover a broad range in spectral type, and metallicities of solar type stars are typically determined to greater accuracy than the stars on the hotter and cooler ends of the sample.  Due to the complexity of this aspect, we refrain from quoting errors in the isochrone ages and masses.  Representative uncertainty in age and mass can be estimated given a solar type star ($T_{\rm eff} = 5778$~K and $R = 1$~R$_{\odot}$, assuming a very conservative 2\% and 4\% error in $T_{\rm eff}$ and $R$, respectively), are estimated to be $\pm 5$~Gyr, and 5\% in mass. These ages become less reliable for the lowest luminosity stars, as the sensitivity to age from isochrone fitting is minimal.  This ultimately leads to unrealistic ages greater than the age of the universe, and thus this region should be regarded with special caution. On the other hand, the ages and masses of the earlier-type stars will be determined to better precision (uncertainty of 20\% and 1\%, in age and mass, respectively.)

Figures~\ref{fig:Mass_VS_Radius_color} and \ref{fig:Mass_VS_Teff_color} show the data on the mass - radius and mass - temperature planes, where again the color of the data point reflects the metallicity of the star, and the size of the data point reflects the linear radius\footnote{As opposed to isochrone fitting, masses for the low-mass stars studied in DT2 are found using the empirically-based mass - luminosity relation from \citet{hen93} (see text in DT2 for details).}.  Inspection of the Figures~\ref{fig:Mass_VS_Radius_color} and \ref{fig:Mass_VS_Teff_color} clearly show that for more massive stars, stellar evolution has broadened the correlation between these parameters with stellar age. 

On the mass - luminosity plane however, broadening due to evolution is not observed (e.g., see \citealt{boh89}, Chapter 9.6): the data in the top panel of Figure~\ref{fig:Mass_VS_Lumin_VS_Radius_log_color} shows the stellar mass versus luminosity, where the size of each data point reflects the linear size of the corresponding star.  The bottom panel plots the data without radius information (and thus making the data points smaller), to illustrate more clearly that only the stellar metallicity is a contributing factor in the correlation between the stellar mass and luminosity.  Note that the masses for the low-mass stars were derived using empirically-based mass-luminosity relations (as described in DT2), which are currently independent of metallicity, whereas masses for the higher mass stars described here were found by isochrone fitting, with metallicity as a valid input parameter.


\section{Color - Temperature Relations}
\label{sec:discussion}

We use the full range of interferometrically characterized stars to determine relations linking color index to effective temperature. This sample consists of luminosity class V and IV stars, ranging from spectral types A0 to M4, having temperatures of $\sim 3100$ to $10,000$~K, and metallicities of $-0.5 <$~[Fe/H]~$< 0.4$. The anthology of stellar parameters for the earlier-type stars is presented in Table~\ref{tab:diameters_literature}.  Data for the later-type stars are taken from DT2 \citep{boy12b}.  

Photometry from various sources was collected to derive the color-temperature relations. There are a total of 125 stars, however some have no measured magnitudes for some of the photometric bandpasses we use.  All of the sources have Johnson $B$ and $V$ magnitudes.  Near-infrared colors from {\it 2MASS} in the $J, H$ and $K$ bands are saturated and unreliable due to the fact that these stars are quite bright.  Therefore, alternative sources for $JHK$ measurements when possible, keeping to the bandpass of the Johnson system. The cases where alternate Johnson $JHK$ magnitudes are not available, {\it 2MASS} $JHK$ colors are used, and we discuss the implications of this in Section~\ref{sec:discussion_2mass}.  For the stars having only {\it 2MASS}~$JHK$ magnitudes, we convert them to the Johnson system. This is done by combining the transformations in \citet{car01} for {\it 2MASS} to Bessell \& Brett with the transformations in \citet{bes88} for Bessell \& Brett to Johnson\footnote{We use the updated \citet{car01} transformations available at http://www.astro.caltech.edu/\~jmc/2mass/v3/transformations/}.  
Table~\ref{tab:phot_forfits} designates the magnitudes that use the transformation with a superscript $c$.  

Where available, we collect $R$ and $I$ magnitudes from the systems of Johnson $R_J, I_J$ (e.g., \citealt{joh66}), Cousins $R_C, I_C$ (e.g., \citealt{cou80}), and Kron $R_K, I_K$ (e.g., \citealt{kro57}).  The most prevalent under sampling of photometric data is within the Cousins system and the Kron system, where of the 125 stars, only 34 and 64 stars have such measurements (for Cousins and Kron, respectively).

Magnitudes from the All-Sky Release Source Catalog from the WISE mission \citep{wri10} are available for most stars with the W4 filter ($22.1~\mu$m), as it saturates on stars brighter than W4$=-0.4$ mags.  Approximately half of the stars in our sample have unsaturated WISE W3 ($11.6~\mu$m) magnitudes, where the saturation limit is for stars brighter than W3$=3.8$~mags.  The WISE W1 and W2 systems have much fainter magnitude limits, and are completely saturated for all stars in this sample\footnote{http://wise2.ipac.caltech.edu/docs/release/allsky/expsup/sec6\_3d.html}. 

Synthetic Sloan $g,r,i,z$ magnitudes are also available for the majority of stars through the works of \citet{ofe08, pic10}.  Although these synthetic magnitudes are carefully calibrated, we caution that some of the calculations rely on measurements from the {\it 2MASS} catalog, that are saturated for most stars in this sample (as mentioned above).  For our sample of stars, we find no statistically significant differences in the published magnitudes from the two references (the accuracies of these synthetic magnitudes are not tested here), and thus we chose to use the average of the \citet{ofe08} and \citet{pic10} values when constructing the color - temperature relations.  All magnitudes used in the color - temperature relations are listed in Table~\ref{tab:phot_forfits} for each star.

We use MPFIT, a non-linear, least-squares fitting routine in IDL \citep{mar09} to fit the observed color index to the measured temperatures of the stars in each bandpass. All stars with available photometry in said bandpass  are fit to a $3^{rd}$ order polynomial in the form of:

\begin{equation}
\label{eq:poly3c}
T_{\rm eff} = a_0 + a_1 X + a_2 X^2 + a_3 X^3
\end{equation}

\noindent where the variable $X$ represents the color index and $a_0, a_1, a_2, a_3$ are each solution's coefficients.  In Table~\ref{tab:poly3_coeffs}, we list 33 color-indices and their coefficients derived in this manner. Table~\ref{tab:poly3_coeffs} also lists the number of points used in the fit (where the total number of points will be $<125$ if photometry is not available for stars in some bandpasses), the range in color-index where the relation holds true, and the standard deviation about the fit expressed as a normalized percentage, calculated as Std.Dev. ($(T_{i,\rm Obs} - T_{i,\rm Calc})/T_{i,\rm Obs} \times 100$).  Figures~\ref{fig:relations1}, \ref{fig:relations2}, \ref{fig:relations3}, \ref{fig:relations4}, and \ref{fig:relations5} show the solutions and the data for each of the 33 color indices analyzed.  In the discussions that follow, we comment on solutions using varied approaches in detail.


\subsection{Slippery Solutions and Crummy Colors}
\label{sec:discussion_2mass}

\subsubsection{Further Vetting of the Sample}
\label{sec:discussion_astars_evolution}

We investigate whether a portion of the scatter about the best fit color - temperature relations is a consequence of slight differences among the stars in the sample.  Two possible differences that we consider are (1) distortion of the photosphere caused by rapid rotation, and (2) early post-main sequence evolution.  Interferometrically constructed images of rapidly rotating A-stars such as Altair ($v \sin i = 240$~km~s$^{-1}$) show it to be distinctively oblate ($R_{pole} = 1.63$~R$_{\odot}$ versus $R_{equator} = 2.03$~R$_{\odot}$) with severe gravity darkening ($T_{pole} = 8450$~K versus ($T_{equator} = 6860$~K; \citealt{mon07}).  These effects appear to be common among many early-type stars (see review by \citealt{van12}).  This compromises interferometrically determined temperatures because the measured radius is orientation-dependent and the strong temperature gradients lead to the apparent luminosity being inclination dependent.  For instance, \citet{auf06} calculate that the apparent luminosity of the pole on star Vega is 35\% larger than its bolometric luminosity, because of our pole-on line of sight.  

Fortunately, these complicating effects are primarily restricted to mid-F and hotter stars.  These early-type stars lack a convective zone in their outer atmosphere, and thus the ability to generate a magnetic field that could couple to the stellar wind and magnetically brake the star's rotation.  In Section~\ref{sec:observations_literature_averages} we describe the early-type stars that have been omitted from the {\it Anthology} because of the effects of rapid rotation, as determined from interferometric imaging.  The remaining early-type stars included in the {\it Anthology} may nevertheless have biased radii and apparent luminosities, which could introduce additional scatter into the best fit relations.  To eliminate this possible error source, and thus determine even stricter relations for cooler, Sun-like stars, we redo the analysis omitting these early-type stars.  Specifically, stars hotter than $T_{\rm eff} = 6750$~K are excluded, corresponding to spectral type F3, approximately.  As can be seen in the H-R diagram plotted in Figure~\ref{fig:Lumin_VS_Teff_VS_Radius_color}, there exists a natural break in the sample at this point.  A total of 13 early-type stars are removed for the re-analysis.

We use the same approach of fitting the new subset of data as we did fitting the full sample, described in Section~\ref{sec:discussion}.  The results are plotted in Figures~\ref{fig:relations1}, \ref{fig:relations2}, \ref{fig:relations3}, \ref{fig:relations4}, and \ref{fig:relations5} as a red dash-dot line. For each color -temperature relation, Table~\ref{tab:poly3_coeffs} shows the number of points used in the fit, color range where the fit is applicable, coefficients to each polynomial, and the standard deviation ( each row is marked with a superscript $c$ to indicate the fit was made omitting the early-type stars).  For each color index, we document the maximum difference in temperature predicted using the fits with and without early-type stars in Table~\ref{tab:max_diff_omit_hot}. Carefully inspecting the differences, we find that most the new fits do not deviate more than a few tenths of a percent from the full AFGKM star solution. Deviations larger than a few percent are manifested at the endpoints of the fit - where the fits omitting the early-type stars better represent the data in most cases.  Exceptions are the $(R_J-J)$, $(R_J-H)$, and $(R_J-K)$ color relations, which are subject to poor fitting from lack of sampling on the coolest end of the fits, and use of these three relations in this region should be used with caution. Detailed discussion of the $(B-V)$ - temperature fits follow in Section~\ref{sec:discussion_bmv}.

Regarding the latter possible source of error in the color - temperature relations: the {\it Anthology} is restricted to be \textquoteleft~stars on or near the main-sequence (luminosity classes V or IV)\textquoteright.  However, inspection of the H-R diagram in Figure~\ref{fig:Lumin_VS_Teff_VS_Radius_color} hints that there is moderate girth in the band of the main-sequence for stars greater than a few tenths of a solar luminosity. While these stars are far off from being giants - and we do not claim to re-classify them as such - their less-than-ZAMS surface gravity could lead to a distinctively different temperature scale than the truly qualified ZAMS population.  We do not think that this is a source of error in our analysis for several reasons.  The first clue to this not being an issue is that the sample of low-mass stars (i.e., the KM dwarfs from DT2) do not have any less-than-ZAMS surface gravity interlopers, since they are all low-mass enough to be considered un-evolved over the lifetime of the galaxy.  Regarding the residuals of the color - temperature relations in this region of low-mass stars, we see that the residuals are of comparable magnitude to the higher-mass stars whom are tainted with evolutionary effects.  

We apply a more quantitative approach by inspecting the temperature residuals as a function of surface gravity $\log g$.  For this exercise, we derive $\log g$ by the equation $g = G M R^{-2}$, where $G$ is the gravitational constant, $R$ is the interferometrically measured radius, and $M$ is the mass derived from isochrone fitting. This yields surface gravities for the sample ranging from $\log g = 3.3$ to 5.0, with a mean value of 4.3 and standard deviation of 0.3~dex.  By comparing the fractional residuals of the color - temperature fits with surface gravity, we find no evidence that stars with lower $\log g$'s will bias the color - temperature fits.

Although the luminosity classes IV and V do not differentiate the evolutionary state of the stars very well, as pointed out in Section~\ref{sec:evolution}, we also checked for correlation with luminosity class in the residuals of the color-temperature fits and found none.

\subsubsection{Improvement on $(B-V)$ Relations}
\label{sec:discussion_bmv}

The robustness of the $(B-V)$ - temperature solution suffers from two artifacts: (1) the need for a higher order polynomial to properly model the data and (2) trends in the residuals with respect to metallicity.  Pertaining to the first issue, the residuals in the $(B-V)$ - temperature relation shown in Figure~\ref{fig:relations1} show that the solution using a $3^{rd}$ order polynomial does not model the inflection point in the data ($\sim 0.2 < (B-V) < 0.5$; $\sim 6500 < T_{\rm eff} < 7500$) well, thus yielding temperatures $\sim 5$\% cooler than observed in this range. In fact, the $(B-V)$ - temperature fit omitting the early-type stars produces temperatures 5\% different from the $3^{rd}$ order polynomial solution (Table~\ref{tab:max_diff_omit_hot}, Figure~\ref{fig:relations1}). Thus, in order to model the full AFGKM sample correctly, we use the approach in DT1 and apply a $6^{th}$-order polynomial in order to remove this artifact. The form of this equation is expressed as: 

\begin{eqnarray}
\label{eq:poly6c}
T_{\rm eff} & = &	a_0 + a_1 (B-V) + a_2 (B-V)^2 +  a_3 (B-V)^3 + \nonumber \\
			&	&	a_4 (B-V)^4 + a_5 (B-V)^5 + a_6 (B-V)^6
       \end{eqnarray}

\noindent where the coefficients are:

\begin{eqnarray}
      a_0	& =	&  9552 \pm   19\nonumber \\
      a_1 & = 	&       -17443 \pm  350 \nonumber	\\
      a_2 & = 	&         44350 \pm 1762	\nonumber \\
      a_3 & = 	&       -68940 \pm 3658	\nonumber \\
      a_4 & = 	&       57338 \pm 3692	\nonumber \\
      a_5 & = 	&     -24072 \pm 1793	\nonumber \\
      a_6 & = 	&    4009 \pm  334	\nonumber .
       \end{eqnarray}

\noindent The standard deviation of the fit for Equation~\ref{eq:poly6c} is 3.1\%, and data are plotted along with the solution and residuals in the top panel of Figure~\ref{fig:Temp_VS_BmV_p6}. The solution is also displayed in Figure~\ref{fig:relations1} as a dashed line.  Although this standard deviation is only slightly smaller than the solution using the $3^{rd}$ order polynomial fit (3.3\%), we note that it maps the region of concern containing the inflection point more accurately. This solution using this $6^{th}$-order polynomial fitting the whole sample produces temperatures identical to the solution found by eliminating the early-type stars from the fitting in Section~\ref{sec:discussion_astars_evolution}.    

With the exception of the $(B-V)$ - temperature relation, the residuals in each color-temperature relation, plotted in Figures~\ref{fig:relations1}, \ref{fig:relations2}, \ref{fig:relations3}, \ref{fig:relations4}, \ref{fig:relations5}, and \ref{fig:relations6}, reveal no pattern with respect to the metallicity of the star.  Attempts to fit functions dependent on color and metallicity, such as the one developed for the low-mass stars in DT2, were shown not to improve the fits.  We note that even within the results of DT2, only mild dependence on metallicity was detected in the multi-variable, color - metallicity - temperature fits, prevalent only for the latest-type stars ($T_{\rm eff} < 4000$~K, see section~4.1 in DT2).  The likely causes for this lack of apparent connection in the data on the color - metallicity - temperature plane are that (1) large photometric errors are prominent throughout the data set and (2) significant errors in the metallicity (especially systematics) may exist.  This attribute is complicated further when combined with the fact that the range of metallicity in the sample (roughly $-0.5 <$~[Fe/H]~$< 0.4$, with two low metallicity outliers) might not be broad enough to detect such an effect observationally. 

The metallicity dependence on the $(B-V)$ colors is strongest of all color-indices analyzed. For instance, both the $3^{rd}$ and $6^{th}$ order $(B-V)$ - temperature solutions show residuals correlated with the stellar metallicity (Figures~\ref{fig:relations1} and \ref{fig:Temp_VS_BmV_p6}, where stars with higher than solar metallicity have higher temperatures than those with lower metallicities at the same $(B-V)$ color.  For this reason, we construct a $2^{nd}$ order multi-variable function dependent on both the $(B-V)$ color index and metallicity [Fe/H] expressed as:

 \begin{eqnarray}
 \label{eq:poly2cm}
 T_{\rm eff} & = &	a_0 + a_1 (B-V) + a_2 (B-V)^2 + \nonumber \\
 			&	&	a_3 [Fe/H] +  a_4 [Fe/H]^2 + \nonumber \\ 
 			&	&	a_5 (B-V) [Fe/H] .
        \end{eqnarray}

We use the sample that omits early-type stars (easily modelled by a lower-order polynomial) in order to remove effects due to metallicity.  The fit produces the coefficients:

\begin{eqnarray}
      a_0	& =	&	 7978 \pm   16	\nonumber \\
      a_1 & = 	&  -3811 \pm   36 \nonumber	\\
      a_2 & = 	&  636 \pm   17	\nonumber \\
      a_3 & = 	&      479 \pm   26	\nonumber \\
      a_4 & = 	&      -126  \pm  19	\nonumber \\
      a_5 & = 	&       -150  \pm  22	\nonumber 
       \end{eqnarray}

\noindent The fit uses a total of $n=111$ points, is valid for $0.32 < (B-V) < 1.73$, and gives a standard deviation about the fit of $\sigma = 2.6$\%, a value now comparable to the best solutions of the other color-indices analyzed.   We show the data and solution in Figure~\ref{fig:Temp_VS_BmV_p6} plotted for three metallicities [Fe/H]$=-0.25, 0.0$, and $+0.25$ (green, black, and red lines, respectively).  

The addition of metallicity as a variable to model the $(B-V)$ color - metallicity - temperature connection eliminates the pattern of residuals with respect to metallicity (see bottom panels in Figure~\ref{fig:Temp_VS_BmV_p6}). Due to the metallicity range of the data, the relation only holds for $-0.25<$~[Fe/H]~$<+0.25$, where the data are heavily sampled (Figure~\ref{fig:FeH_histogram}). For stars with $(B-V)=0.6$, the calculated temperatures at the high and low metallicity boundaries show a difference of $\sim 350$~K (or $\sim 5$\%), so therefore a necessary correction is needed for accurate conversions of the stellar temperature from $(B-V)$ colors.  

\subsubsection{Infrared Colors}
\label{sec:discussion_ircolors}

As previously mentioned, transformed {\it 2MASS} {\it JHK} magnitudes are used for stars that do not have {\it JHK} magnitudes from an alternate source, and these {\it 2MASS} magnitudes are sketchy due to saturation and have large errors associated with them (e.g., magnitude errors of saturated stars are $\sim 0.2$~mags).  In each color - temperature relation that uses {\it JHK} magnitudes, we perform an additional fit that includes only stars with alternate {\it JHK} magnitudes, omitting all stars with saturated {\it 2MASS} colors.  Filtering the data in this way decreases the available number of points used in each fit in all cases (up to a 20\% drop in sample size).    

These solutions are plotted in Figures~\ref{fig:relations1}, \ref{fig:relations2}, \ref{fig:relations3}, \ref{fig:relations4}, \ref{fig:relations5}, and \ref{fig:relations6} as dotted lines along with the relation using the full data set that includes the transformed but saturated {\it 2MASS} photometry (solid line). The solutions are shown to be almost identical, deviating to hotter temperatures by only few tens of Kelvin for the earliest type stars. The fractional residuals show in the bottom panel of each relation mark the stars having {\it 2MASS} colors with an $\times$.  These data points comprise the majority of stars with fractional deviations greater than 5\% from the solution, especially apparent for stars with temperatures $>6500$~K (early F- and A-type stars)\footnote{These stars are also among the brightest in the sample.}.  The coefficients to the solutions derived omitting any {\it 2MASS} photometry are marked with a superscript {\it d} in Table~\ref{tab:poly3_coeffs}.

The removal of outliers due to suspected bad photometry improves the standard deviation of the fit by 0.3\% to 2.6\% (see Table~\ref{tab:poly3_coeffs}). Since the solutions remain almost identical, we estimate that the standard deviations using the modified data sets reflect the true errors of the relations. 

\subsubsection{Comparison to other works}
\label{sec:discussion_teff_others}

The infrared flux method (IRFM, \citealt{bla79}) is a semi-empirical method of determining stellar effective temperature, for which the results are always tested and/or calibrated with interferometric data.  Here we view from the alternate perspective, and compare our interferometrically derived temperatures to temperatures derived from solutions via the IRFM in the two recent works of \citet{cas10} and \citet{gon09}\footnote{Note that the transformations in \citet{cas10} and \citet{gon09} are two-dimensional, $2^{nd}$-order polynomials, dependent on both the stellar color index and metallicity [Fe/H].}.  Figure~\ref{fig:Teff_VS_Cas10_Gon09_multiplot} shows the results of this comparison, displaying side-by-side the effective temperatures using the \citet{cas10} and \citet{gon09} relations (left and right, respectively). For each color index, each panel displays the fractional deviation in temperature for the stars with available photometry, allowing only the effective color ranges for each IRFM reference.  Each plot also displays the average offset in the deviation of temperature (expressed in \%) as well as the standard deviation of the data, also expressed in \%.  Note that the IRFM temperature scales in both \citet{gon09} and \citet{cas10} are based off the {\it 2MASS} bandpasses. As such, to make the comparison of the visual to infrared colors ($V-J, V-H, V-K$; bottom six panels in Figure~\ref{fig:Teff_VS_Cas10_Gon09_multiplot}), the infrared magnitudes of the {\it Anthology} stars were transformed to the {\it 2MASS} system by the expressions in \citet{bes88} and \citet{car01}. 

Agreement is within a couple percent for both references, where the offsets in the \citet{gon09} temperature scale are $\sim$ half those from the \citet{cas10} scale.  We find that in all cases, the temperatures derived via the IRFM are higher than those presented here  (refer to the $<\Delta T / T>$ value in Figure~\ref{fig:Teff_VS_Cas10_Gon09_multiplot}). 

An effective temperature scale based on the Sloan photometric system was recently evaluated and revised in \citet{pin12}. \citet{pin12} use YREC isochrones in addition to MARCS stellar atmosphere models to derive their temperature relations, adopting a [Fe/H]~$=-0.2$, and isochrone age on 1~Gyr.  We compare the temperatures derived in \citet{pin12} to ours in Figure~\ref{fig:Teff_VS_Pin12_multiplot} for the ($g-r$), ($g-i$), and ($g-z$) color indices\footnote{The temperatures are based on the Sloan $griz$ filters, and do not apply the zero-point shifts described in \citet{pin12} to transcribing the magnitudes to the Kepler Input Catalog $griz$ system.}.  Similar to the temperature comparisons above of the IRFM, we truncate each panel in color and temperature range to only contain data where the \citet{pin12} relations hold (refer to table caption in their table~2). Within each panel is printed the average offset in the deviation of temperature and the standard deviation of the data. We find agreement of the two temperature scales $<2$\%, where \citet{pin12} temperatures are systematically higher than interferometric temperatures. This agreement improves to much less than a percent offset in all color - temperature relations if stars with temperatures $>5100$~K are compared.  On the other hand, if we consider adjusting the temperatures to bring the \citet{pin12} scale (assumed [Fe/H]~$=-0.2$~dex) to the characteristic metallicity of our sample of close to solar (mean [Fe/H]~$=-0.02$~dex, median [Fe/H]~$=-0.01$~dex; in the range that overlaps), the \citet{pin12} model temperatures produce even higher temperature values at a given color index (see their table~3) typically on the order of a few tens of Kelvin. This correction for metallicity would produce a larger offset in temperature, and thus is not a source of the disagreement.

Both the polynomial relations in \citet{cas10} and \citet{gon09} are metallicity dependent, while the \citet{pin12} polynomials take the same form as our own without defining a dependence on metallicity.  In each subpanel for each reference / color index in Figures~\ref{fig:Teff_VS_Cas10_Gon09_multiplot} and \ref{fig:Teff_VS_Pin12_multiplot} we display the residual scatter in comparing our interferometrically determined temperatures to those derived using each polynomial relation.  We find that for every instance, this scatter is equivalent to within a couple tenths of a percent to the scatter of our derived relations (Table~\ref{tab:poly3_coeffs}).  This supports our approach that the inclusion of metallicity as an additional variable in color - temperature relations is not a necessary factor, with the exception of the ($B-V$) - temperature relation.  While this is true based on the sample employed, it must also be pointed out that the metallicity range encompassed by the interferometric sample is relatively limited. Therefore, any metallicity dependence could still show up in other bands, should the metallicity range be larger.

The Spectroscopic Properties of Cool Stars (SPOCS) catalog \citep{val05} presents spectroscopic temperature measurements of 68 stars in common with the interferometric sample collected here.   In the top panel of Figure~\ref{fig:Compare_spocs_feh_diff_multi}, we compare the data sets in the same manner as in Figure~\ref{fig:Teff_VS_Cas10_Gon09_multiplot}, showing excellent agreement with spectroscopic temperatures as well, with only a 1.7\% offset to spectroscopic temperatures preferring higher temperatures compared to interferometric values.  The stars at the hot and cool ends of the the plot hint that a linear trend could arise with an upward slope towards hotter temperatures.  

Figure~\ref{fig:Compare_spocs_feh_diff_multi} also shows the radii published in the SPOCS catalog versus those with direct interferometric measurements presented here.  The radii values for stars in the SPOCS Catalog are computed with the Stephan-Boltzmann law: $R \sim L^{0.5} T^{-2}$.  The calculation uses the spectroscopically derived temperature $T$ and the luminosity $L$, a function of the stellar distance, $V$-band magnitude, and bolometric correction from \citet{van03}.   This comparison is shown in the middle panel of Figure~\ref{fig:Compare_spocs_feh_diff_multi}, where the average deviation in radii is 3.4\%, about double that of the offset in temperature. We find that for stars with radii $<1.3$~R$_{\odot}$, the offset averages $\sim$2\%, whereas most stars larger than this radius are offset in the positive direction, with an average offset of $\sim$5\%. 

The bottom panel in Figure~\ref{fig:Compare_spocs_feh_diff_multi} compares the masses we derive $M_{\rm Iso}$ to those in the isochrone masses in the SPOCS Catalog $M_{\rm Iso,SPOCS}$, which are also derived using the same set of $Y^2$ isochrones. The SPOCS values are derived by fitting their spectroscopically determined effective temperature, metallicity, alpha element enhancement, and the bolometric correction based luminosity. We find an average offset of $-3.9$\%, where the majority of the low-offset outliers lie between 0.9~$M_{\odot}$ and 1.3~$M_{\odot}$.


\section{Conclusion and Summary of Future Prospects}
\label{sec:conclusion}

Using the CHARA Array, we measure the angular diameters of 23 nearby, main-sequence stars, with an average precision of a couple of percent. Five of these stars were previously measured with LBOI, and our new values show an average of 4.3 times improvement in measurement errors, as well as showing consistency through less direct methods of estimating the stellar angular size.  These measurements are added to a collection dubbed as an {\it Angular Diameter Anthology}, which reports a collection of diameter measurements published in the literature until present time (Table~\ref{tab:diameters_literature}). According to our research, the current census totals 125 main-sequence or near main-sequence stars with diameters measured to better than 5\%.

We use the interferometrically measured angular diameter in combination with the star's measured bolometric flux and distance to derive the stellar radius (linear), effective temperature, and absolute luminosity. These absolute quantities are used to derive ages and masses from model isochrones. Using the observed photometric properties of the sample, we are able to build transformations to the stellar effective temperature that are precise to a few percent. The empirical temperatures compared to those derived via models, the IRFM and spectroscopy typically agree within a couple of percent, where the temperatures derived via indirect methods have a tendency to predict higher temperatures compared to those with interferometric observations. 

Currently our group is using this interferometric dataset to develop formula to robustly predict stellar angular sizes using broad-band photometry (e.g. see \citealt{ker08, van99b, bar76}).  Such methods of determining stellar sizes are applicable to the interferometry community in search of the perfect calibrator to observe \citep{bon06, bon11}.  The broader impacts on the astronomical community point to such empirically based calibrations enabling the use of eclipsing binaries as standard candles \citep{south05}.

\begin{figure}										
  \centering
         \epsfig{file=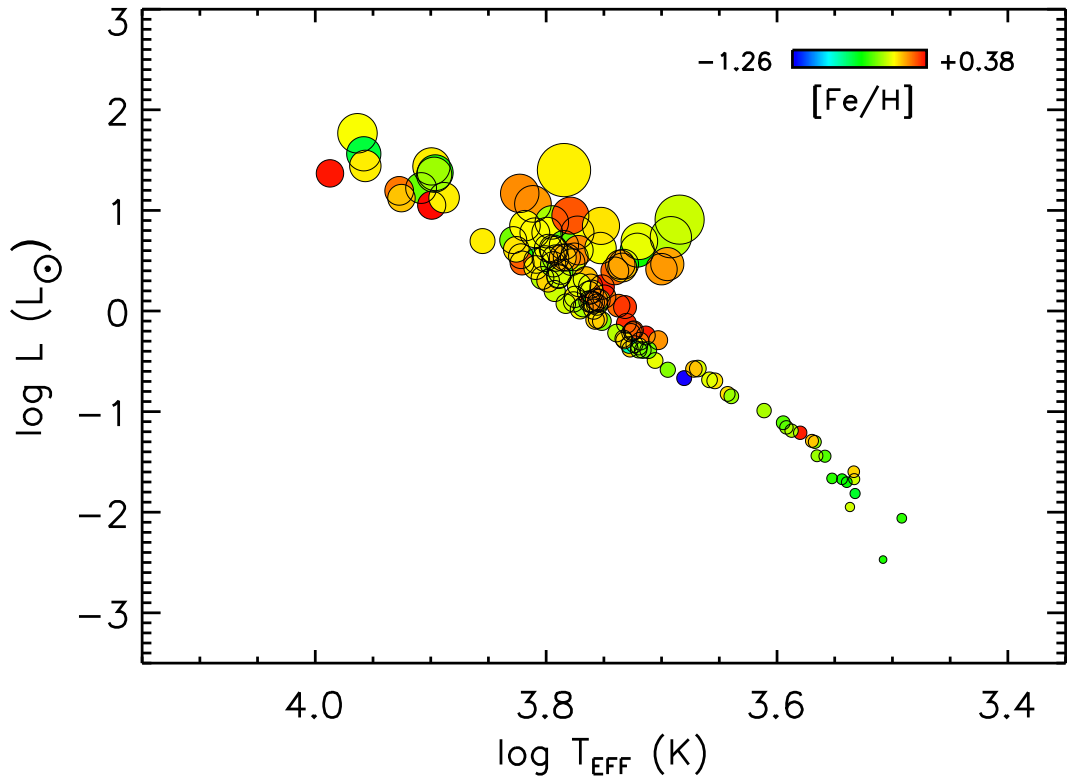, width=.9\linewidth, clip=} 
  \caption[ ] {H-R diagram on the luminosity - temperature plane for all stars in Table~\ref{tab:diameters_literature} plus the collection of low-mass star measurements in DT2.  The color and size of the data point reflects the metallicity and linear size of the star, respectively.  See Section~\ref{sec:stellar_params} for details.}
   \label{fig:Lumin_VS_Teff_VS_Radius_color}
   \end{figure}

\begin{figure}										
  \centering
         \epsfig{file=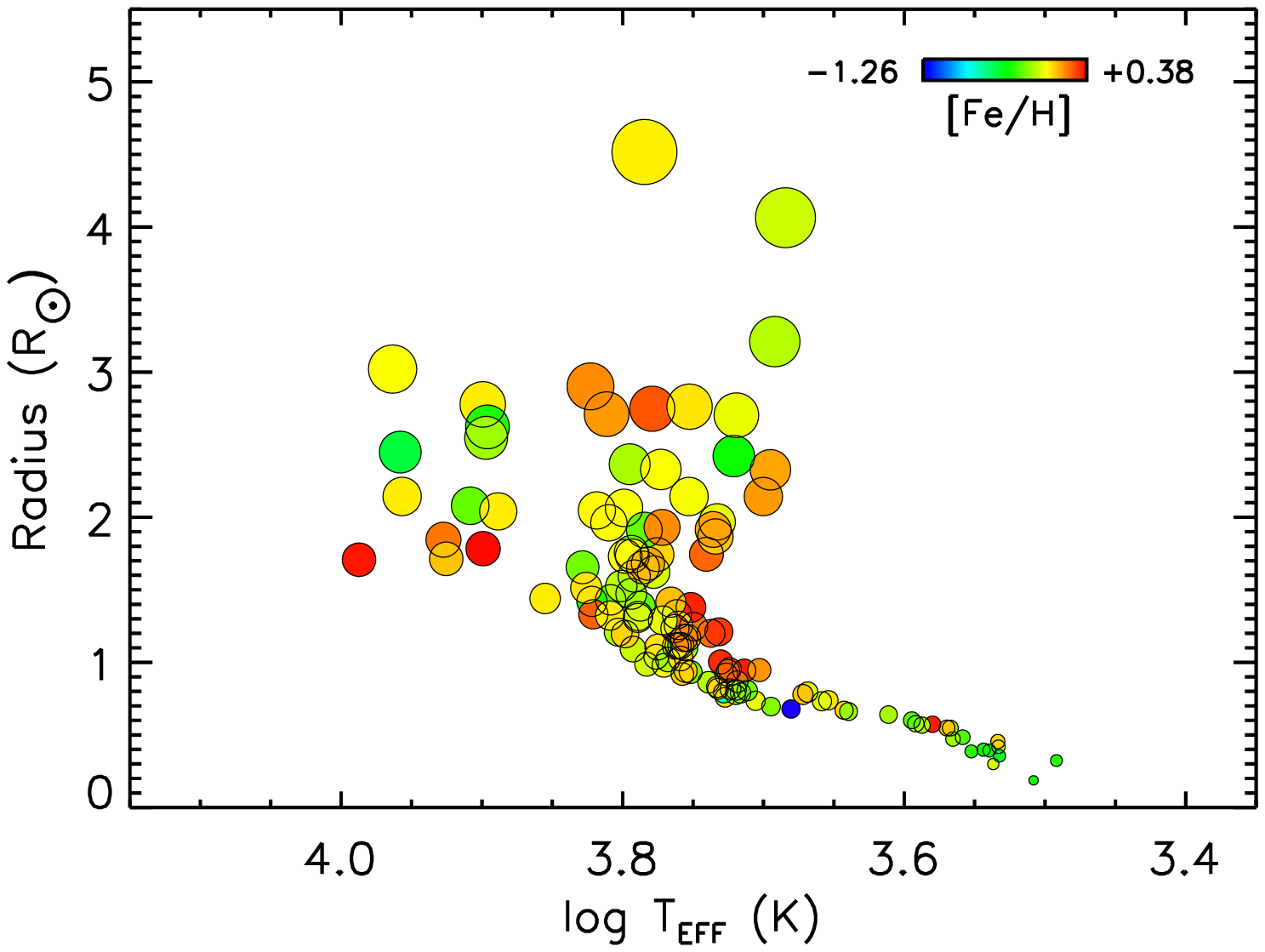, width=.9\linewidth, clip=} 
  \caption[ ] {Stellar temperature versus radius for all stars in Table~\ref{tab:diameters_literature} plus the collection of low-mass star measurements in DT2.  The color and size of the data point reflects the metallicity and linear size of the star, respectively.  See Section~\ref{sec:stellar_params} for details.}
   \label{fig:Teff_VS_Radius_color}
   \end{figure}

\begin{figure}										
  \centering
         \epsfig{file=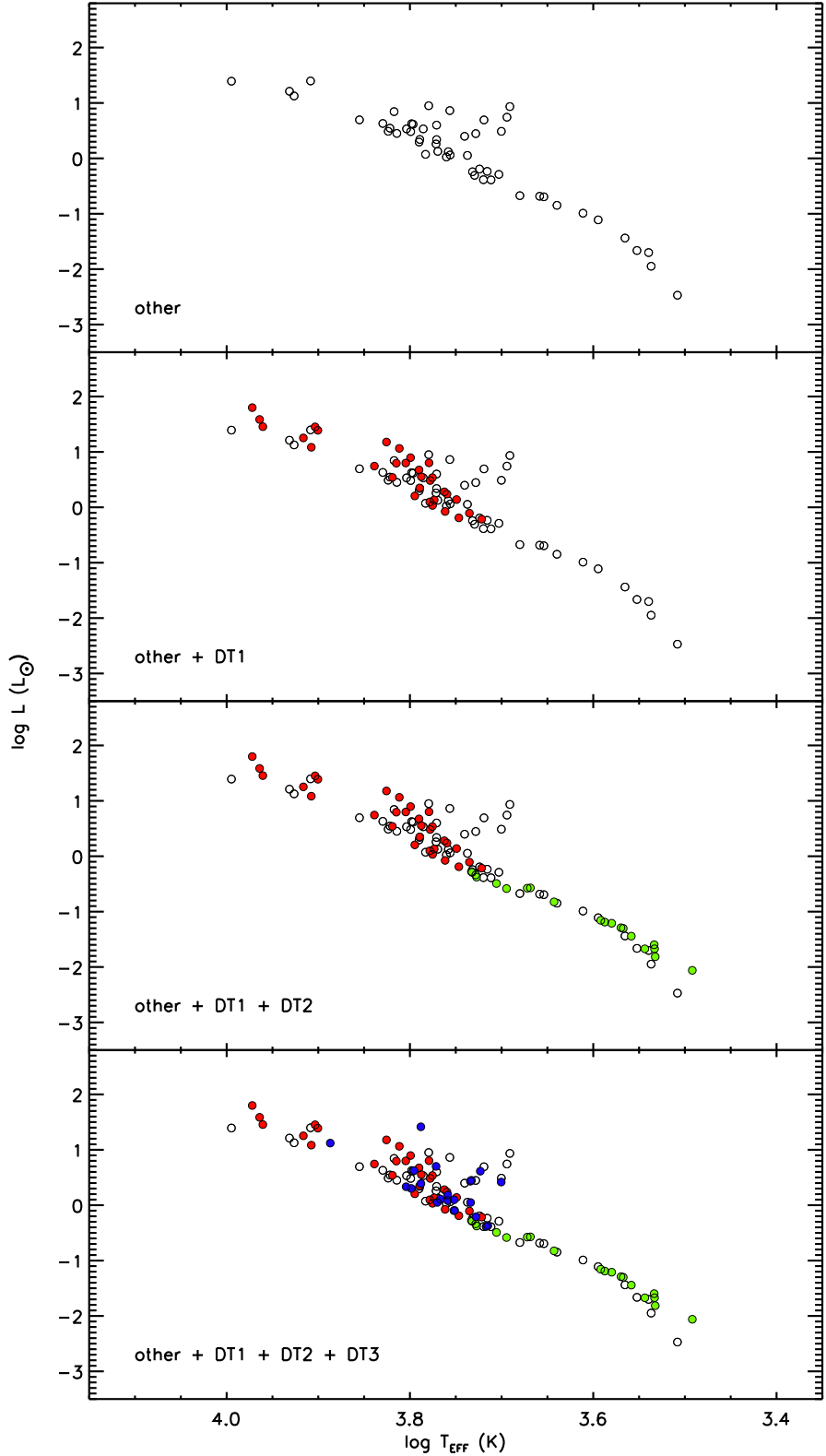, width=.7\linewidth, clip=} 
  \caption[ ] {Each panel in this figure shows our progress in furnishing measurements to build a fundamentally determined H-R diagram on the luminosity - temperature plane. All published measurements in Table~\ref{tab:diameters_literature} plus the previously published low-mass star measurements collection of in table~7 in \citet{boy12b} ({\it other}; black points) are shown in all panels.  The second panel down adds the stars from \citet{boy12a} (DT1; red points).  The third panel down adds the stars from \citet{boy12b} (DT2; green points). The bottom panel adds the stars from this work (DT3; blue points). Stars with multiple measurements (i.e., marked with a $^{\dagger}$ in Table~\ref{tab:diameters_literature} or with a $^{\dagger}$ or $^{\dagger\dagger}$ in table~7 of \citet{boy12b} are fall under the {\it other} category, since they are not unique contributions from our DT1, DT2, or DT3 interferometric surveys). See Section~\ref{sec:stellar_params} for details.}
   \label{fig:Lumin_VS_Teff_multiplot}
   \end{figure}

\begin{figure}										
  \centering
         \epsfig{file=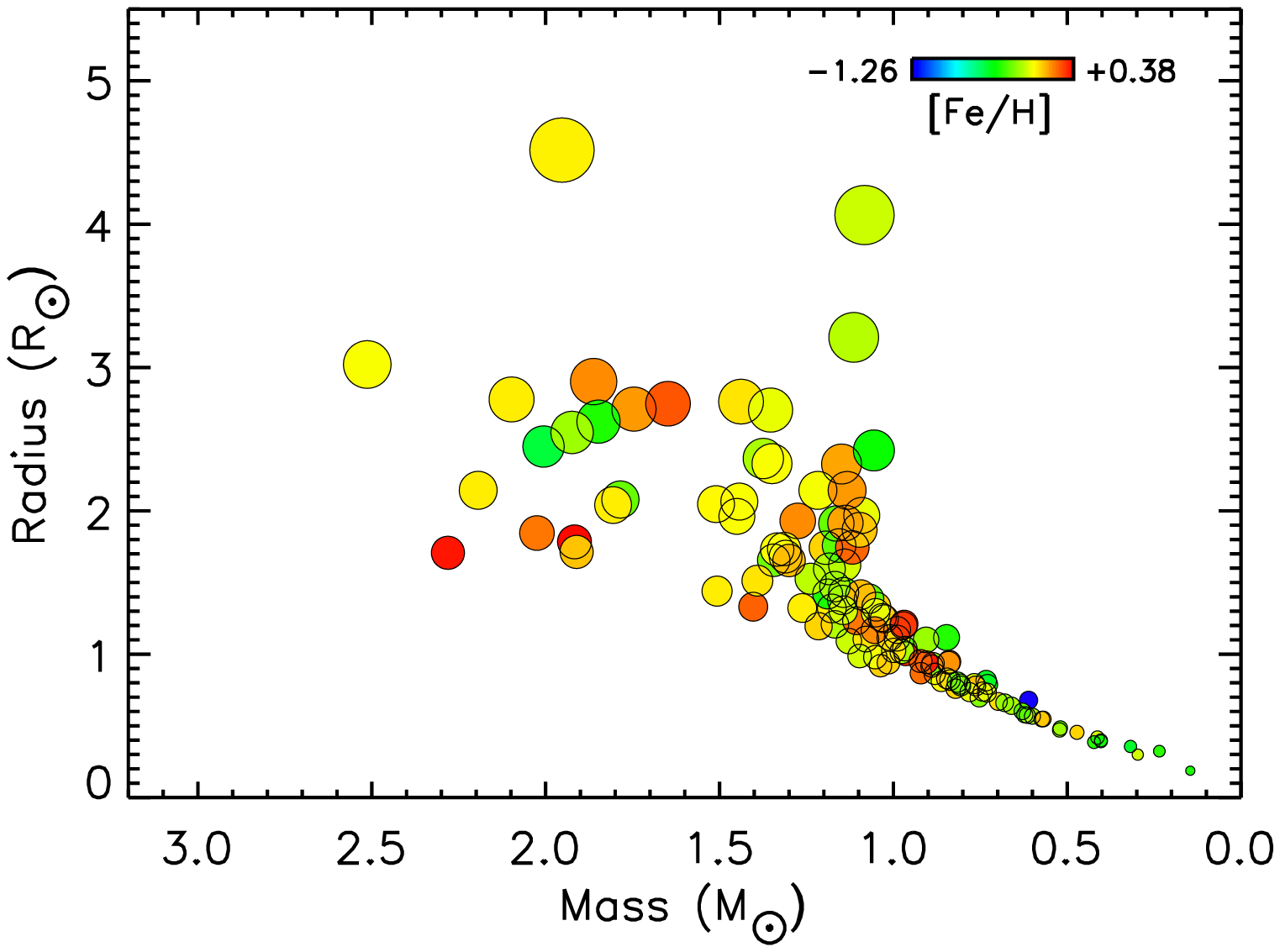, width=.9\linewidth, clip=} 
  \caption[ ] {Stellar mass versus radius plotted for all stars in Table~\ref{tab:diameters_literature} plus the collection of low-mass star measurements in DT2.  The color and size of the data point reflects the metallicity and linear size of the star, respectively.  See Section~\ref{sec:evolution} for details.}
   \label{fig:Mass_VS_Radius_color}
   \end{figure}

\begin{figure}										
  \centering
         \epsfig{file=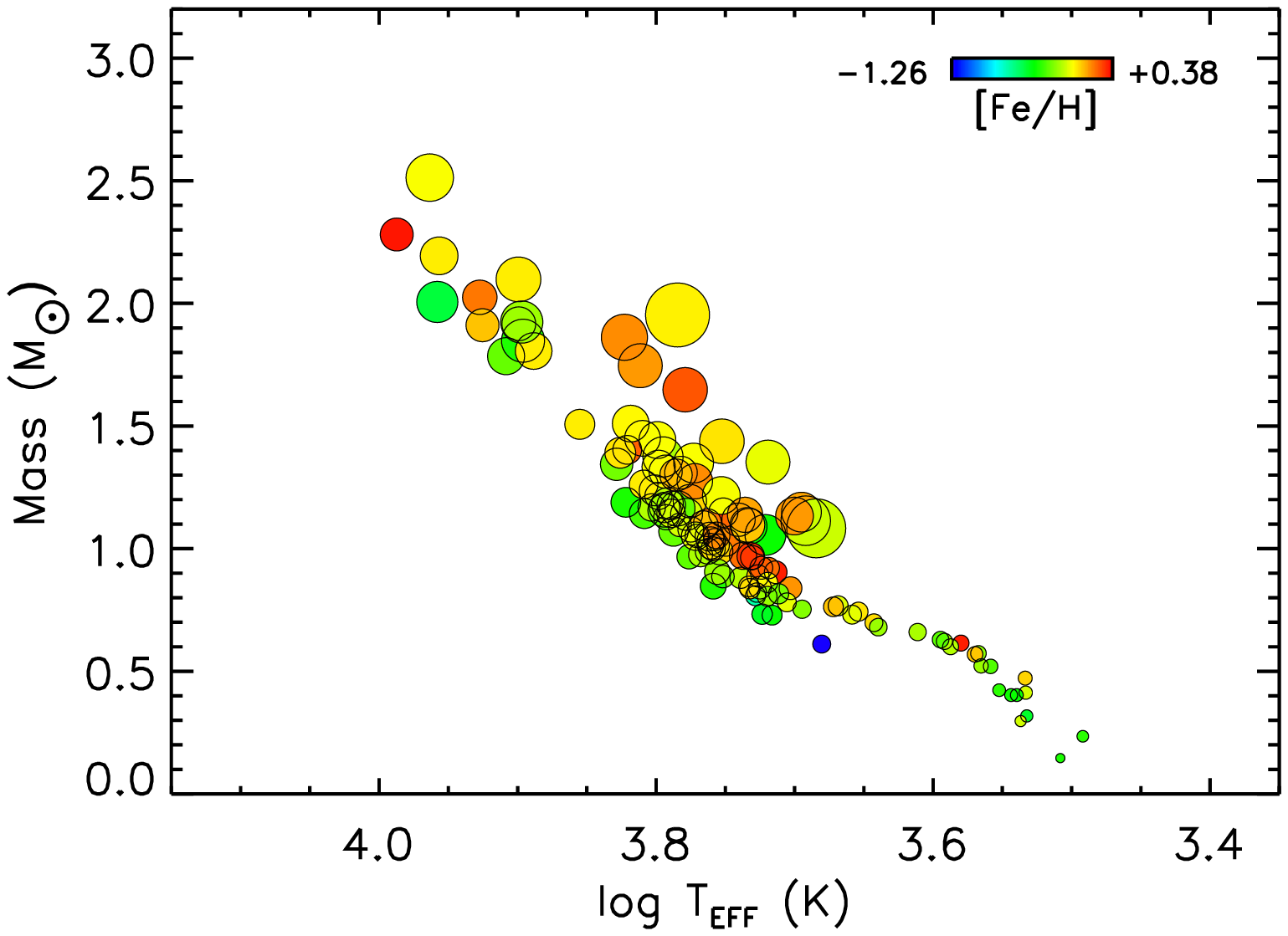, width=.9\linewidth, clip=} 
  \caption[ ] {Stellar mass versus radius plotted for all stars in Table~\ref{tab:diameters_literature} plus the collection of low-mass star measurements in DT2.  The color and size of the data point reflects the metallicity and linear size of the star, respectively.  See Section~\ref{sec:evolution} for details.}
   \label{fig:Mass_VS_Teff_color}
   \end{figure}

\begin{figure}										
  \centering
  \begin{tabular}{c}
         \epsfig{file=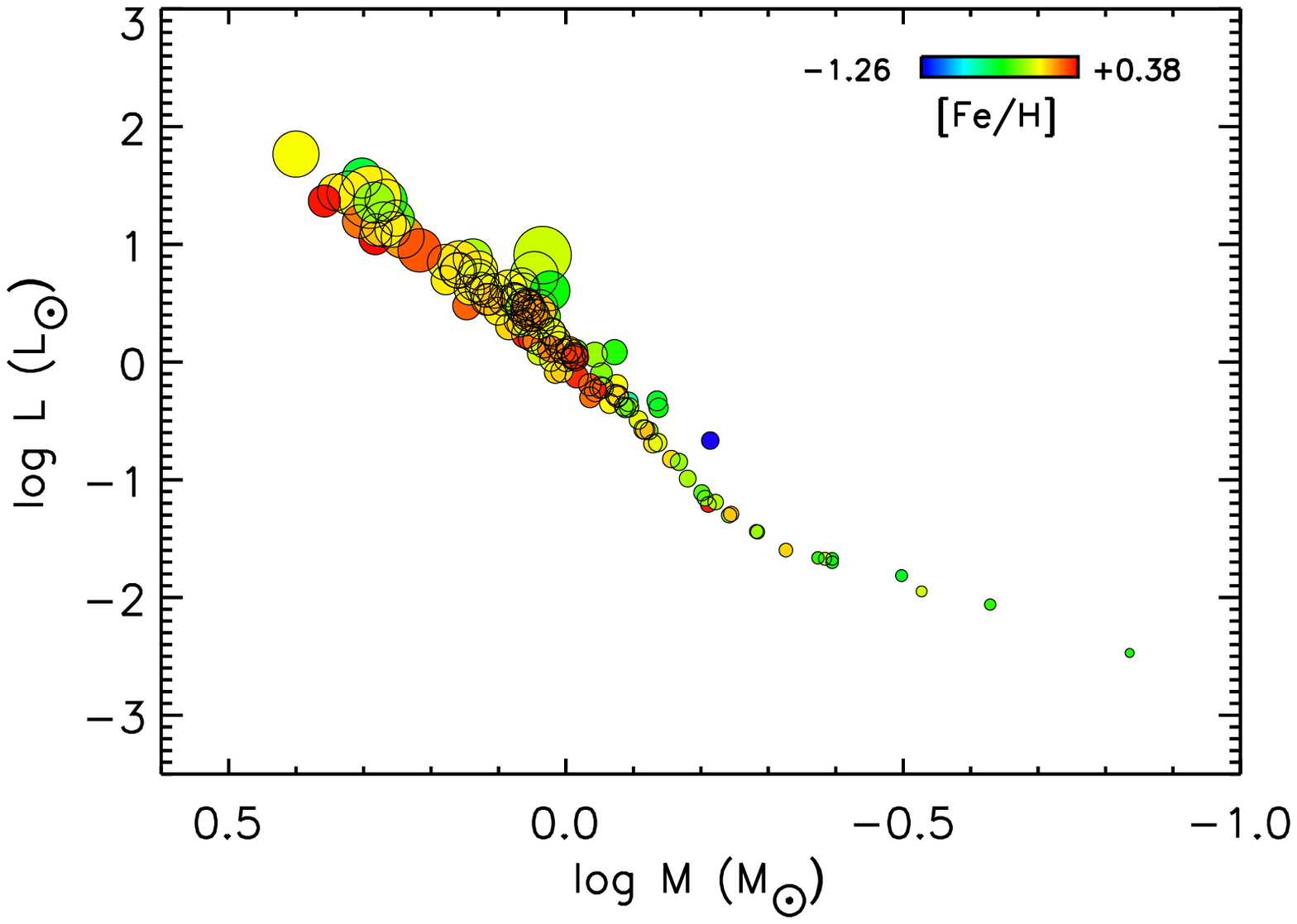, width=.8\linewidth, clip=} \\
 		\epsfig{file=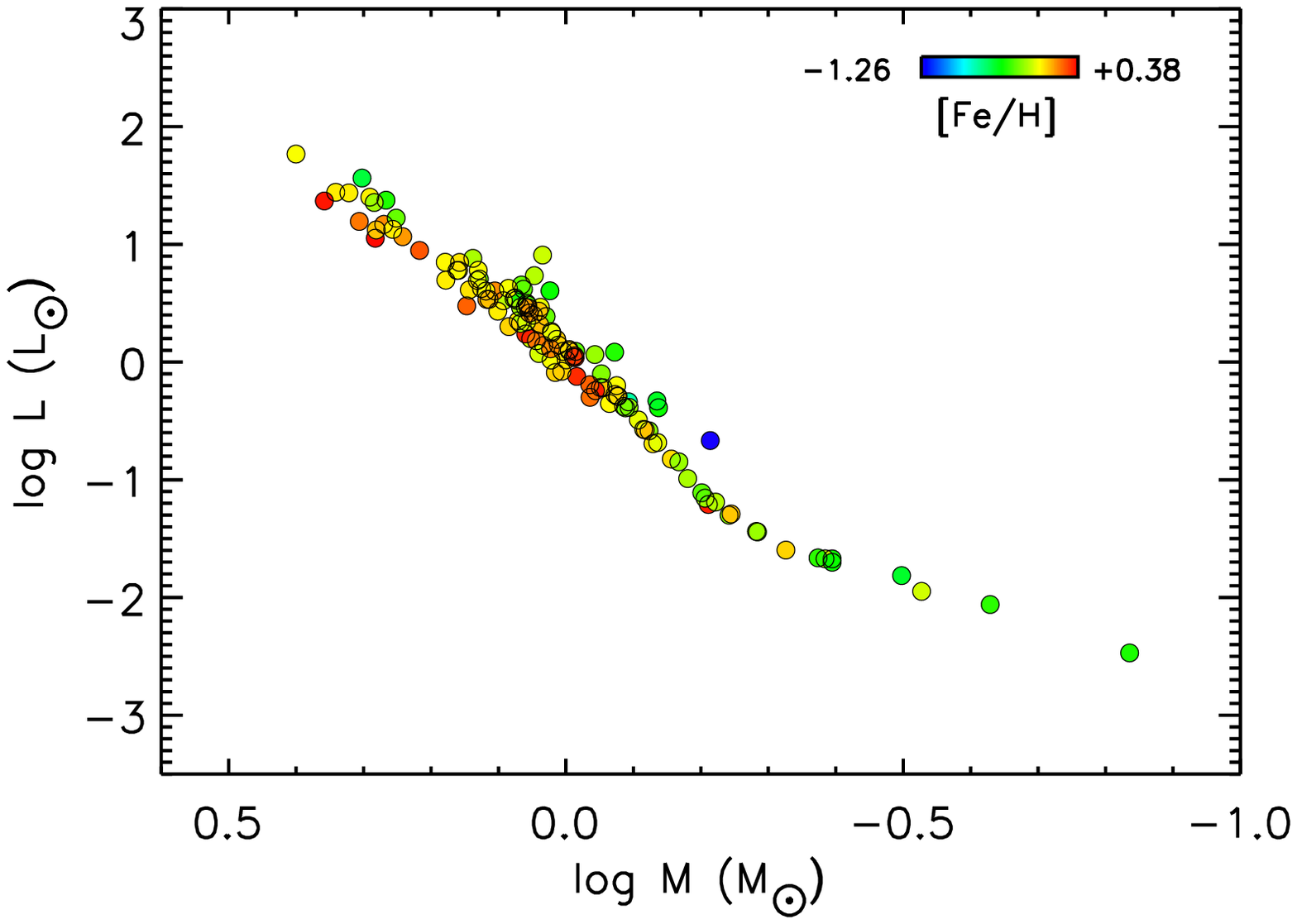, width=.8\linewidth, clip=} 
   \end{tabular}
  \caption[ ] {Stellar mass versus luminosity plotted for all stars in Table~\ref{tab:diameters_literature} plus the collection of low-mass star measurements in DT2.  The color of the data point reflects the metallicity of the star.  The size of the points in the top panel are proportional to the linear size of the star. The data points in the bottom panel are all of equal size in order to more clearly visualize the splitting in the mass - luminosity plane for stars of different metallicities. See Section~\ref{sec:evolution} for details.}
   \label{fig:Mass_VS_Lumin_VS_Radius_log_color}
   \end{figure}

\begin{figure}										
\centering
\begin{tabular}{cc}
\epsfig{file=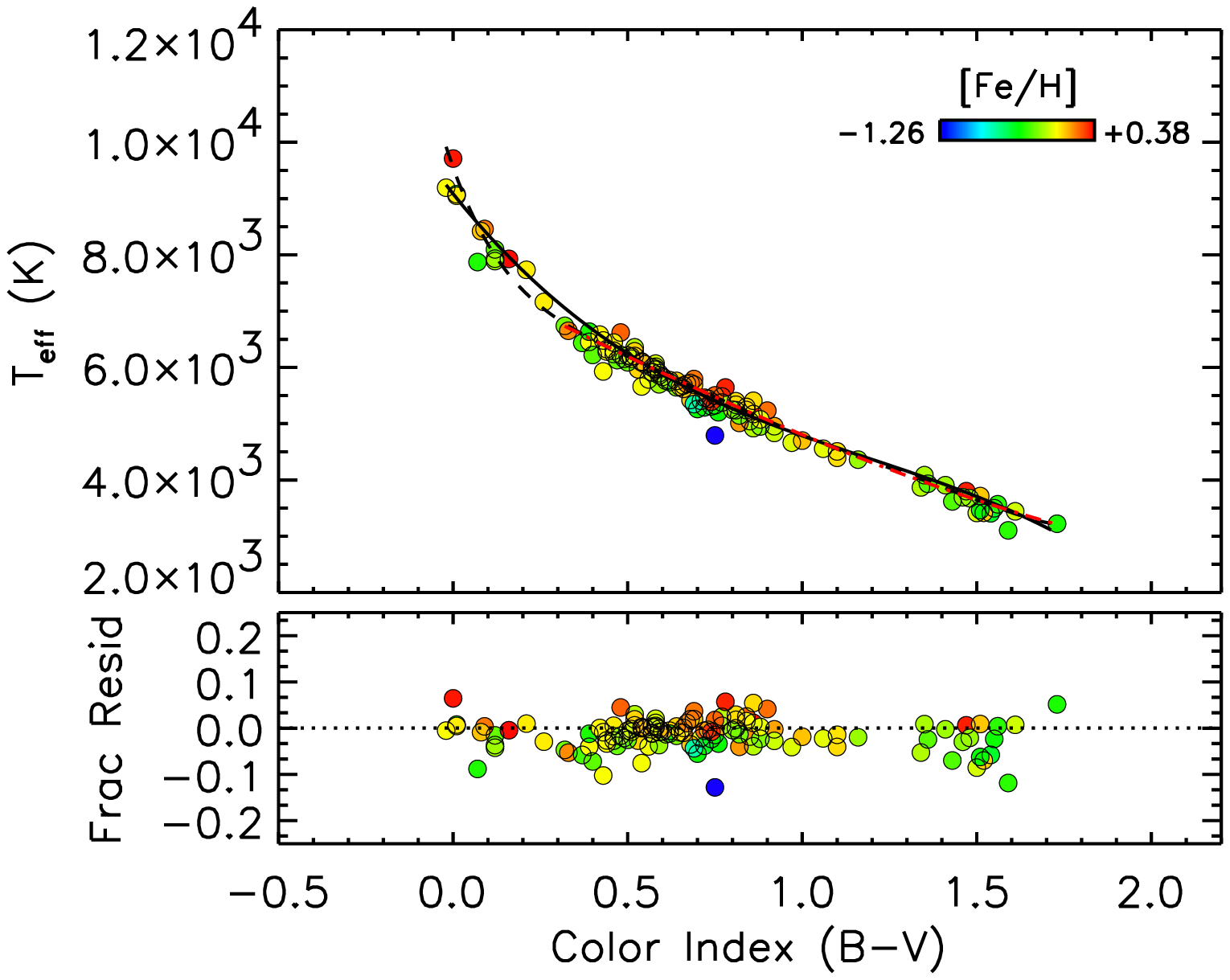, width=0.5\linewidth, clip=} &
\epsfig{file=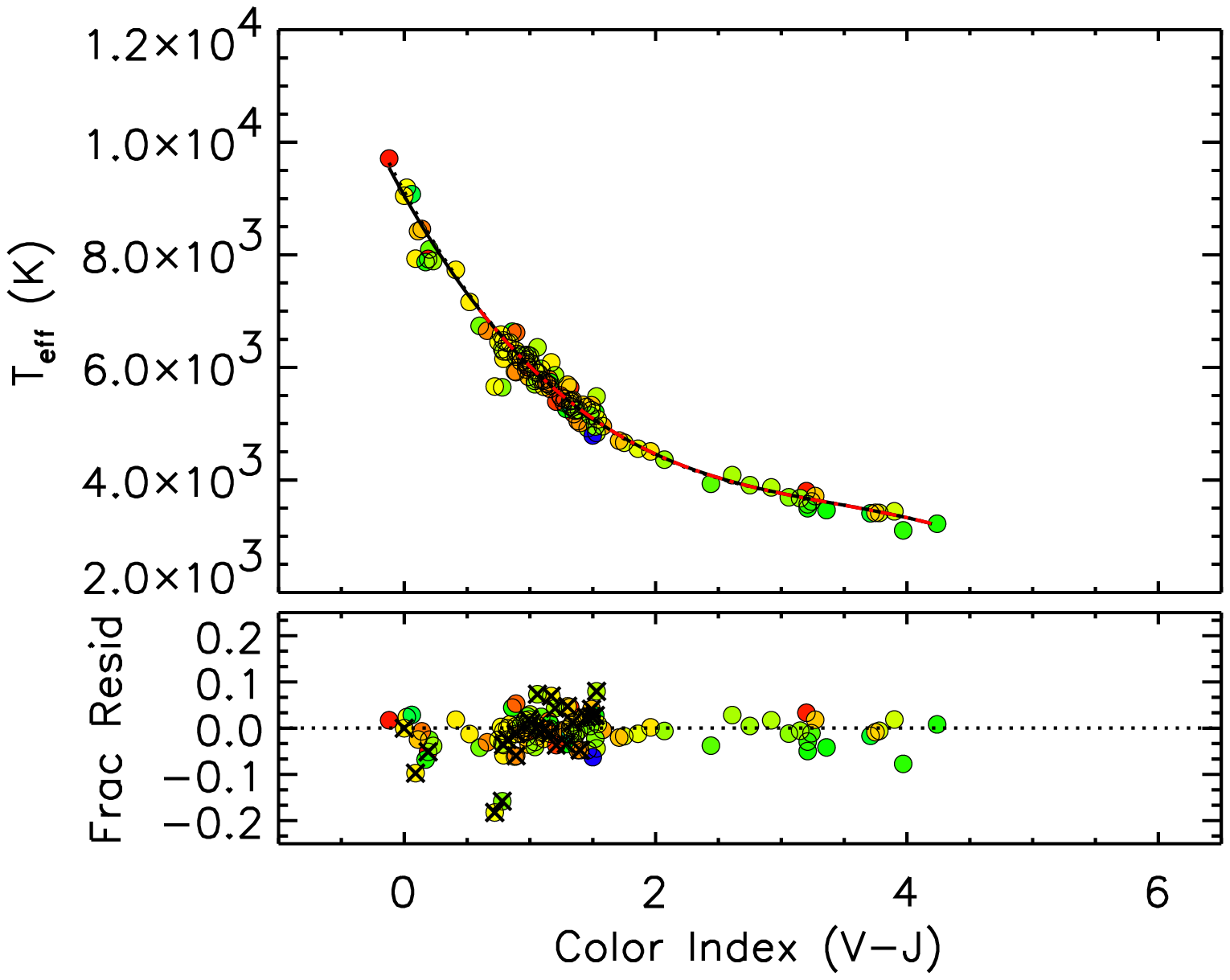, width=0.5\linewidth, clip=} \\
\epsfig{file=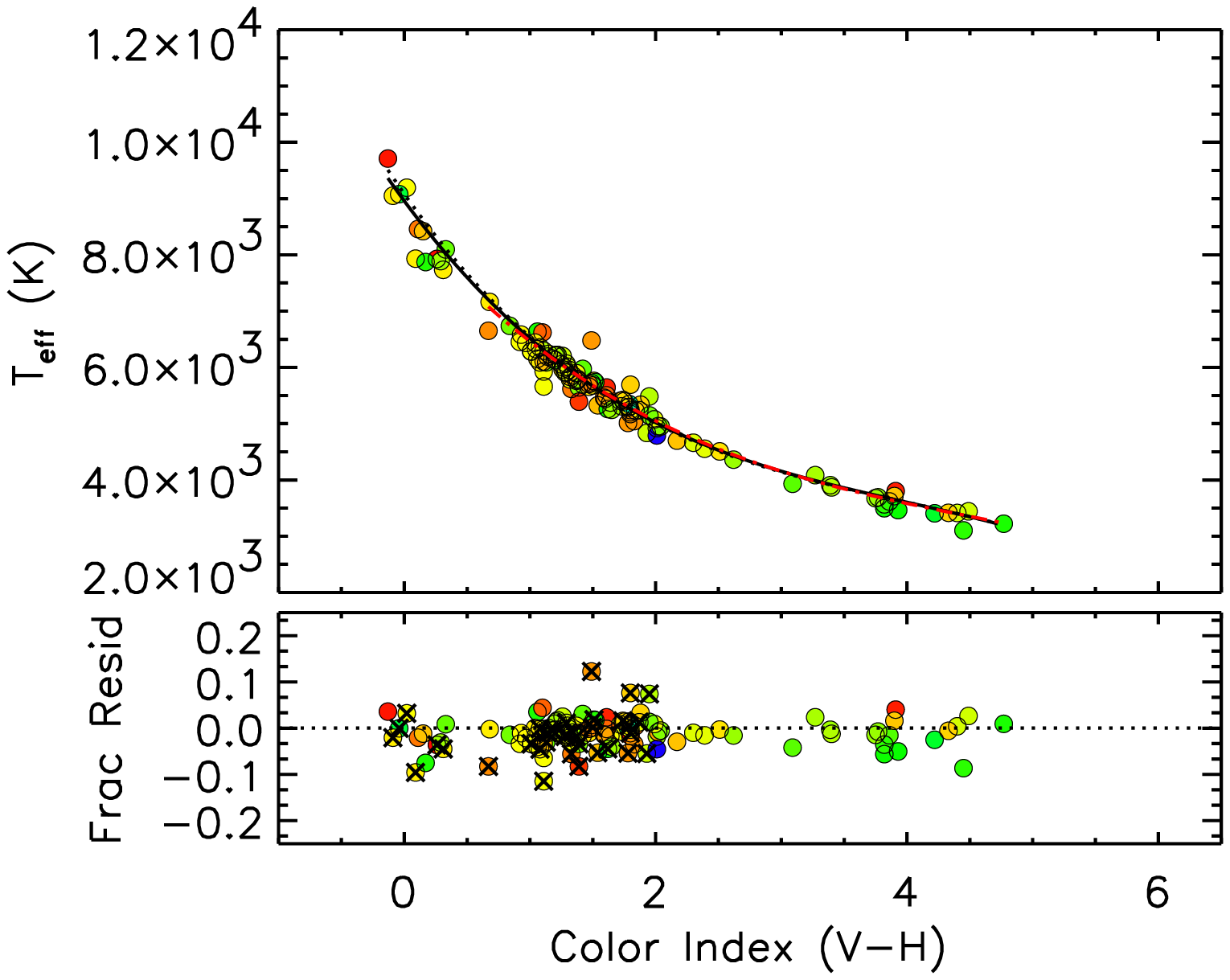, width=0.5\linewidth, clip=} &
\epsfig{file=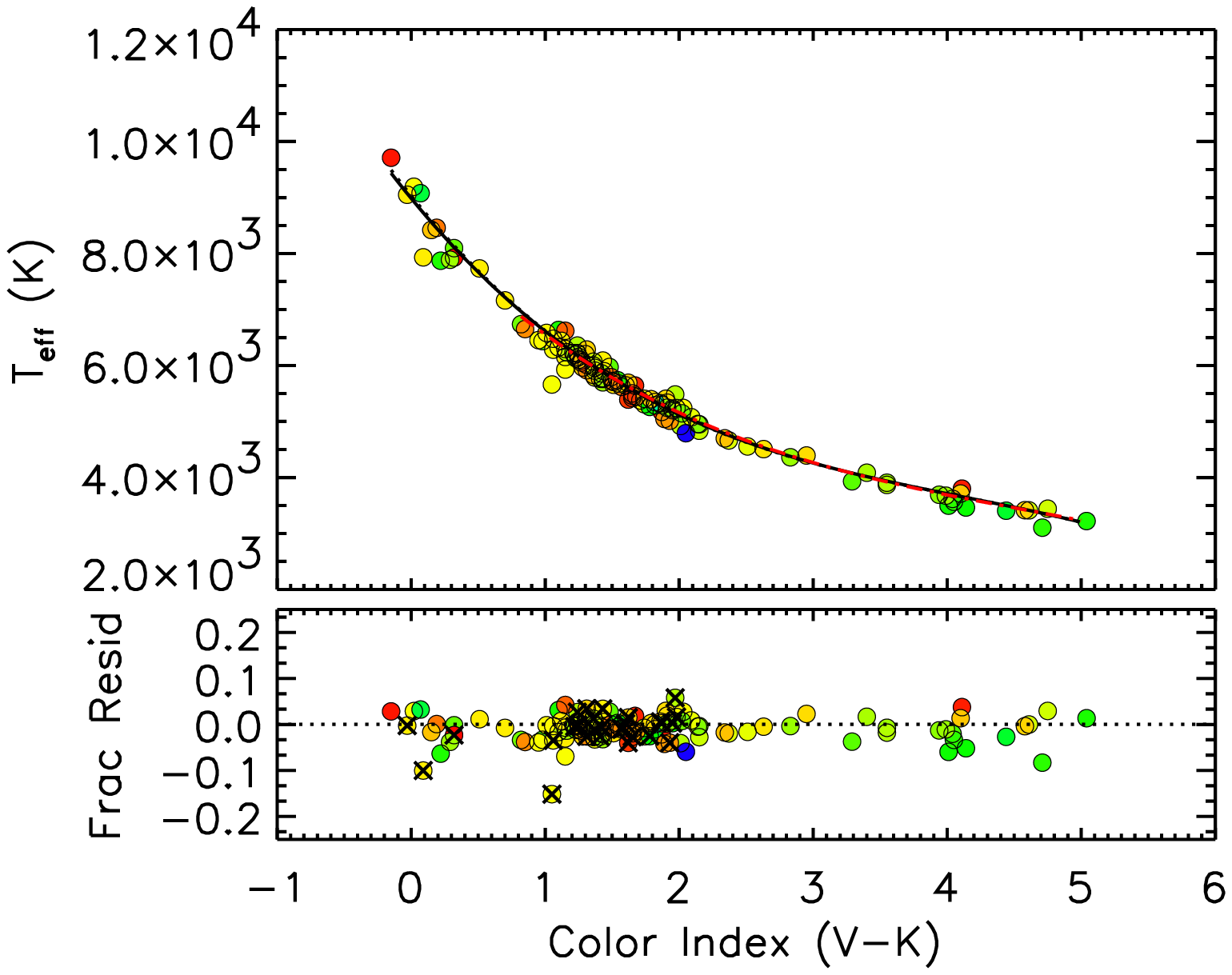, width=0.5\linewidth, clip=} \\
\epsfig{file=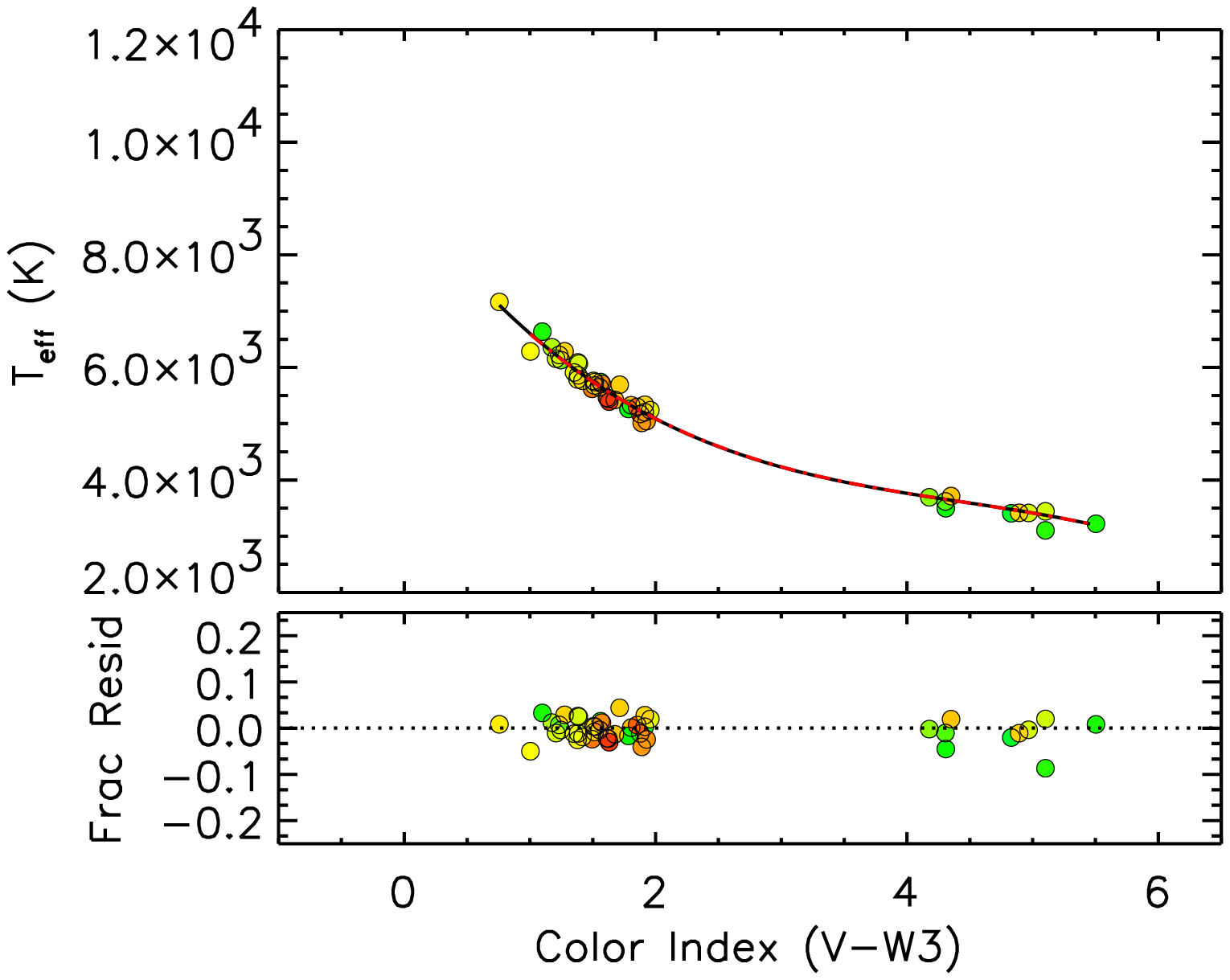, width=0.5\linewidth, clip=} &
\epsfig{file=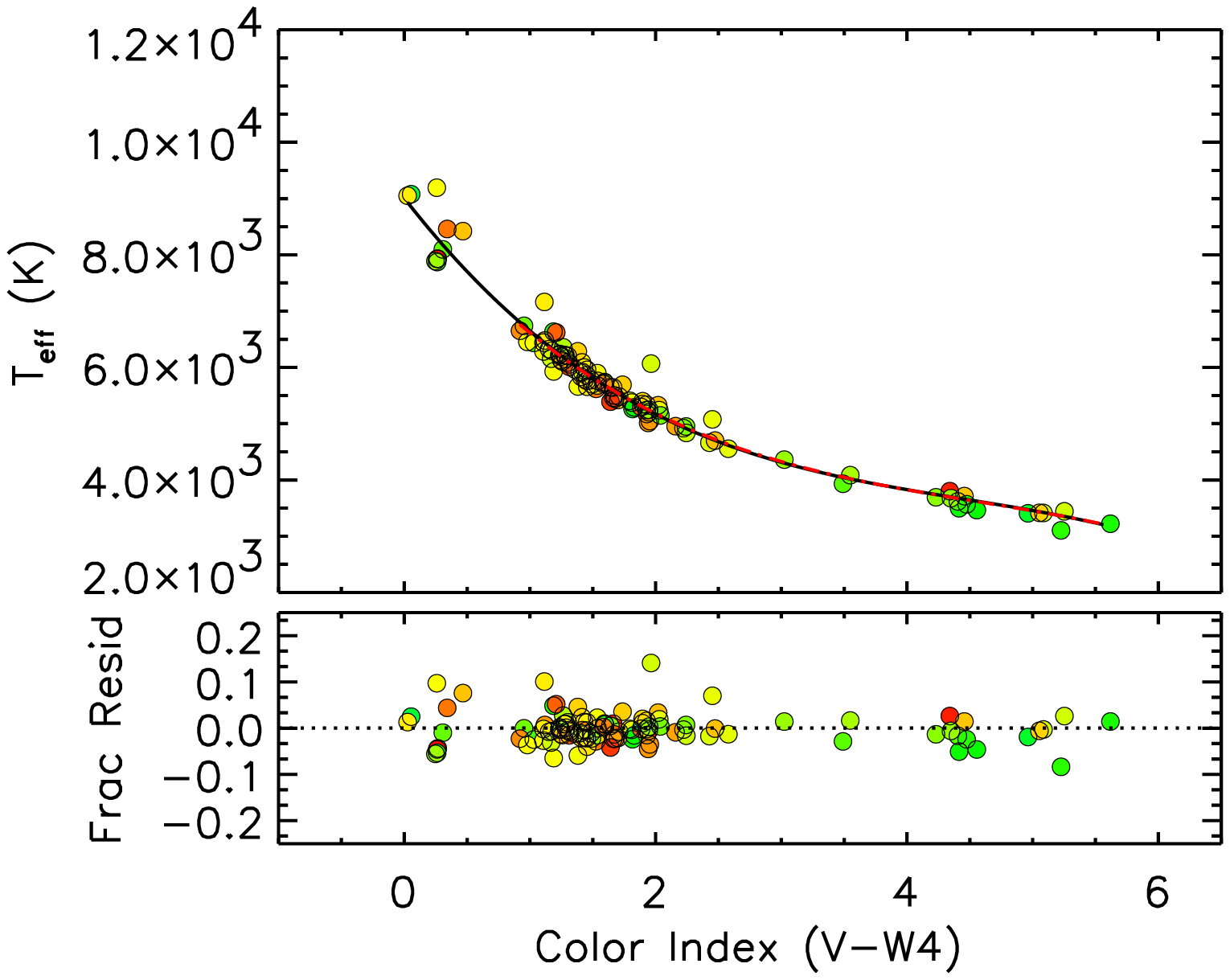, width=0.5\linewidth, clip=} 
 \end{tabular}
 \caption[ ] {The solid black line represents the solution to the color-temperature relation (expressed as Equation~\ref{eq:poly3c} and reported in Table~\ref{tab:poly3_coeffs}). The red dash-dot line represents the solution omitting the early-type stars (Section~\ref{sec:discussion_astars_evolution}, Equation~\ref{eq:poly3c}, Table~\ref{tab:poly3_coeffs}).  The color of the data point reflects the metallicity of the star, and temperature errors are not shown but typically are smaller than the data point.  Those panels that involve infrared {\it JHK} colors have a second solution plotted as a dotted line (mostly eclipsed by the solid line solution). The dashed line in the panel showing the $(B-V)$ relation  is the solution using a $6^{th}$ order polynomial (Section~\ref{sec:discussion_bmv}, Equation~\ref{eq:poly6c}, Figure~\ref{fig:Temp_VS_BmV_p6}).  The bottom panel shows the fractional residual ($T_{\rm Obs.} - T_{\rm Fit})/T_{\rm Obs.}$ to the $3^{rd}$ order polynomial fit, where the dotted line indicates zero deviation. Points with saturated {\it 2MASS} photometry are marked with an $\times$ in the bottom panel (see Section~\ref{sec:discussion_ircolors} for details). See Section~\ref{sec:discussion}, Section~\ref{sec:discussion_astars_evolution}, and Section~\ref{sec:discussion_2mass} for details. }
 \label{fig:relations1}
 \end{figure}

\begin{figure}										
\centering
\begin{tabular}{cc}
\epsfig{file=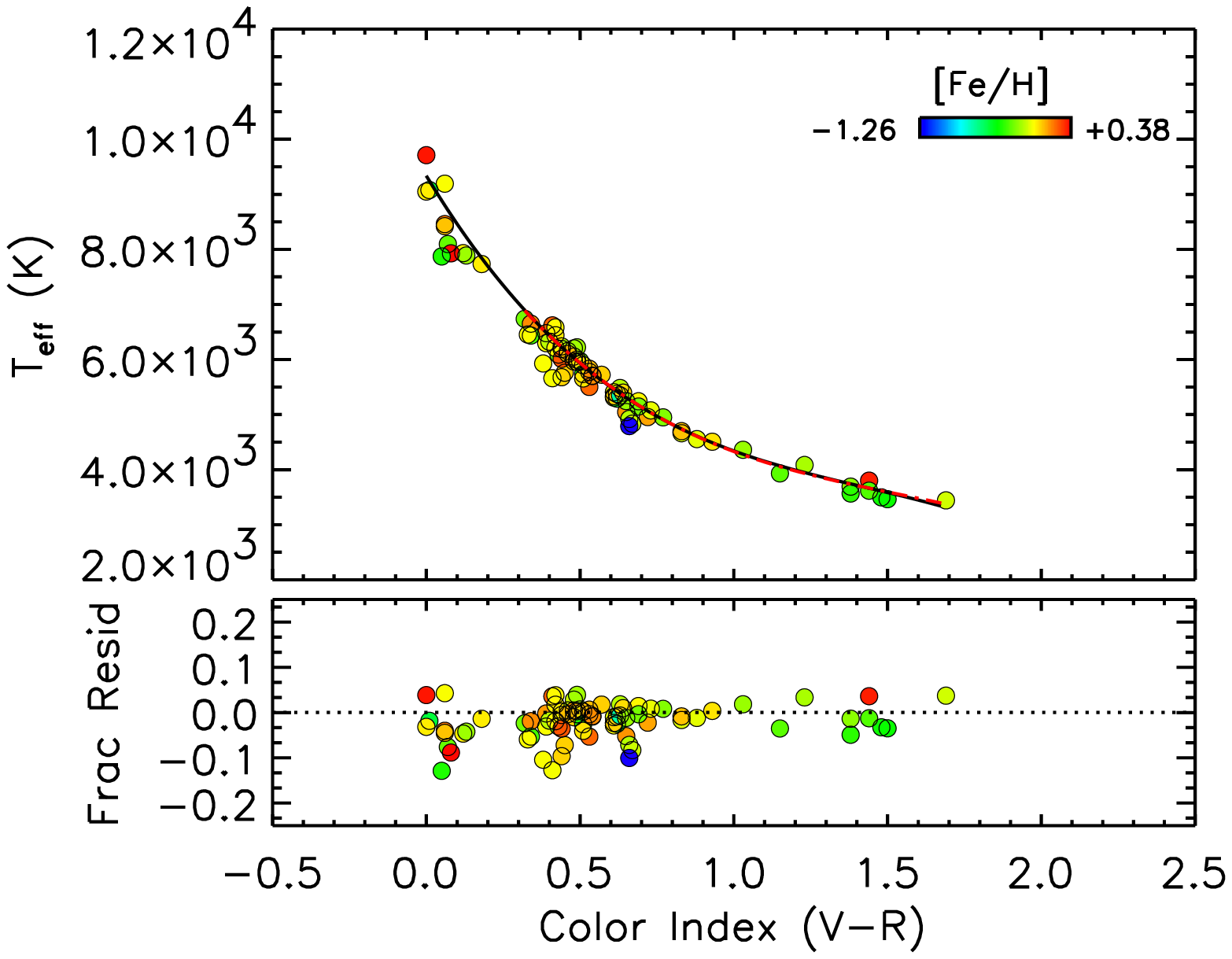, width=0.5\linewidth, clip=} &
\epsfig{file=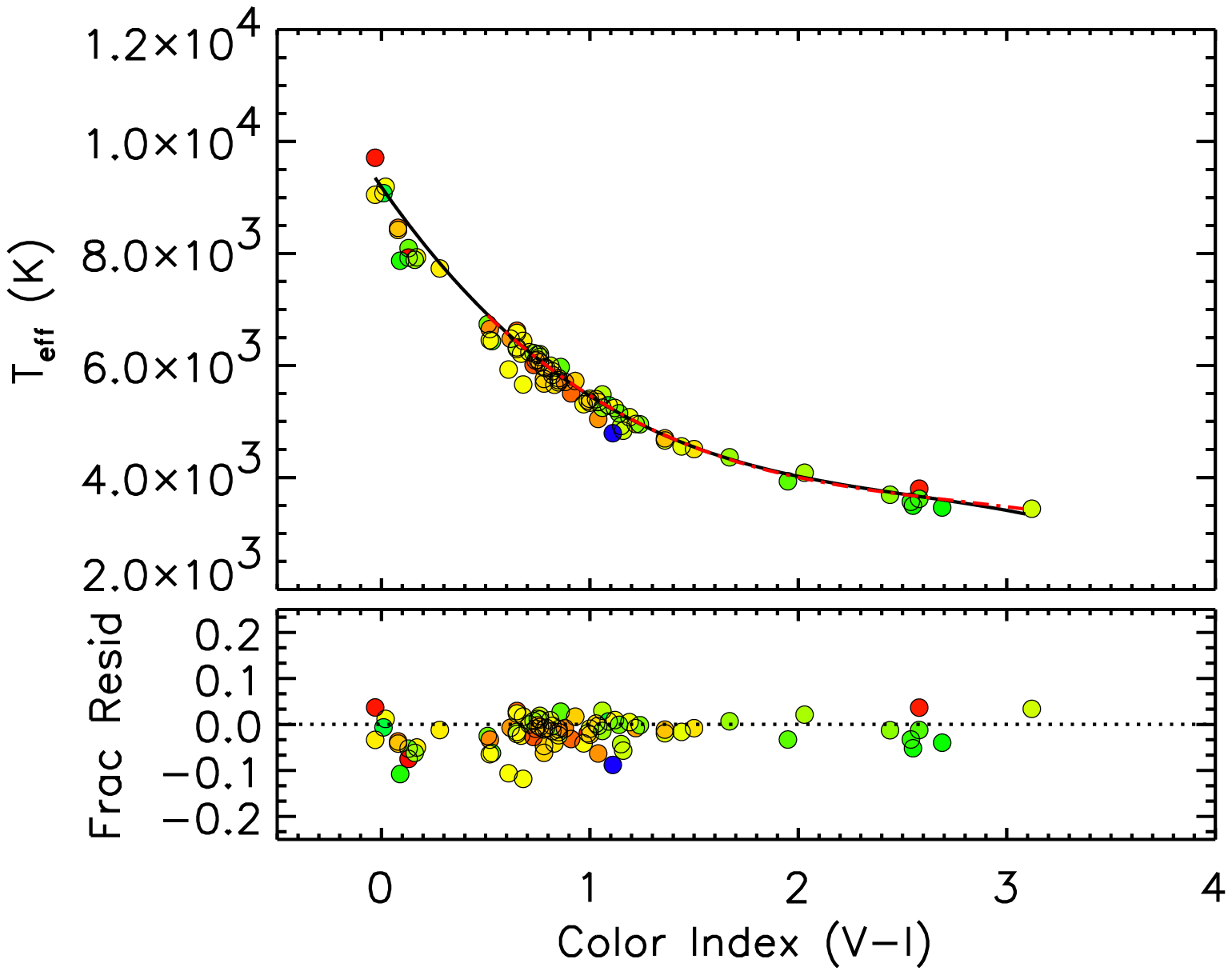, width=0.5\linewidth, clip=} \\
\epsfig{file=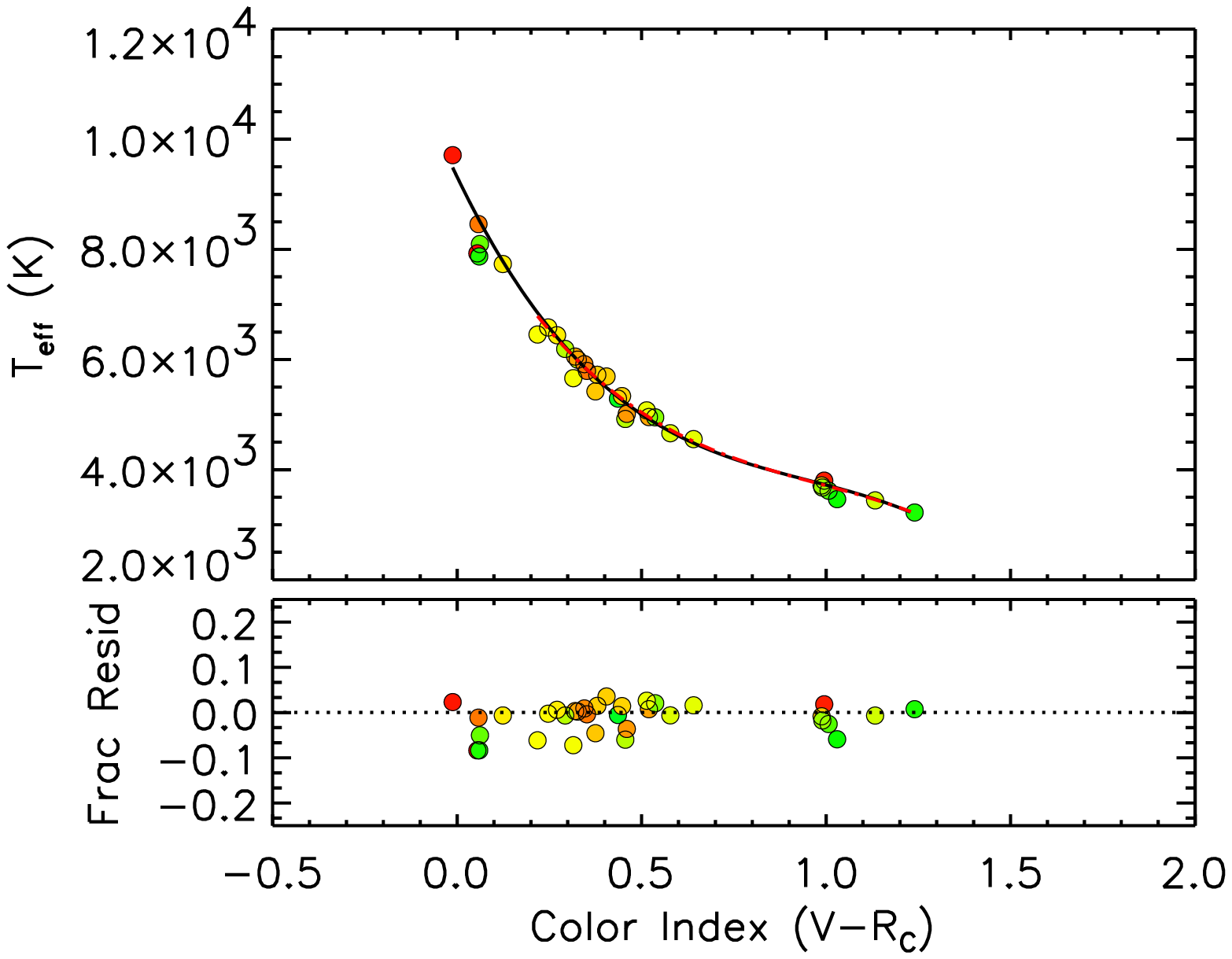, width=0.5\linewidth, clip=} &
\epsfig{file=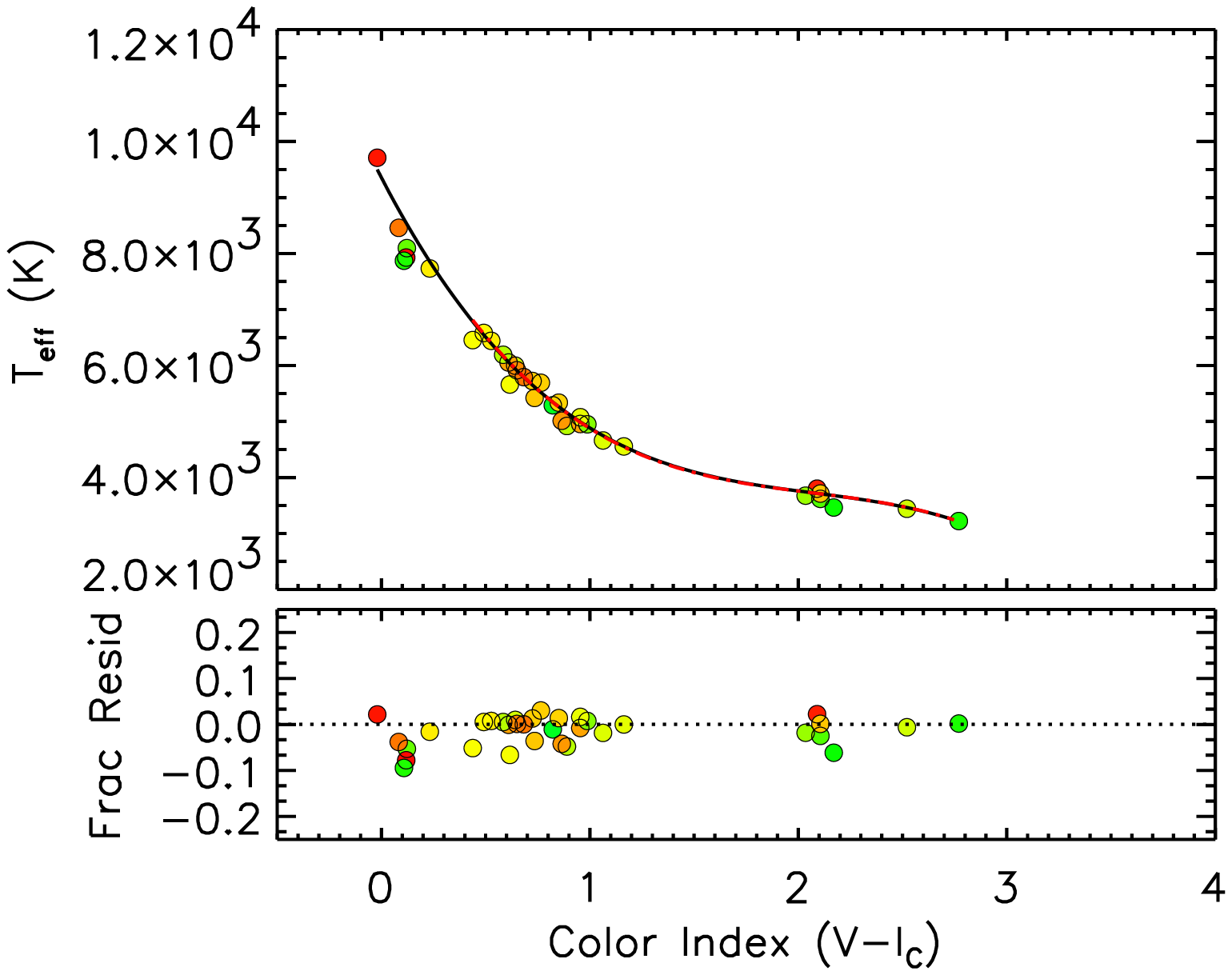, width=0.5\linewidth, clip=} \\
\epsfig{file=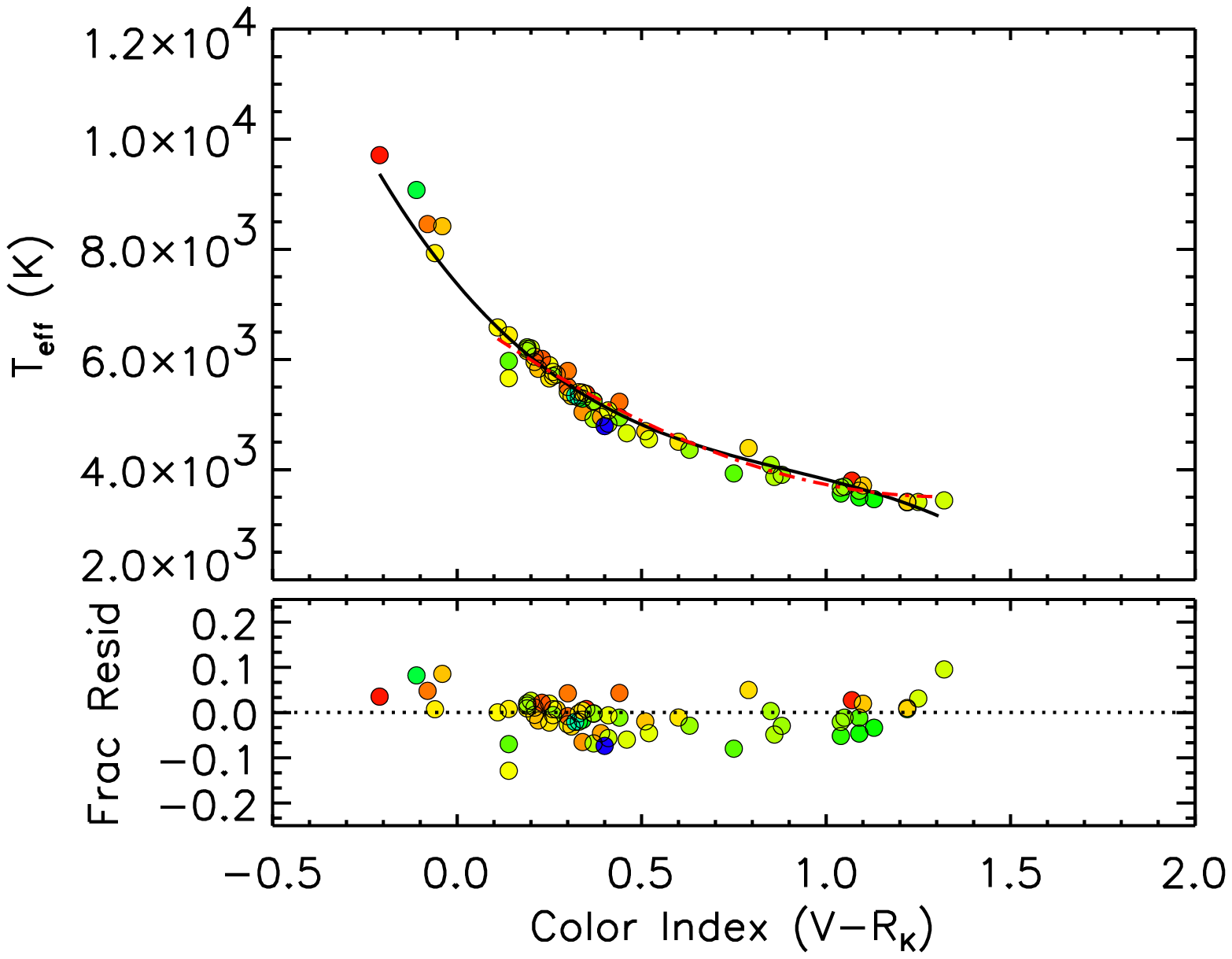, width=0.5\linewidth, clip=} &
\epsfig{file=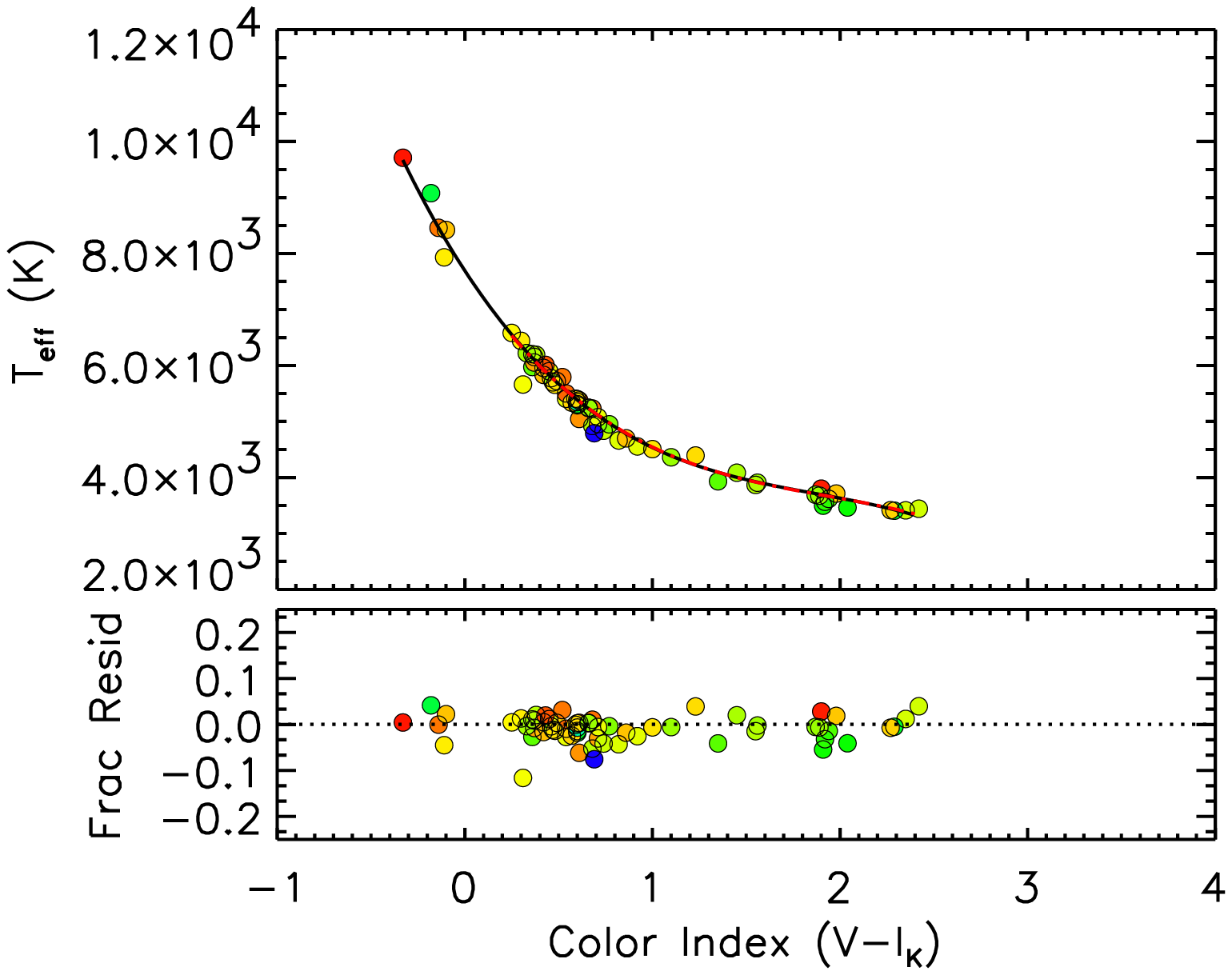, width=0.5\linewidth, clip=} 
 \end{tabular}
 \caption[ ] {The solid black line represents the solution to the color-temperature relation (expressed as Equation~\ref{eq:poly3c} and reported in Table~\ref{tab:poly3_coeffs}). The red dash-dot line represents the solution omitting the early-type stars (Section~\ref{sec:discussion_astars_evolution}, Equation~\ref{eq:poly3c}, Table~\ref{tab:poly3_coeffs}). The color of the data point reflects the metallicity of the star, and temperature errors are not shown but typically are smaller than the data point.  The bottom panel shows the fractional residual ($T_{\rm Obs.} - T_{\rm Fit})/T_{\rm Obs.}$ to the $3^{rd}$ order polynomial fit, where the dotted line indicates zero deviation. See Section~\ref{sec:discussion} and Section~\ref{sec:discussion_astars_evolution} for details. }
 \label{fig:relations2}
 \end{figure}

\begin{figure}										
\centering
\begin{tabular}{cc}
\epsfig{file=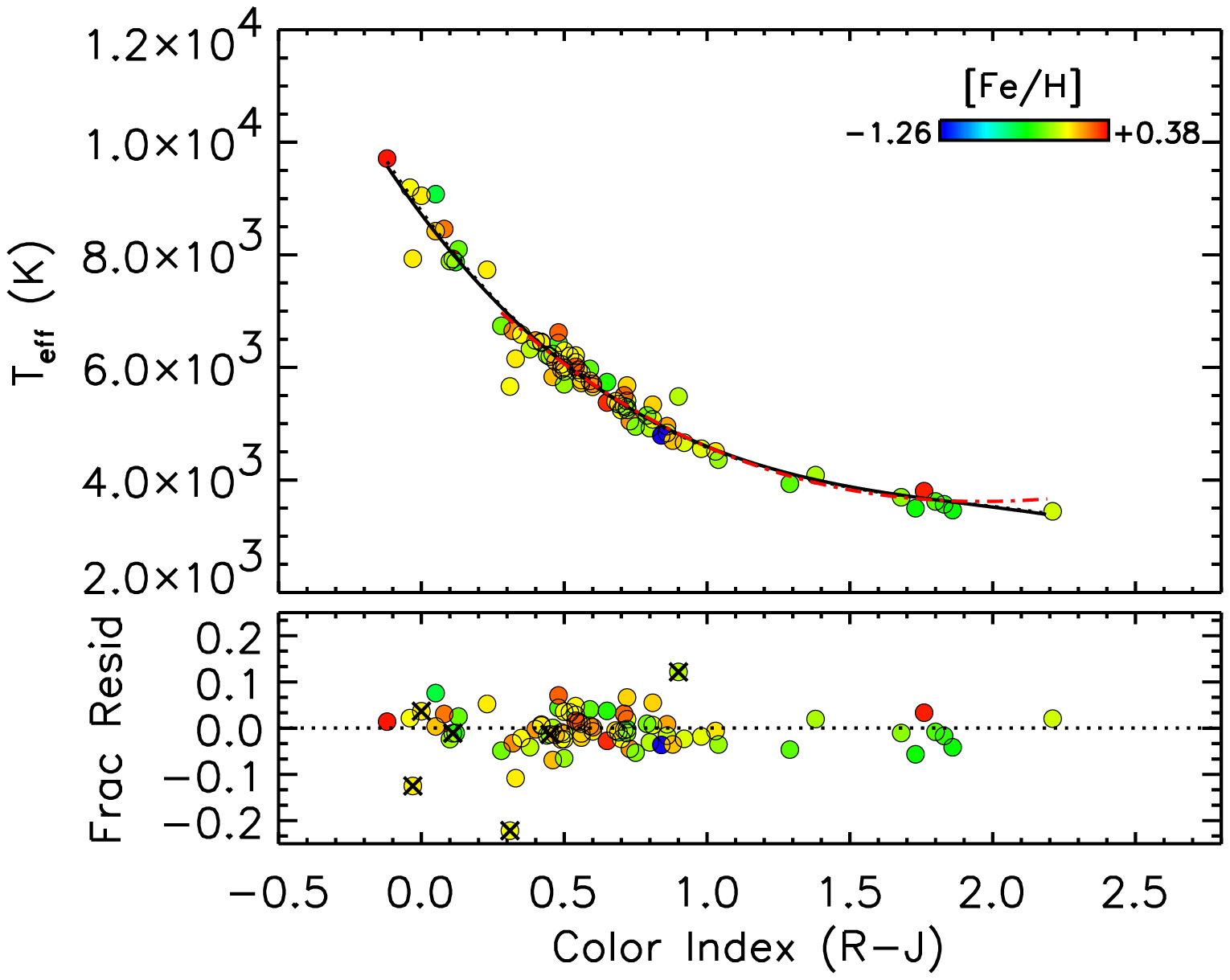, width=0.5\linewidth, clip=} &
\epsfig{file=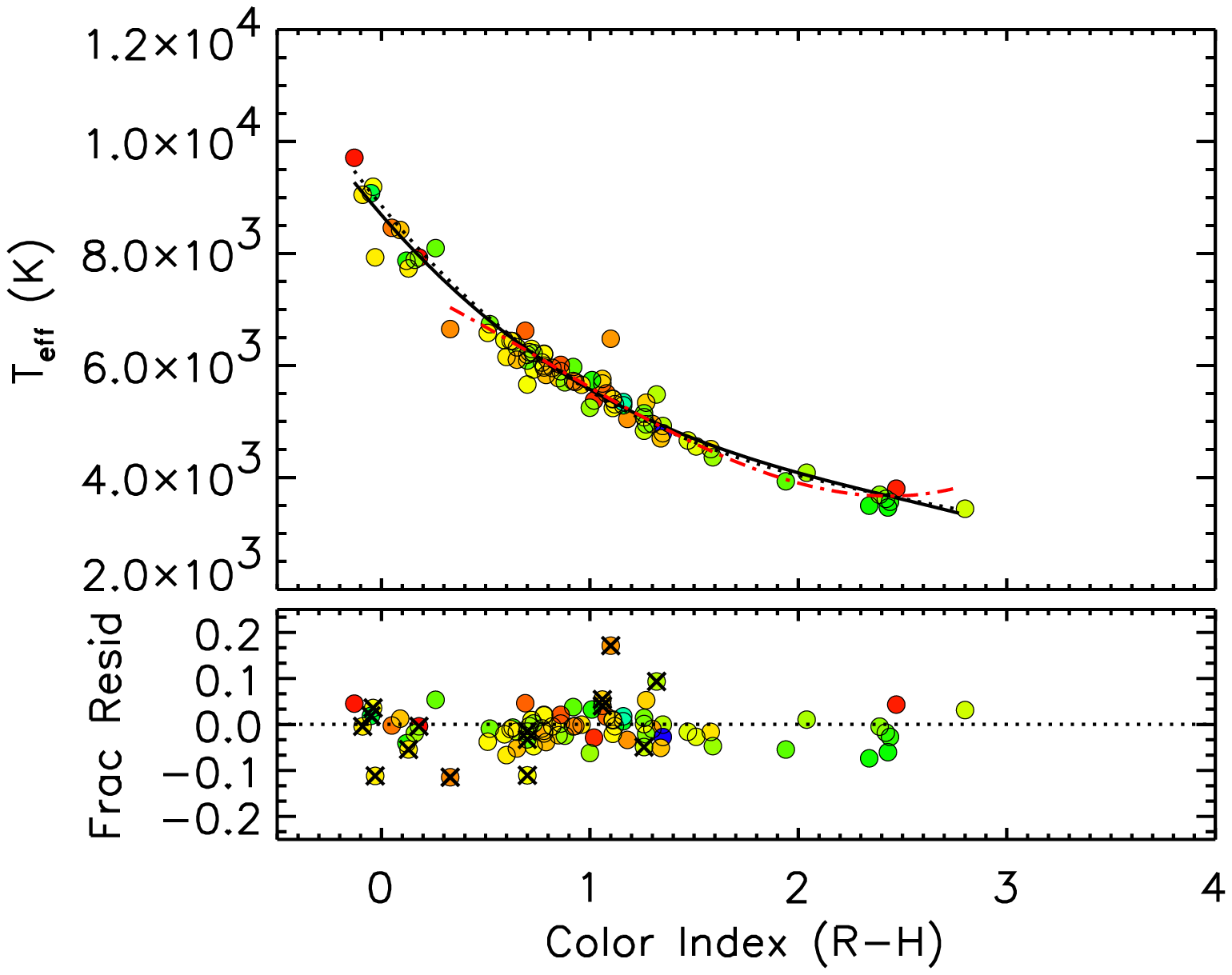, width=0.5\linewidth, clip=} \\
\epsfig{file=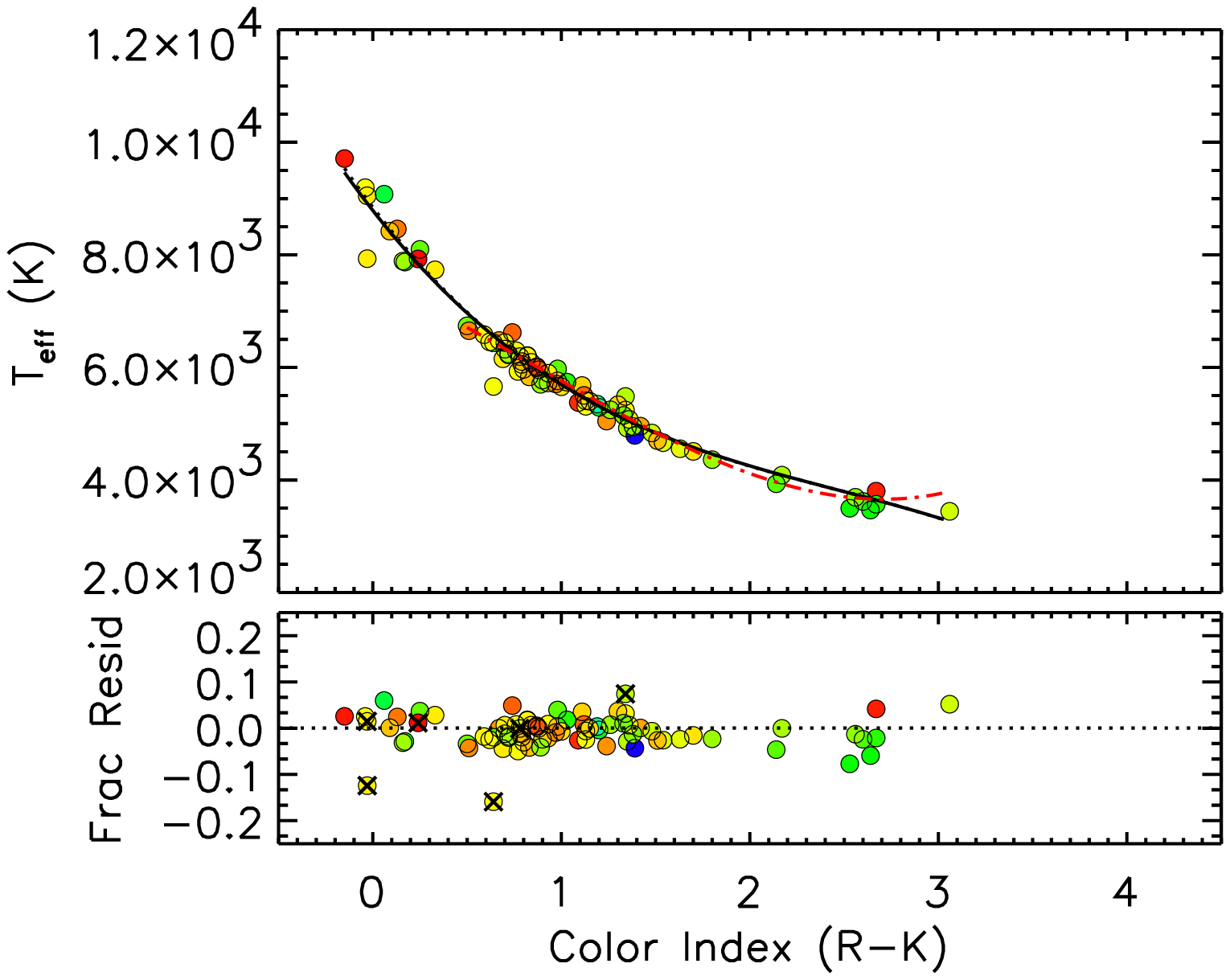, width=0.5\linewidth, clip=} &
\epsfig{file=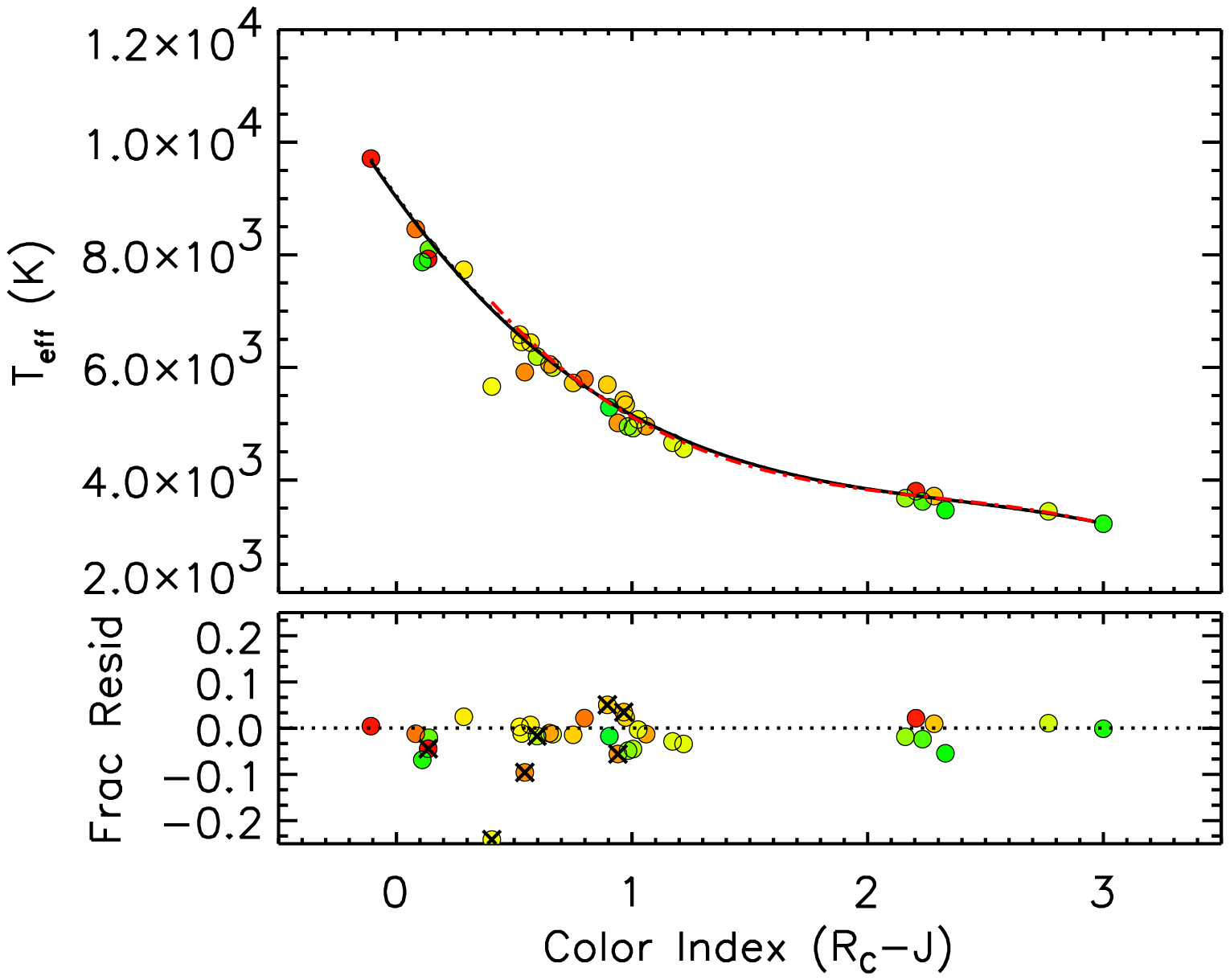, width=0.5\linewidth, clip=} \\
\epsfig{file=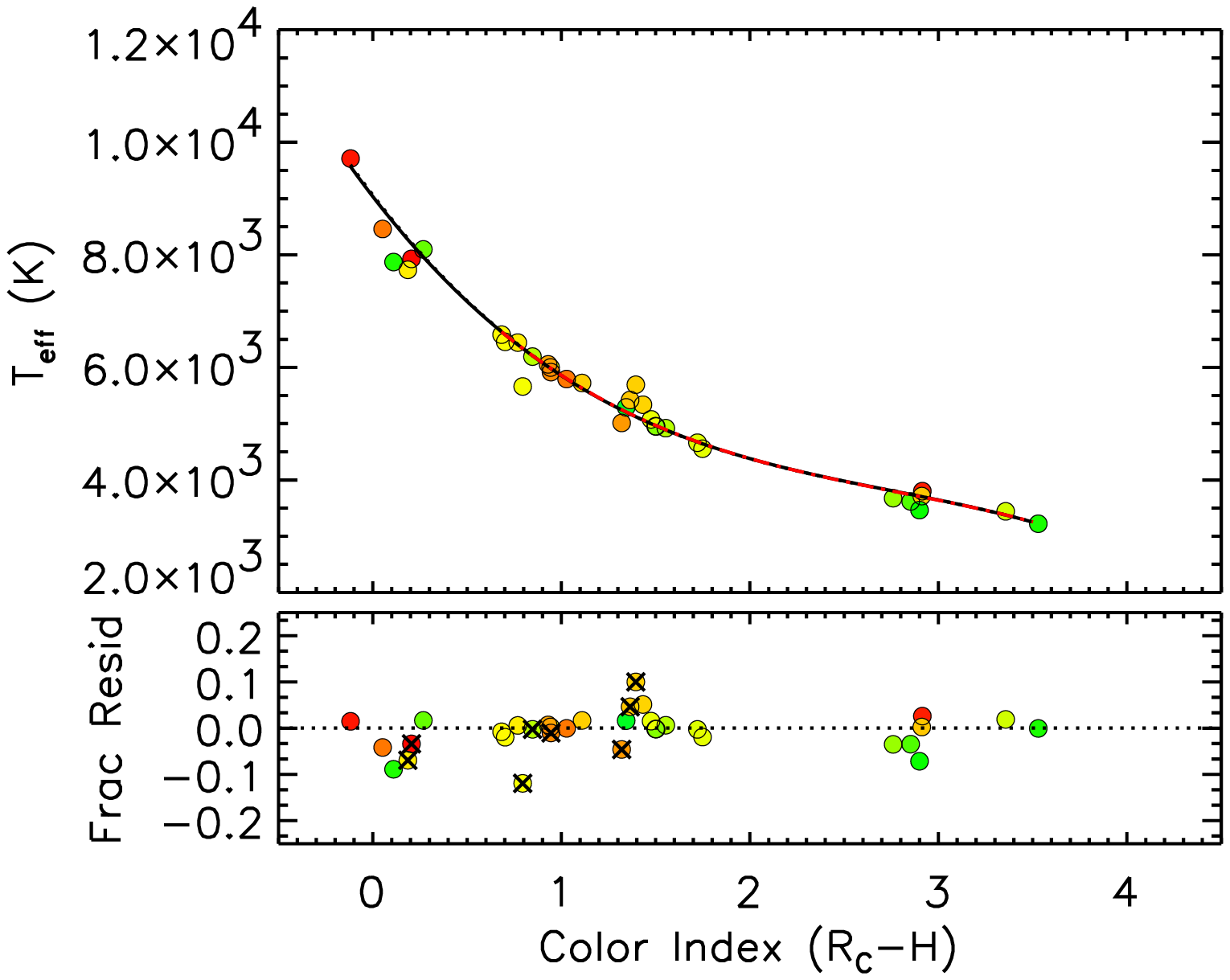, width=0.5\linewidth, clip=} &
\epsfig{file=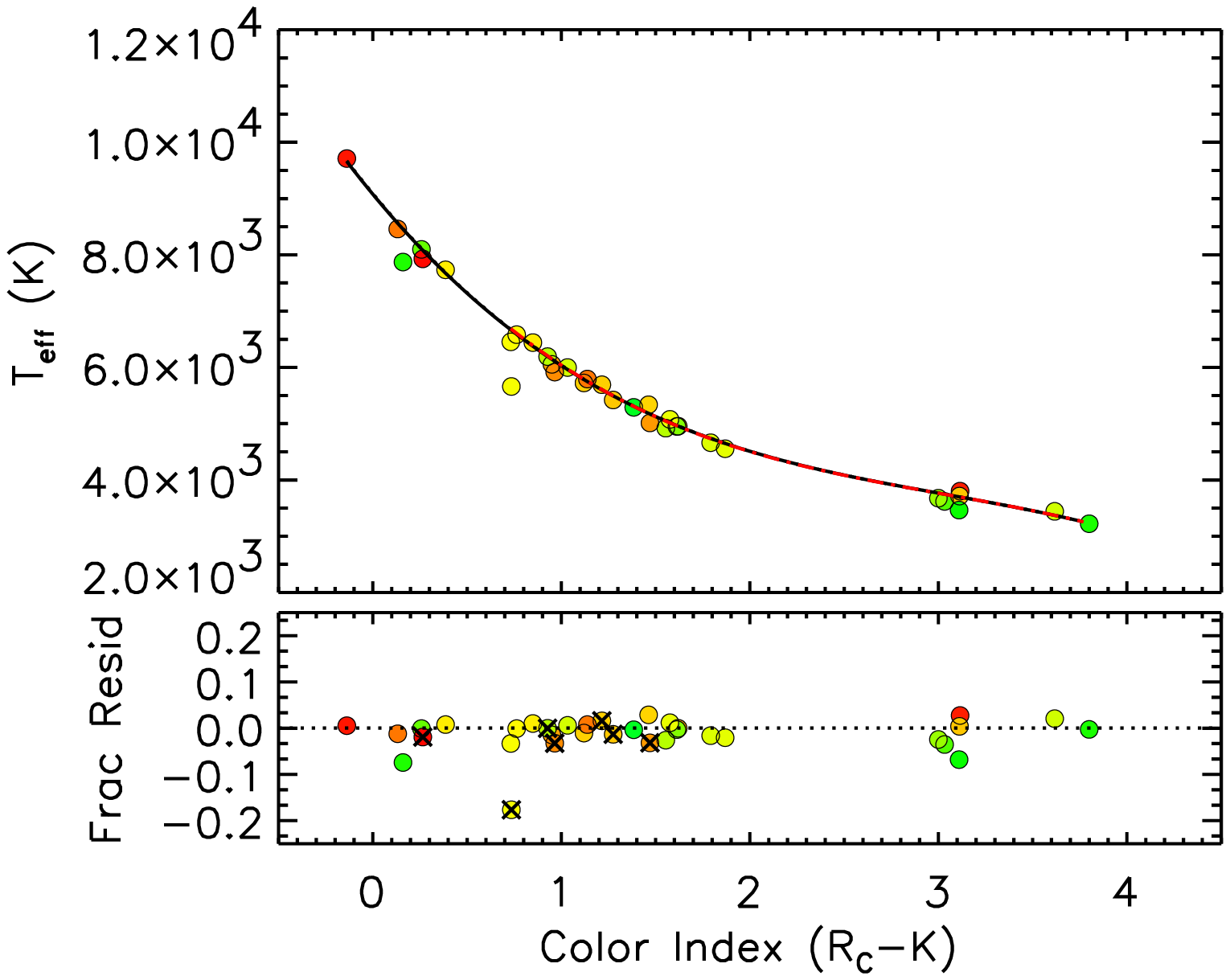, width=0.5\linewidth, clip=} 
 \end{tabular}
 \caption[ ] {The solid black line represents the solution to the color-temperature relation (expressed as Equation~\ref{eq:poly3c} and reported in Table~\ref{tab:poly3_coeffs}).  The red dash-dot line represents the solution omitting the early-type stars (Section~\ref{sec:discussion_astars_evolution}, Equation~\ref{eq:poly3c}, Table~\ref{tab:poly3_coeffs}).  The color of the data point reflects the metallicity of the star, and temperature errors are not shown but typically are smaller than the data point.  Those panels that involve infrared {\it JHK} colors have a second solution plotted as a dotted line (mostly eclipsed by the solid line solution). The bottom panel shows the fractional residual ($T_{\rm Obs.} - T_{\rm Fit})/T_{\rm Obs.}$ to the $3^{rd}$ order polynomial fit, where the dotted line indicates zero deviation. Points with saturated {\it 2MASS} photometry are marked with an $\times$ in the bottom panel (see Section~\ref{sec:discussion_ircolors} for details). See Section~\ref{sec:discussion}, Section~\ref{sec:discussion_astars_evolution}, and Section~\ref{sec:discussion_ircolors} for details. }
 \label{fig:relations3}
 \end{figure}

\begin{figure}										
\centering
\begin{tabular}{cc}
\epsfig{file=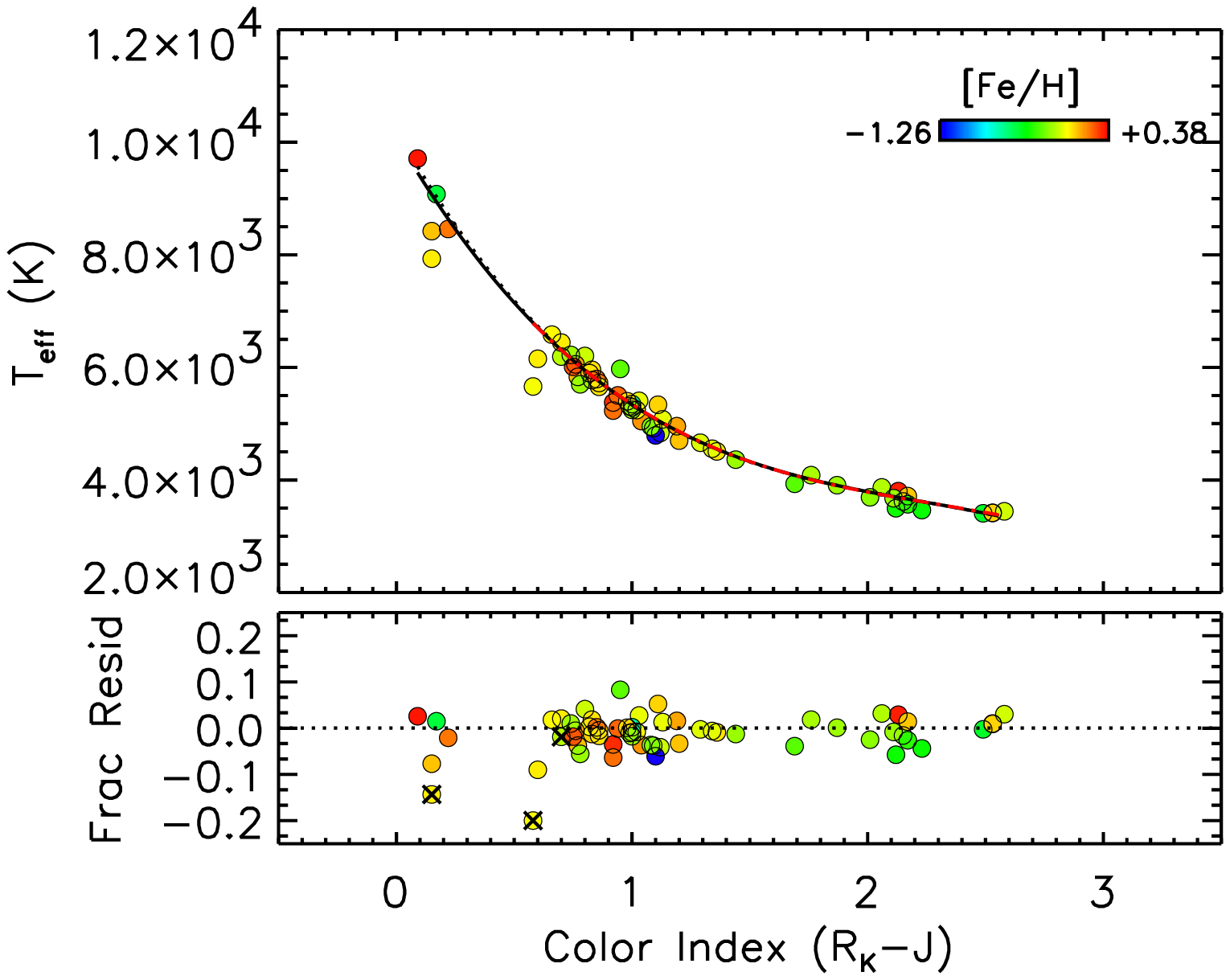, width=0.5\linewidth, clip=} &
\epsfig{file=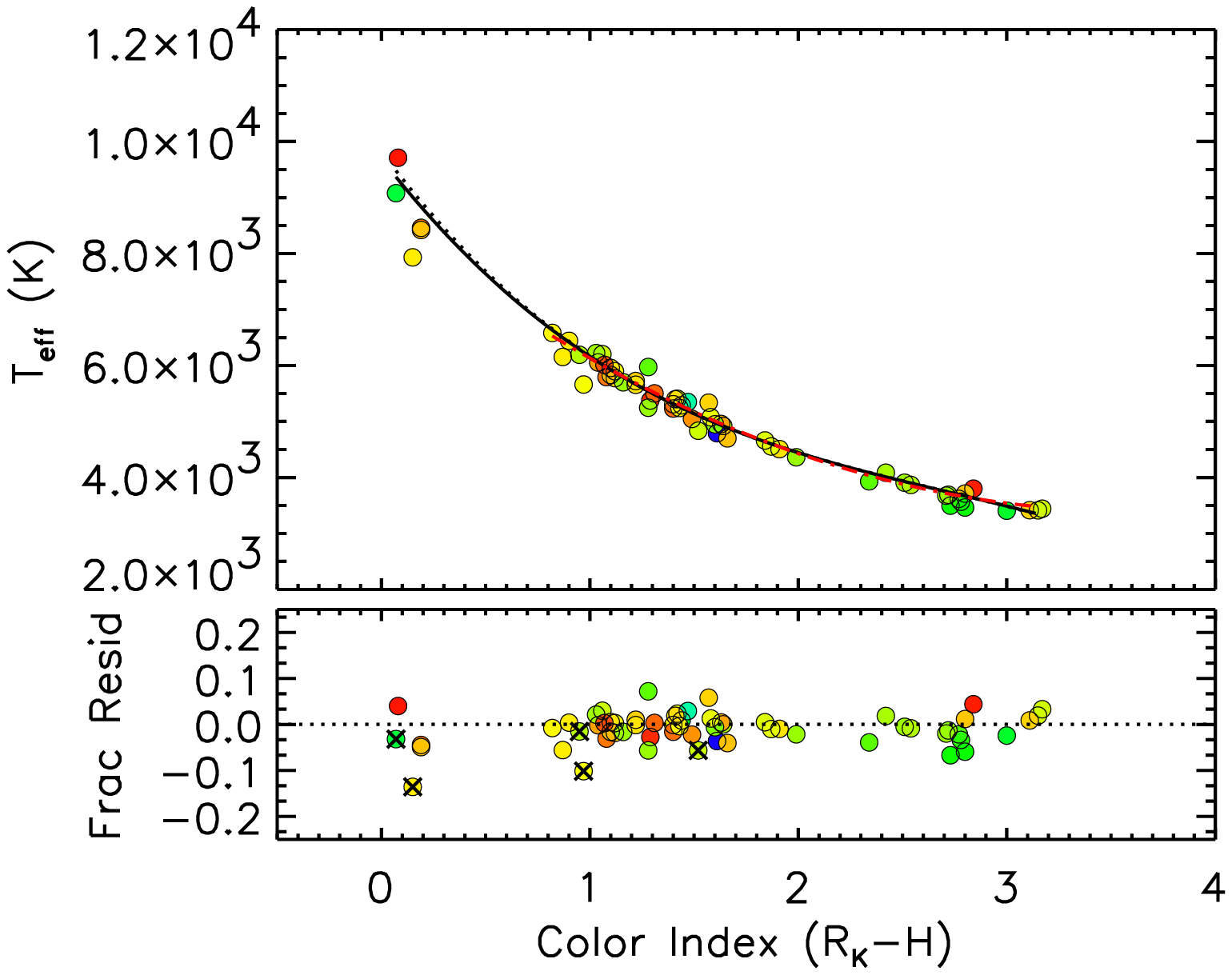, width=0.5\linewidth, clip=} \\
\epsfig{file=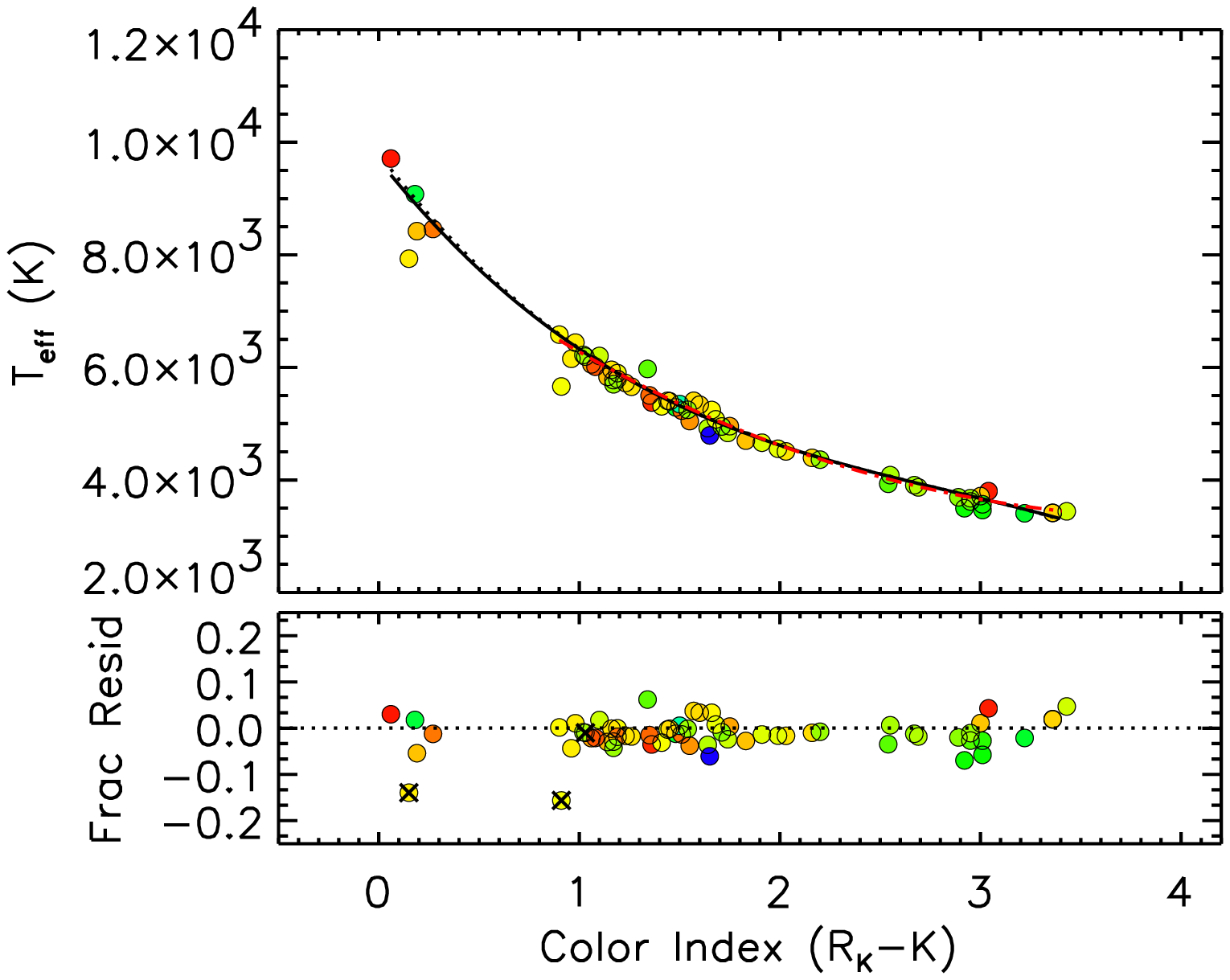, width=0.5\linewidth, clip=} 
 \end{tabular}
 \caption[ ] {The solid black line represents the solution to the color-temperature relation (expressed as Equation~\ref{eq:poly3c} and reported in Table~\ref{tab:poly3_coeffs}).  The red dash-dot line represents the solution omitting the early-type stars (Section~\ref{sec:discussion_astars_evolution}, Equation~\ref{eq:poly3c}, Table~\ref{tab:poly3_coeffs}).  The color of the data point reflects the metallicity of the star, and temperature errors are not shown but typically are smaller than the data point.  Those panels that involve infrared {\it JHK} colors have a second solution plotted as a dotted line (mostly eclipsed by the solid line solution). The bottom panel shows the fractional residual ($T_{\rm Obs.} - T_{\rm Fit})/T_{\rm Obs.}$ to the $3^{rd}$ order polynomial fit, where the dotted line indicates zero deviation. Points with saturated {\it 2MASS} photometry are marked with an $\times$ in the bottom panel (see Section~\ref{sec:discussion_ircolors} for details). See Section~\ref{sec:discussion}, Section~\ref{sec:discussion_astars_evolution}, and Section~\ref{sec:discussion_ircolors} for details. }
 \label{fig:relations4}
 \end{figure}

\begin{figure}										
\centering
\begin{tabular}{cc}
\epsfig{file=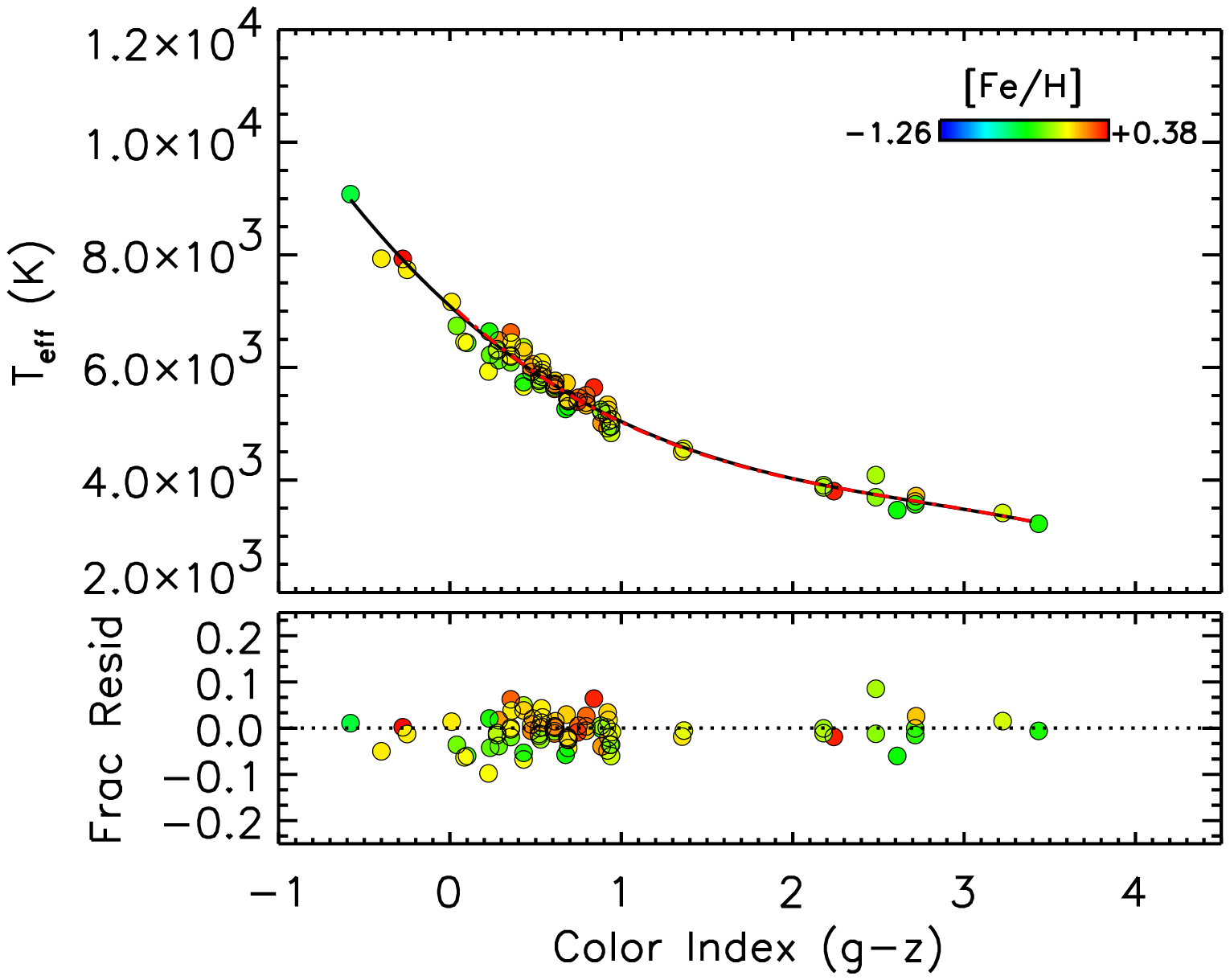, width=0.5\linewidth, clip=} &
\epsfig{file=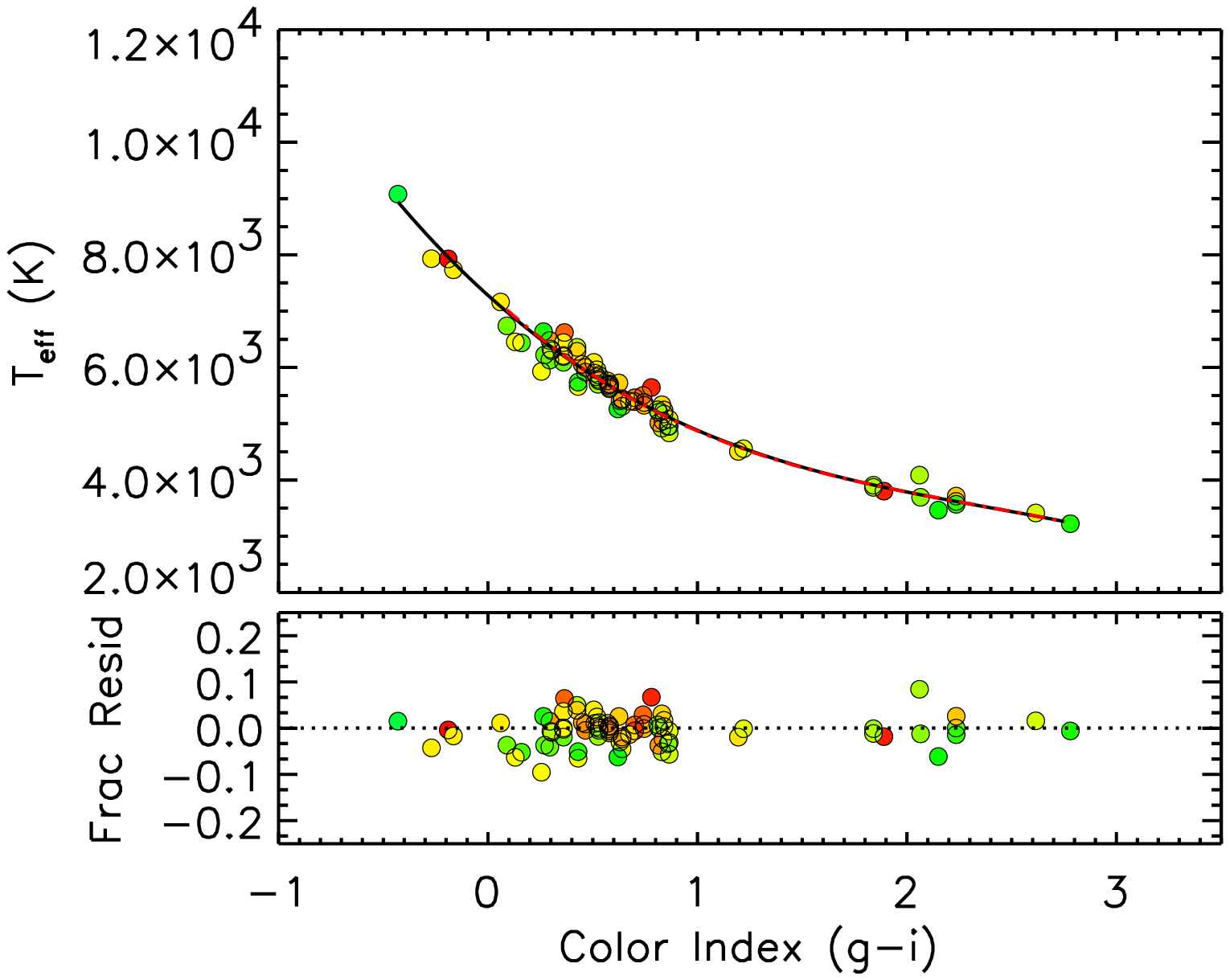, width=0.5\linewidth, clip=} \\
\epsfig{file=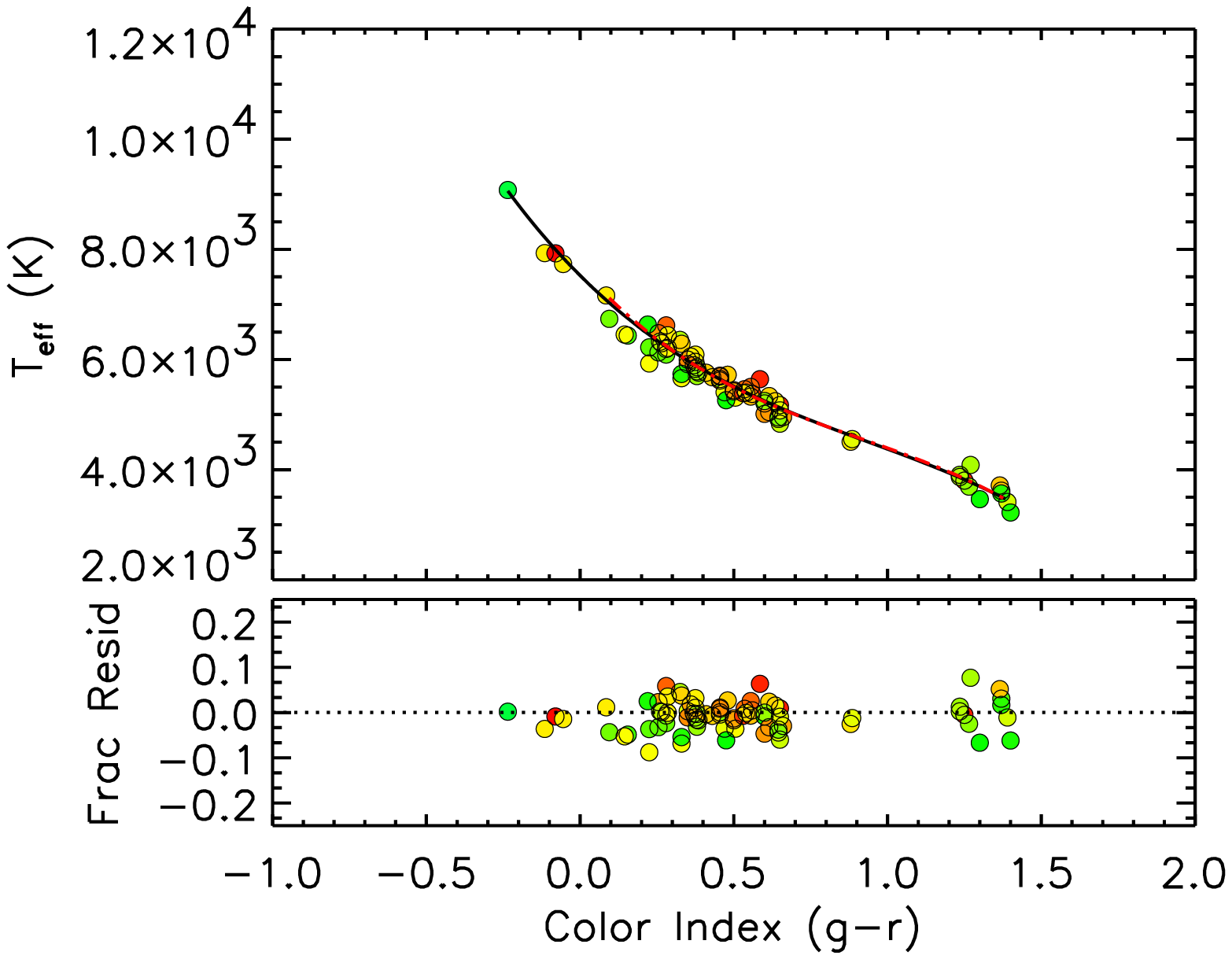, width=0.5\linewidth, clip=} &
\epsfig{file=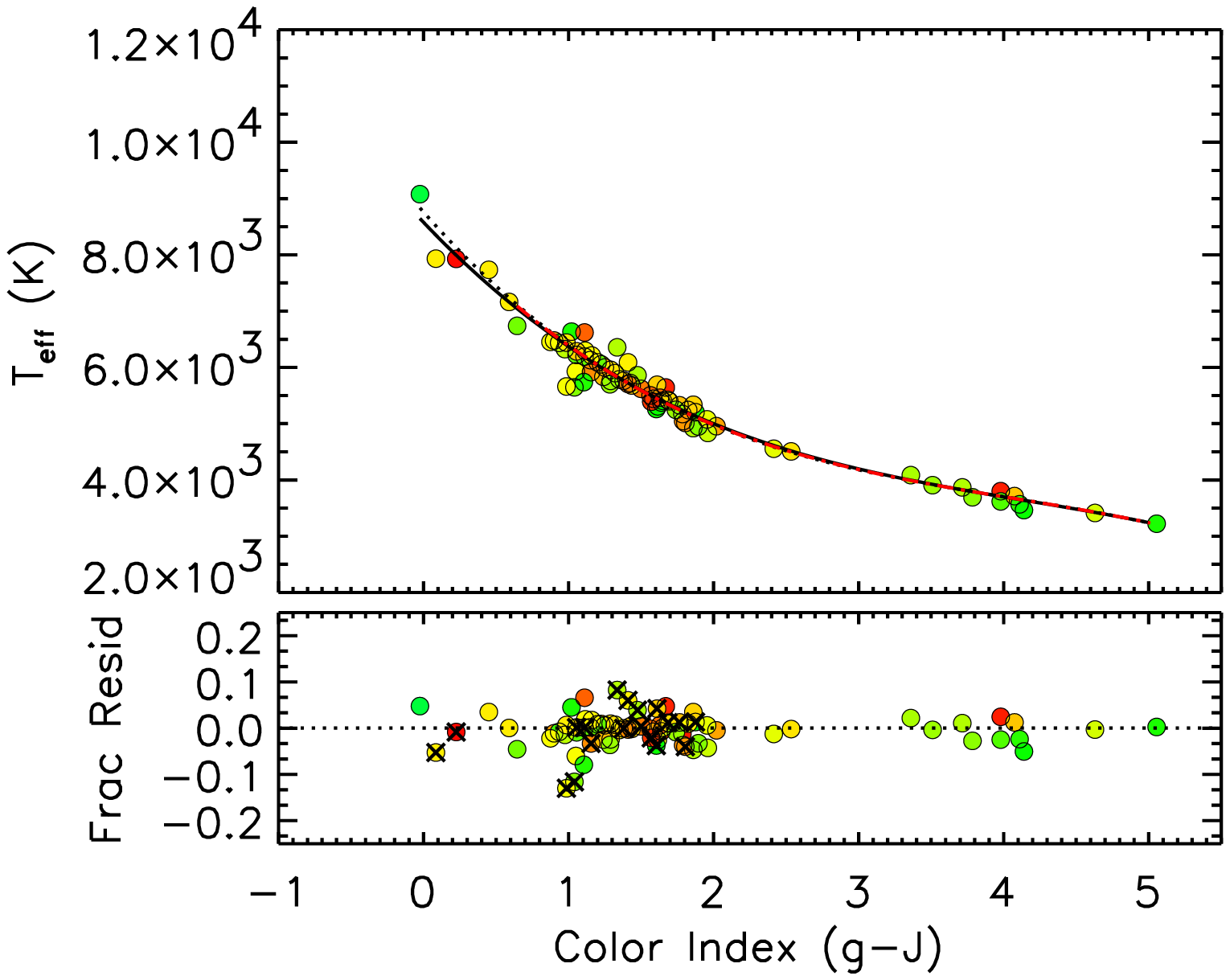, width=0.5\linewidth, clip=} \\
\epsfig{file=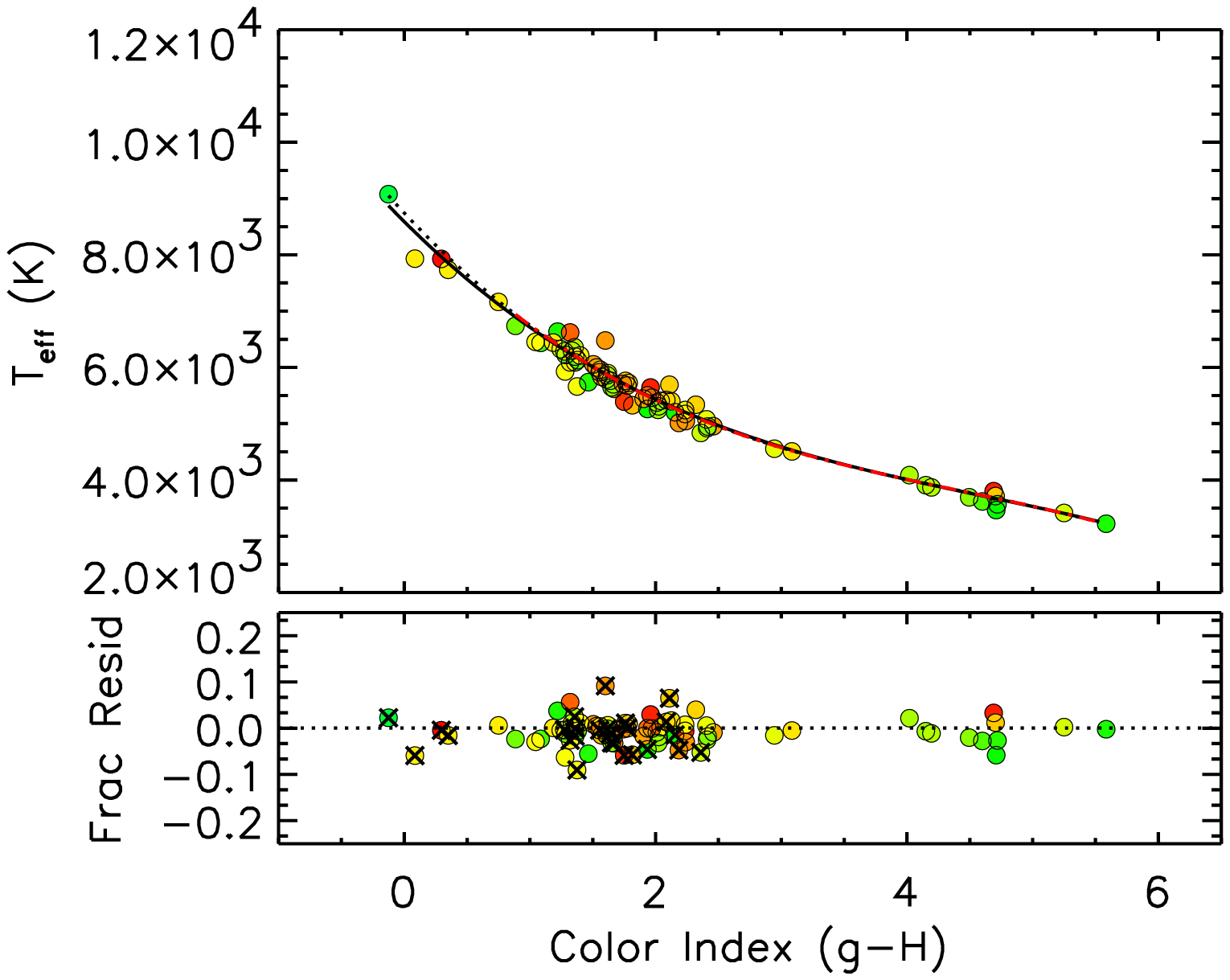, width=0.5\linewidth, clip=} &
\epsfig{file=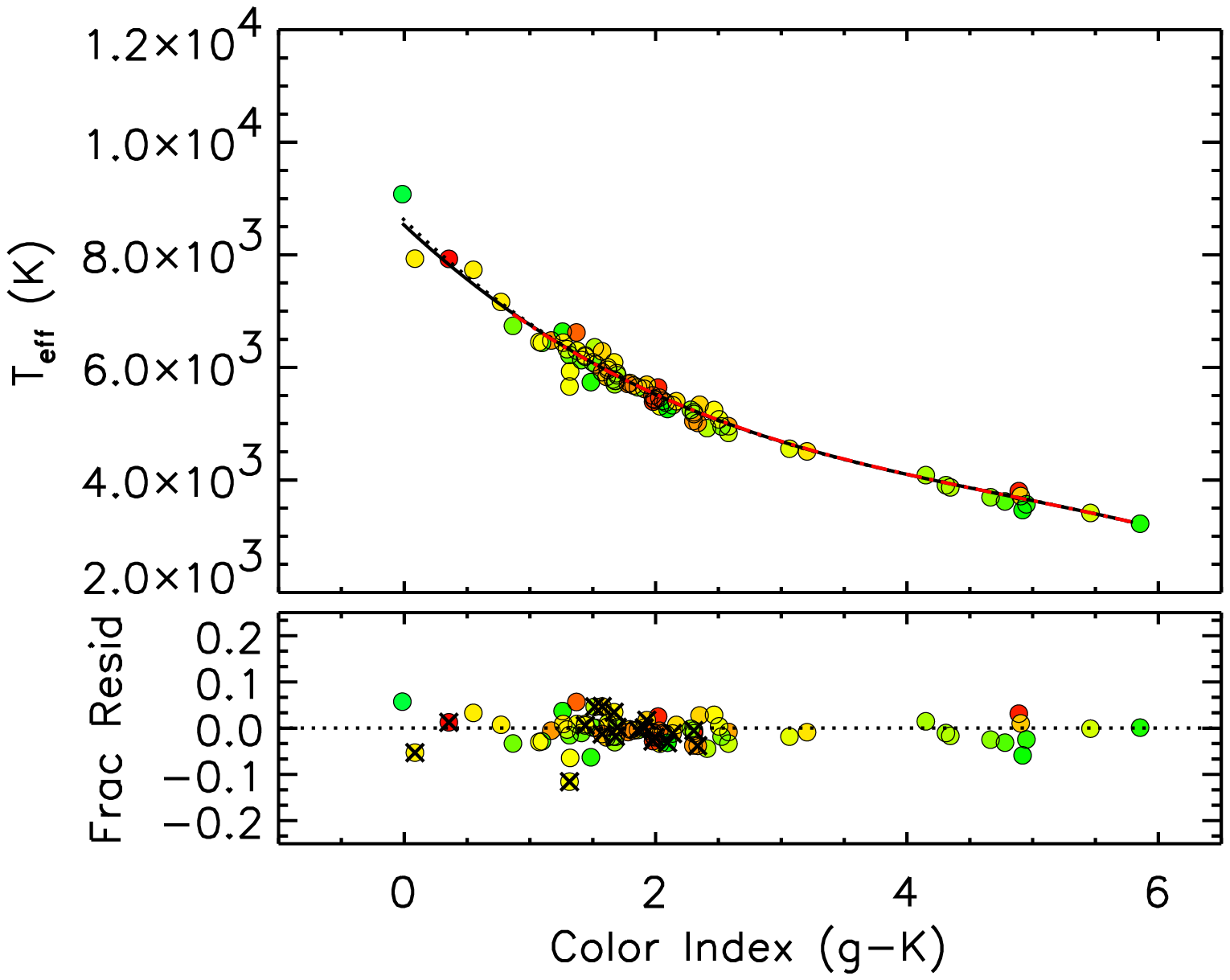, width=0.5\linewidth, clip=} 
 \end{tabular}
 \caption[ ] {The solid black line represents the solution to the color-temperature relation (expressed as Equation~\ref{eq:poly3c} and reported in Table~\ref{tab:poly3_coeffs}).  The red dash-dot line represents the solution omitting the early-type stars (Section~\ref{sec:discussion_astars_evolution}, Equation~\ref{eq:poly3c}, Table~\ref{tab:poly3_coeffs}).  The color of the data point reflects the metallicity of the star, and temperature errors are not shown but typically are smaller than the data point.  Those panels that involve infrared {\it JHK} colors have a second solution plotted as a dotted line (mostly eclipsed by the solid line solution). The bottom panel shows the fractional residual ($T_{\rm Obs.} - T_{\rm Fit})/T_{\rm Obs.}$ to the $3^{rd}$ order polynomial fit, where the dotted line indicates zero deviation. Points with saturated {\it 2MASS} photometry are marked with an $\times$ in the bottom panel (see Section~\ref{sec:discussion_ircolors} for details). See Section~\ref{sec:discussion}, Section~\ref{sec:discussion_astars_evolution}, and Section~\ref{sec:discussion_ircolors} for details. }
 \label{fig:relations5}
 \end{figure}

\begin{figure}										
\centering
\begin{tabular}{cc}
\epsfig{file=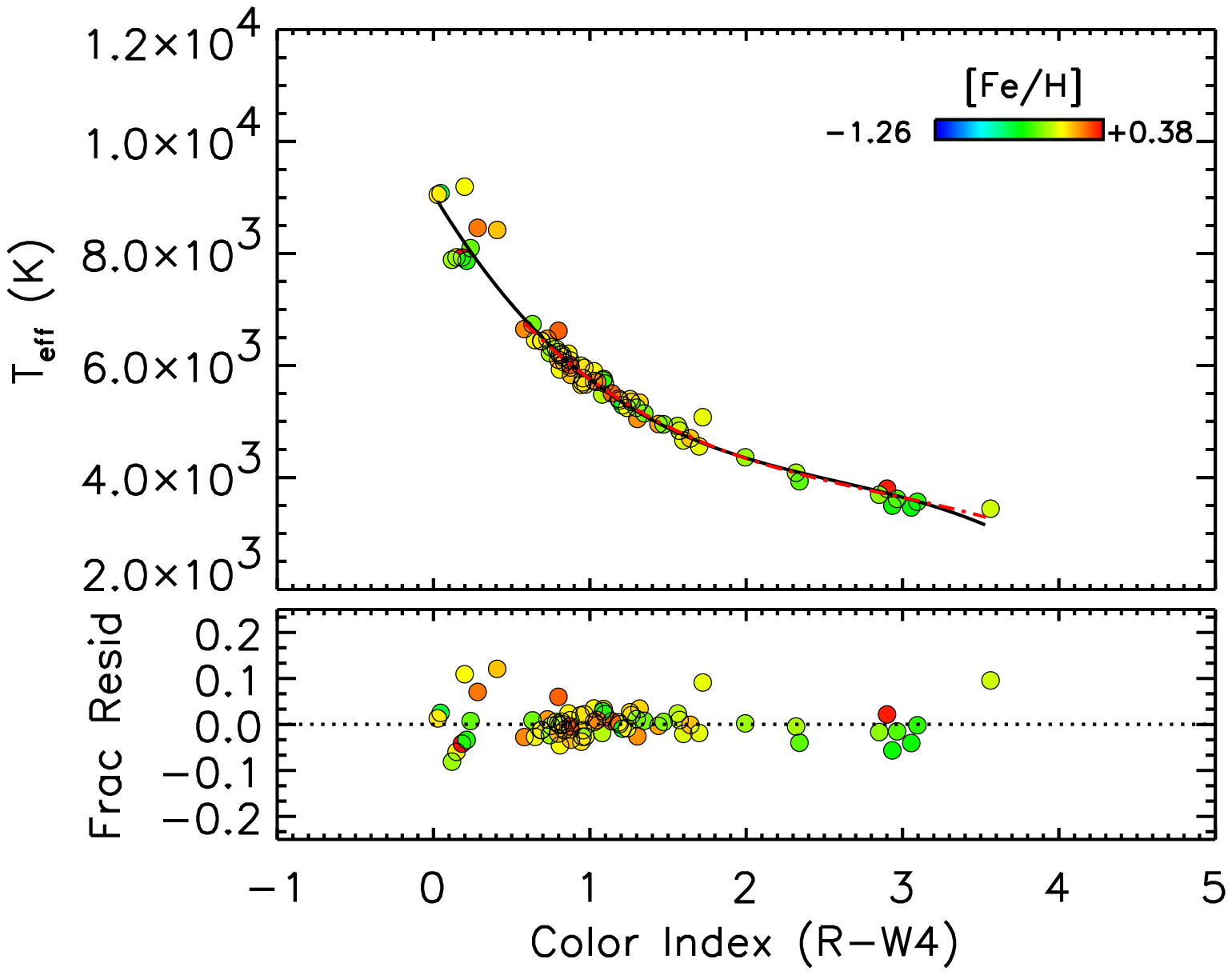, width=0.5\linewidth, clip=} &
\epsfig{file=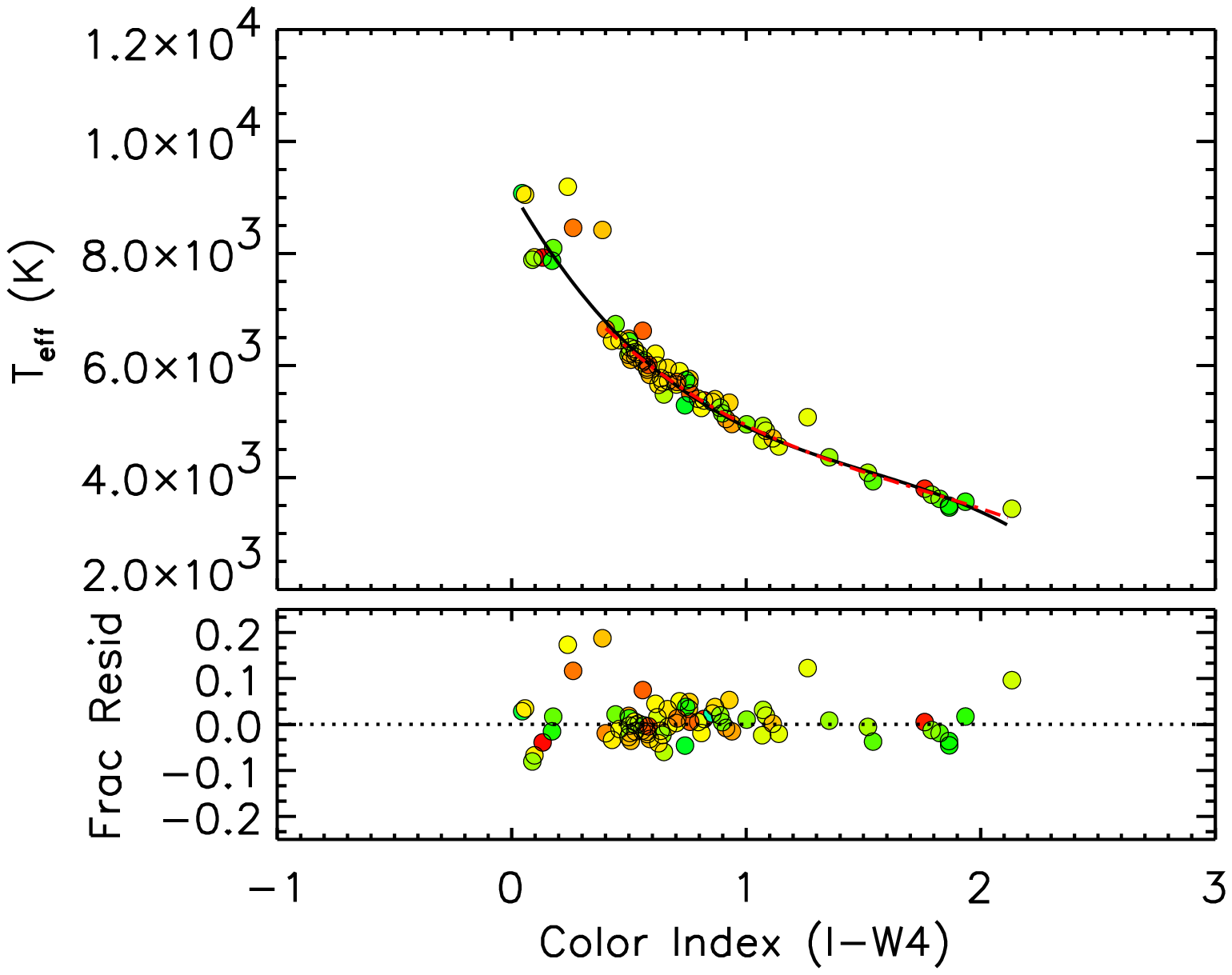, width=0.5\linewidth, clip=} \\
\epsfig{file=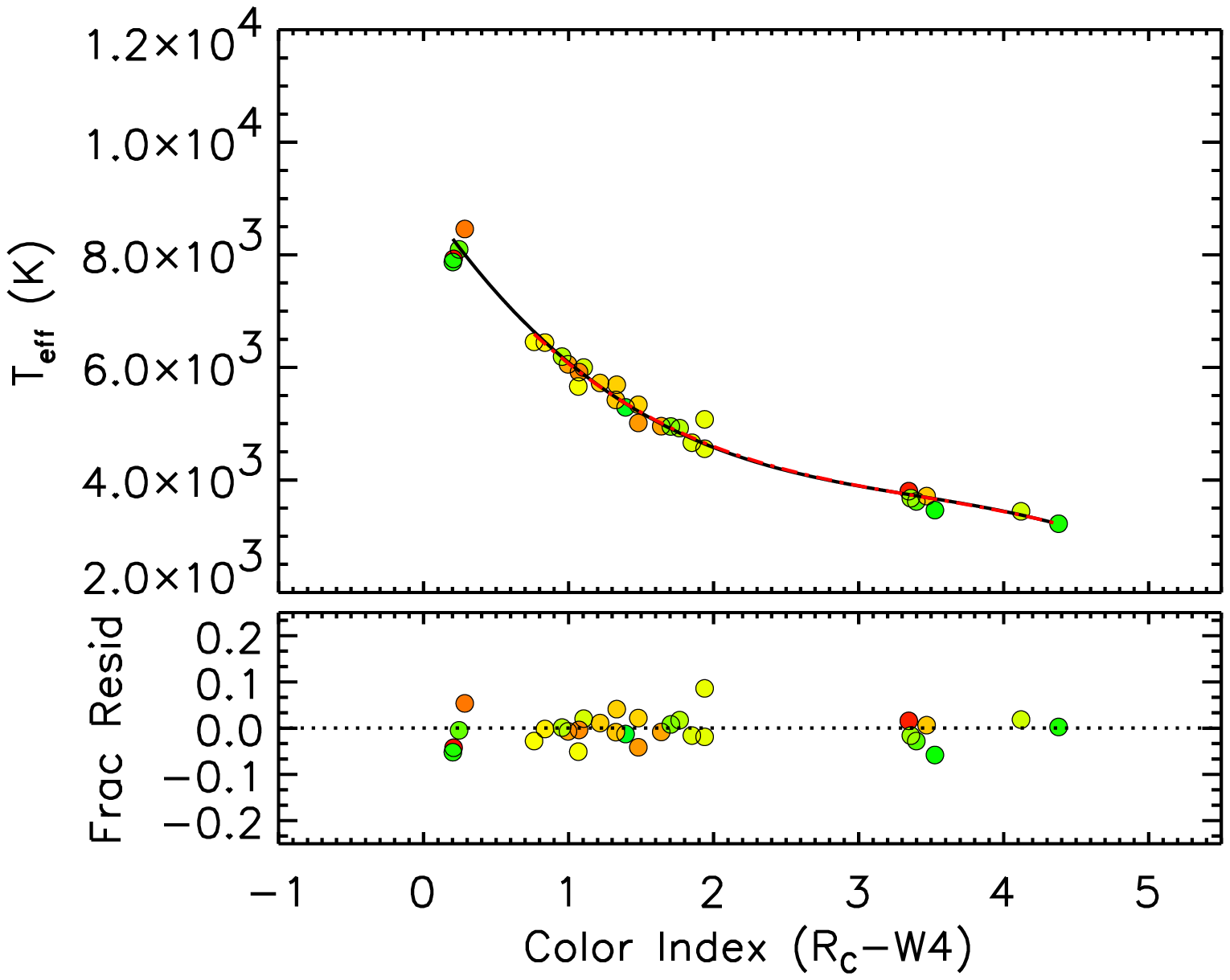, width=0.5\linewidth, clip=} &
\epsfig{file=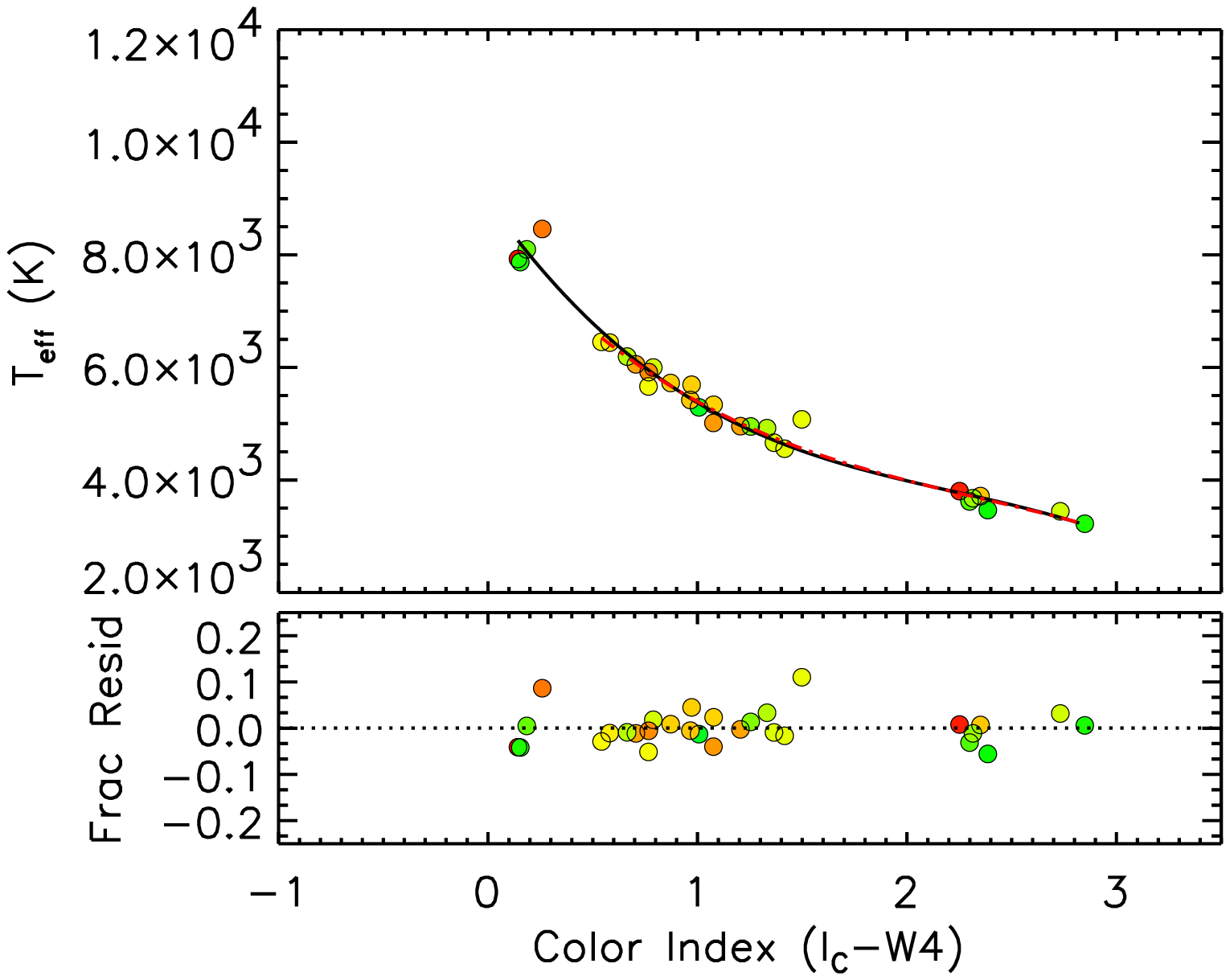, width=0.5\linewidth, clip=} \\
\epsfig{file=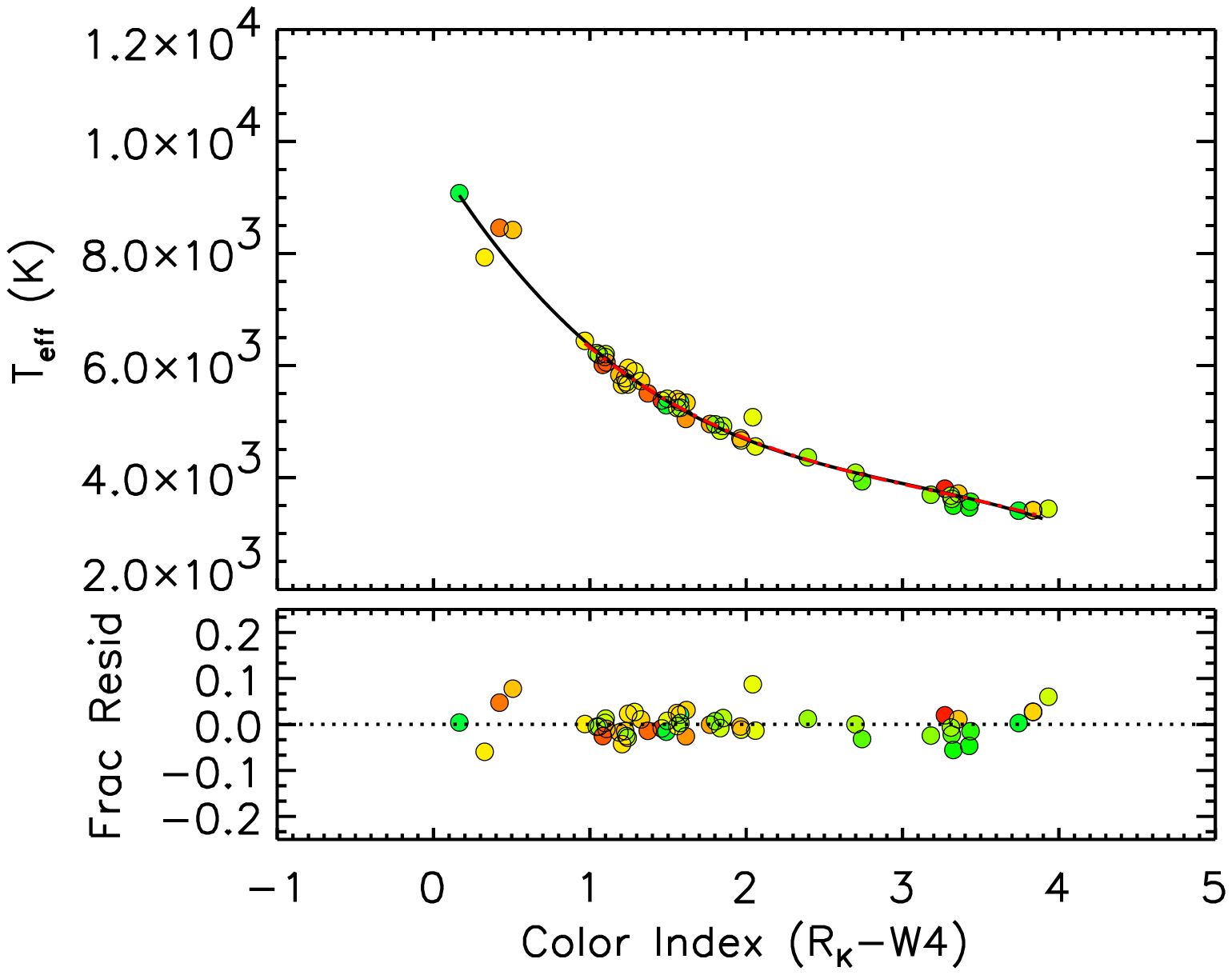, width=0.5\linewidth, clip=} &
\epsfig{file=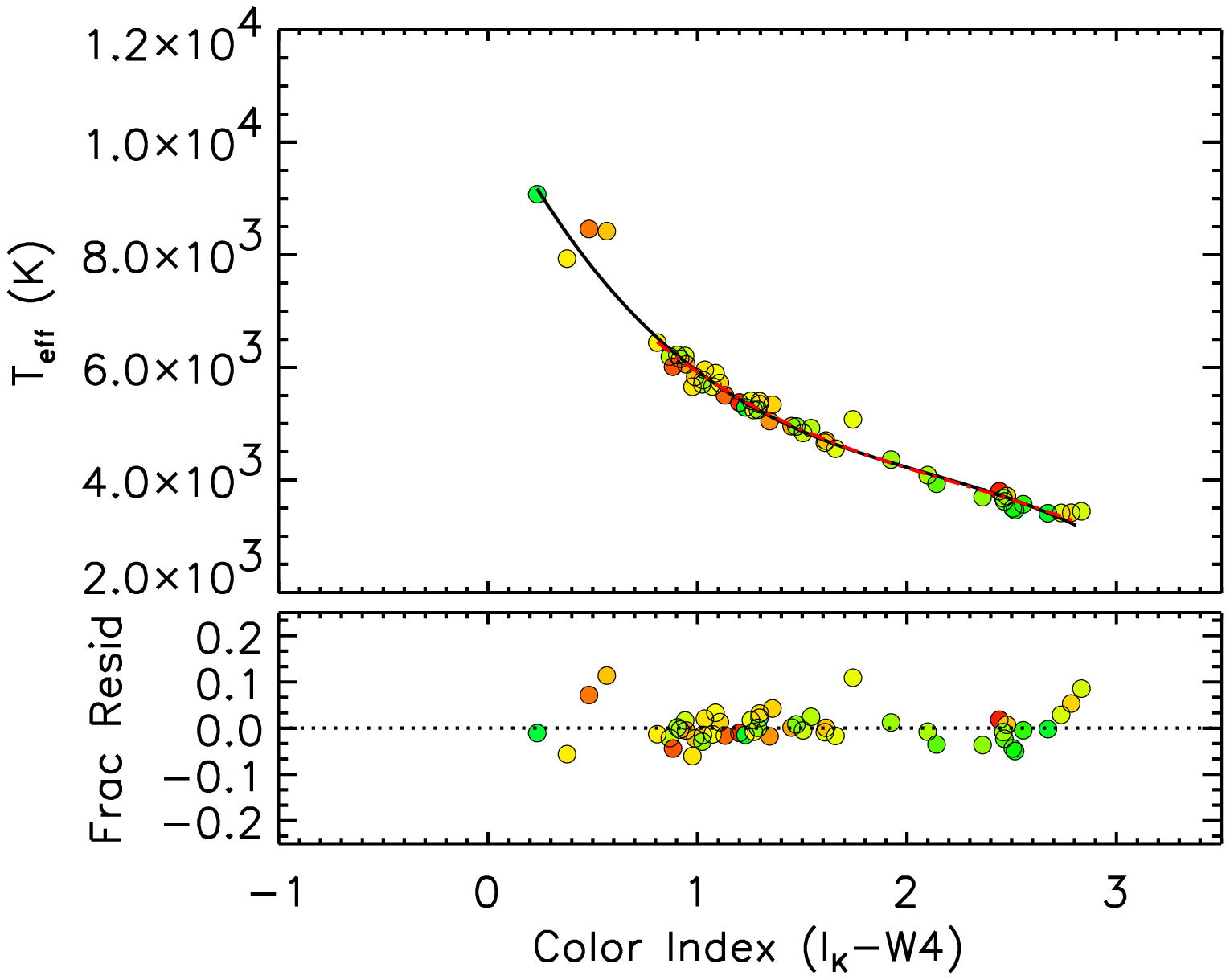, width=0.5\linewidth, clip=} 
 \end{tabular}
 \caption[ ] {The solid black line represents the solution to the color-temperature relation (expressed as Equation~\ref{eq:poly3c} and reported in Table~\ref{tab:poly3_coeffs}).  The red dash-dot line represents the solution omitting the early-type stars (Section~\ref{sec:discussion_astars_evolution}, Equation~\ref{eq:poly3c}, Table~\ref{tab:poly3_coeffs}).  The color of the data point reflects the metallicity of the star, and temperature errors are not shown but typically are smaller than the data point.  The bottom panel shows the fractional residual ($T_{\rm Obs.} - T_{\rm Fit})/T_{\rm Obs.}$ to the $3^{rd}$ order polynomial fit, where the dotted line indicates zero deviation. See Section~\ref{sec:discussion} and Section~\ref{sec:discussion_astars_evolution} for details. }
 \label{fig:relations6}
 \end{figure}

\begin{figure}										
  \centering
  \begin{tabular}{c}
         \epsfig{file=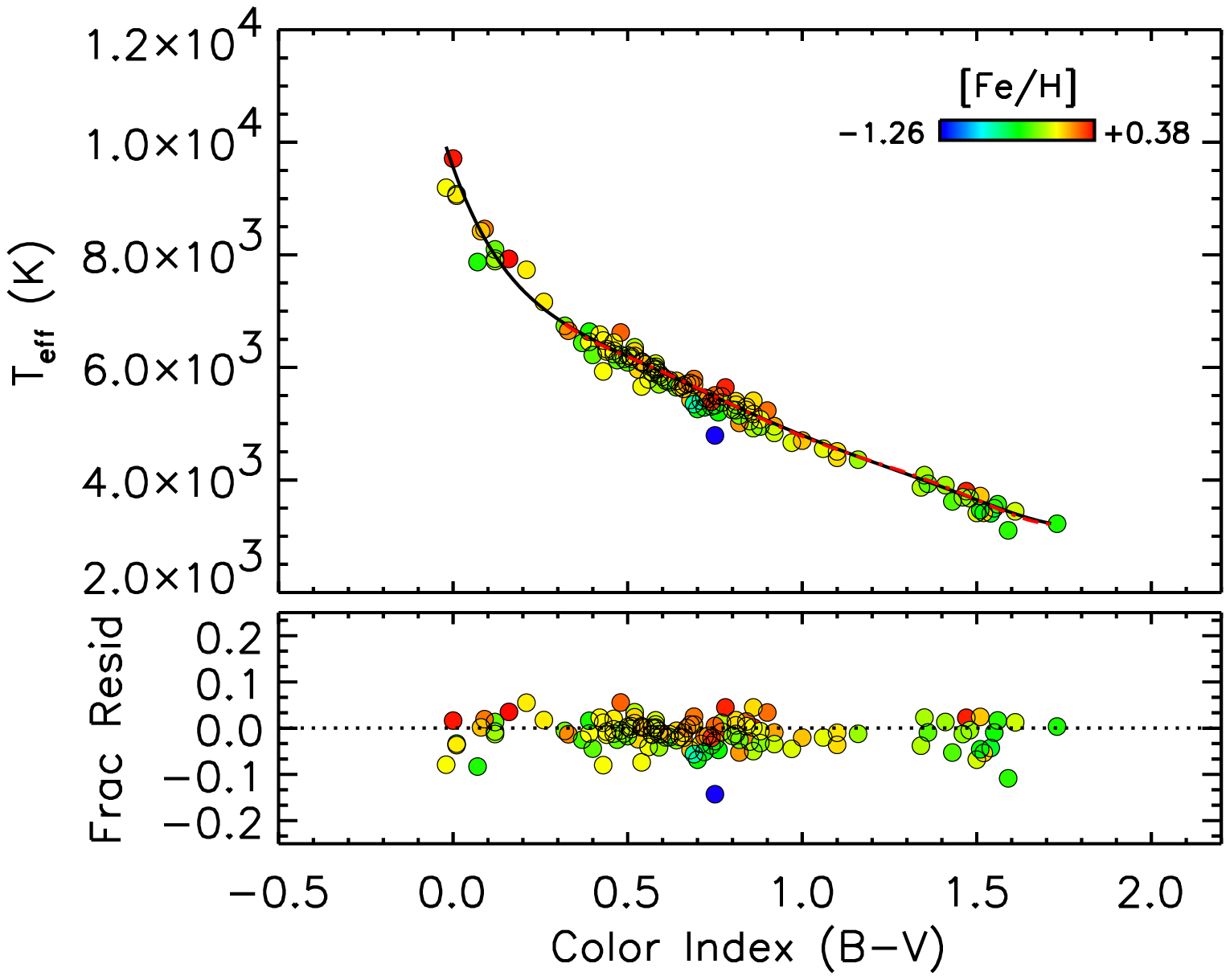, width=0.55\linewidth, clip=} \\
         \epsfig{file=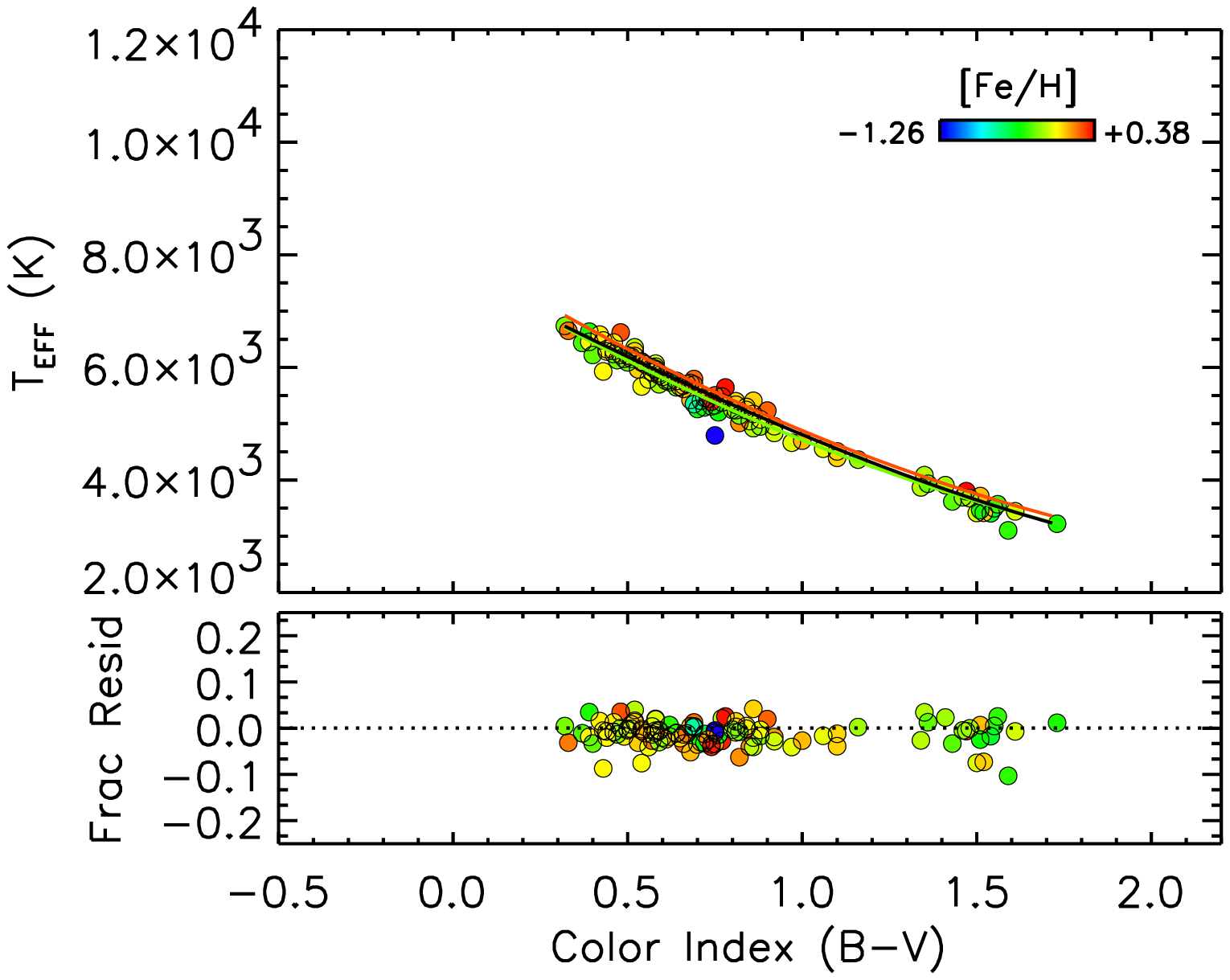, width=0.55\linewidth, clip=} 
  \end{tabular}
  \caption[ ] {Alternate solutions for $(B-V)$ - temperature relations.  The top plot shows the data, fit (solid black line), and fractional residuals to the $6^{th}$ order polynomial function (Section~\ref{sec:discussion_bmv}, Equation~\ref{eq:poly6c}), as well as the fit for the $3^{rd}$ order function omitting the early-type stars (red dash-dot line).  Note the difference in the residuals between $0.3 < (B-V) < 0.5$ for this solution and the ones for the $3^{rd}$ order polynomial fit to the full AFGKM star sample shown in Figure~\ref{fig:relations1}.   The bottom plot shows the solution for $(B-V)$ - metallicity - temperature relation (Equation~\ref{eq:poly2cm}) omitting early-type stars discussed in Section~\ref{sec:discussion_bmv}. Iso-metallicity lines of [Fe/H]~$=-0.25, 0.0, +0.25$ are plotted in green, black, and red, respectively. Note that there are no artifacts in the residuals with respect to metallicity or specific ranges in color index with the solution displayed in the bottom plot. }
   \label{fig:Temp_VS_BmV_p6}
   \end{figure}

\begin{figure}										
  \centering
         \epsfig{file=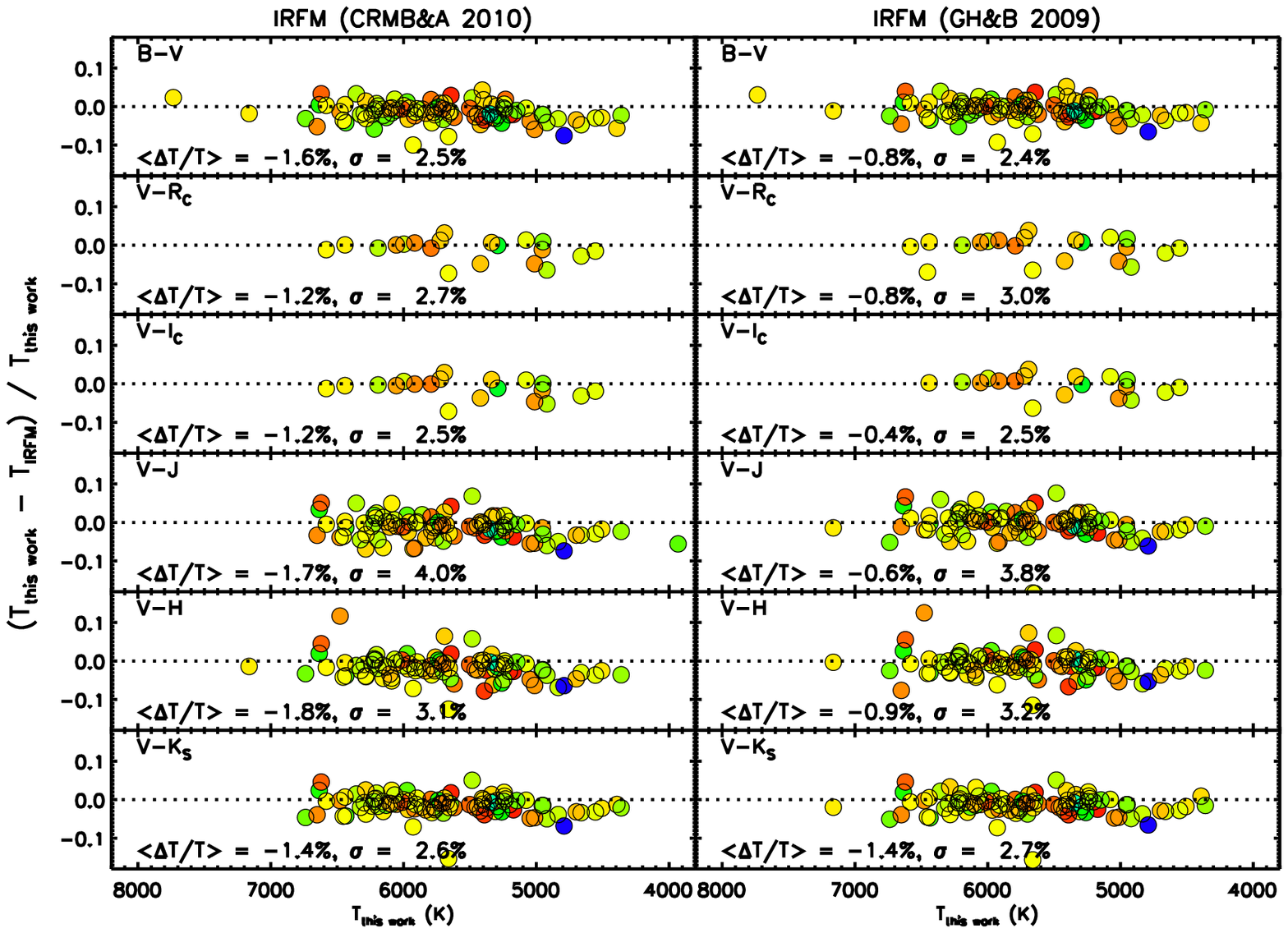, width=.9\linewidth, clip=} 
  \caption[ ] {Fractional deviation in effective temperature for interferometrically determined temperatures ($T_{\rm this~work}$) compared to the effective temperatures derived using the polynomial relations in \citet{cas10} ({\it left}; CRMB\&A 2010) and \citet{gon09} ({\it right}; GH\&B 2009), established via the IRFM ($T_{\rm IRFM}$). The top-left of each panel lists the color index used in the IRFM relation and bottom portion displays the average percentage deviation in temperature (calculated as $(T_{\rm this~work} - T_{\rm IRFM})/T_{\rm this~work})$), and scatter of the data $\sigma$ in percent. The dotted line indicates zero deviation, and the color of the data point reflects the metallicity of the star ranging from [Fe/H]$=-1.26$ to $0.38$ (refer to previous figures for legend).  See Section~\ref{sec:discussion_teff_others} for details.  }
   \label{fig:Teff_VS_Cas10_Gon09_multiplot}
   \end{figure}

\begin{figure}										
  \centering
         \epsfig{file=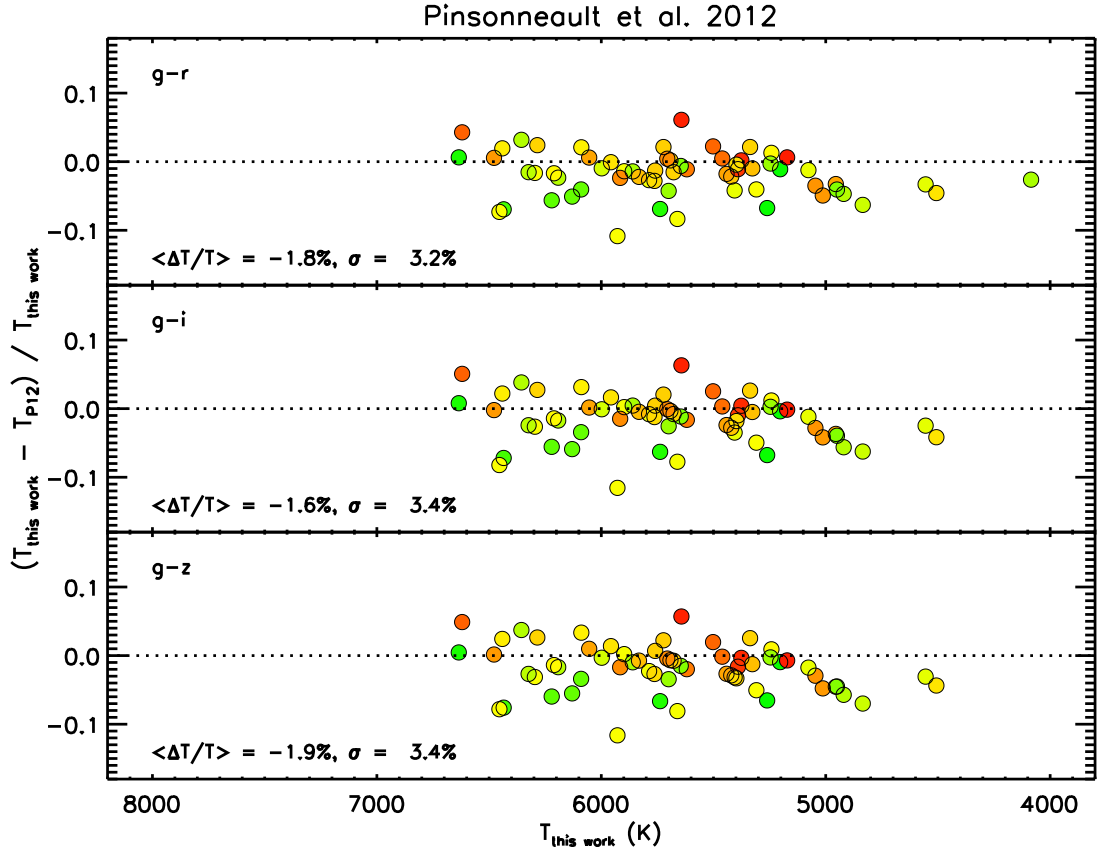, width=.9\linewidth, clip=} 
  \caption[ ] {Fractional deviation in effective temperature for interferometrically determined temperatures ($T_{\rm this~work}$) compared to the effective temperatures derived using the polynomial relations in \citet{pin12} (P12). The top-left of each panel lists the color index used in the relation and bottom portion displays the average percentage deviation in temperature (calculated as $(T_{\rm this~work} - T_{\rm P12})/T_{\rm this~work})$), and scatter of the data $\sigma$ in percent. The dotted line indicates zero deviation, and the color of the data point reflects the metallicity of the star ranging from [Fe/H]$=-1.26$ to $0.38$ (refer to previous figures for legend).  See Section~\ref{sec:discussion_teff_others} for details.  }
   \label{fig:Teff_VS_Pin12_multiplot}
   \end{figure}

\begin{figure}										
  \centering
	\epsfig{file=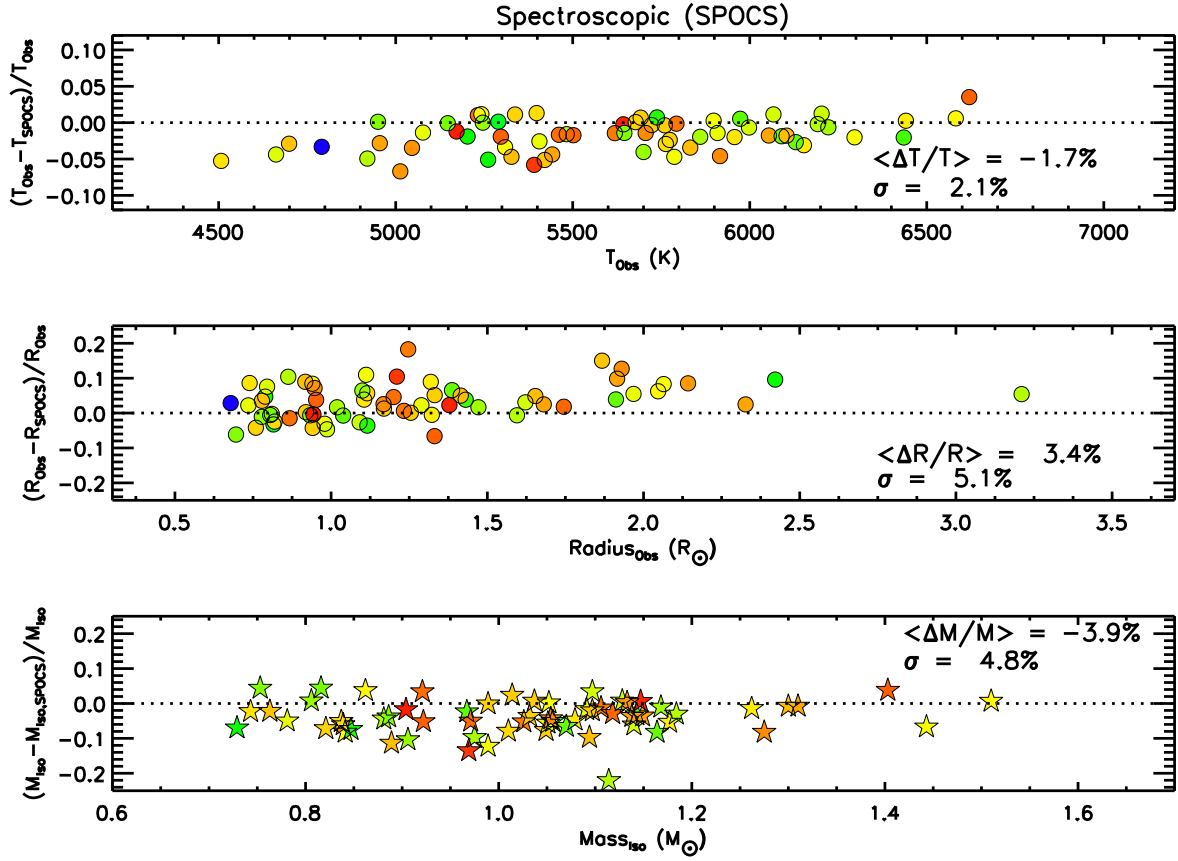, width=.9\linewidth, clip=} 
  \caption[ ] {The top and middle panel show the fractional deviation in interferometrically determined effective temperatures and radii compared to the spectroscopic values in the SPOCS Catalog \citep{val05}. The bottom panel shows the fractional deviation of stellar masses derived in this work versus those derived in the SPOCS Catalog by interpolation within the $Y^2$ isoschrones.  We use different symbols for the points in the bottom panel to accentuate the fact that the origin masses for each are derived from model isochrones. Printed in the left-hand-side of each window are the average percentage deviation for each variable, and the scatter of the data $\sigma$ in percent. The dotted line indicates zero deviation, and the color of the data point reflects the metallicity of the star ranging from [Fe/H]$=-1.26$ to $0.38$ (refer to previous figures for legend).  See Section~\ref{sec:discussion_teff_others} for details. }
   \label{fig:Compare_spocs_feh_diff_multi}
   \end{figure}

Measurements of stellar luminosities and radii are historically among the most difficult fundamental measurements in astronomy.  Access to astrometric surveys from space, and availability of optical/infrared facilities on the ground, have provided a break-through in these measurements.  The status of such studies for bright, nearby main sequence stars is well represented graphically in Figures~\ref{fig:Lumin_VS_Teff_VS_Radius_color}, \ref{fig:Teff_VS_Radius_color}, \ref{fig:Mass_VS_Radius_color}, \ref{fig:Mass_VS_Teff_color}, and \ref{fig:Mass_VS_Lumin_VS_Radius_log_color}.  During the last few years, the number of direct measurements of the class as a whole has grown substantially in size and with increased precision. This improvement can be extended to fainter and more distant starts by using these results to improve the calibration of the IRFM or similar methods.  We also see that the scatter (presumably astrophysical noise) is now greater than our best estimate of the measurement errors.  As shown in Section~\ref{sec:discussion_2mass}, some of this scatter is likely due to metallicity and large photometric errors of such bright and saturated sources. Independent and uniform measures of metallicity will prove to be most informative on the improvement of existing calibrations presented here.  Other sources of scatter must exist, but are difficult to allow for in the study of the full ensemble of targets.

The future of interferometric measurements is promising, where appropriate technical improvements (at CHARA this would involve the use of adaptive optics, at VLTI perhaps a new beam combiner) will lead to single target precision of order 1\%, extending the observable number of targets with interferometry many fold.  While the improvement of diameter precision from $2-3$\% to $\sim 1$\% will have great value, it will soon approach the point where the sample is limited by targets that have distance measurements, absolute photometric calibrations, and measurements of metallicities at this level.  The ability to learn such absolute properties of stars can open the door to study of essential parameters and phenomena such as age, rotation, and magnetic fields, whose impact on evolution may be important but difficult to detect.



\acknowledgments

TSB acknowledges support provided through NASA grant ADAP12-0172.  STR acknowledges partial support from NASA grant NNH09AK731.  We thank the referee for many helpful comments to improve the paper.  We thank Mike Bessell for his helpful advice on photometric transformations. We thank Mike Ireland, Tim White, Vincente Maestro, and Daniel Huber for enlightening discussions on precision bolometric flux measurements. The CHARA Array is funded by the National Science Foundation through NSF grants AST-0908253 and AST 1211129, and by Georgia State University through the College of Arts and Sciences. This research has made use of the SIMBAD literature database, operated at CDS, Strasbourg, France, and of NASA's Astrophysics Data System. This research has made use of the VizieR catalogue access tool, CDS, Strasbourg, France. This publication makes use of data products from the Two Micron All Sky Survey, which is a joint project of the University of Massachusetts and the Infrared Processing and Analysis Center/California Institute of Technology, funded by the National Aeronautics and Space Administration and the National Science Foundation.  This publication makes use of data products from the Wide-field Infrared Survey Explorer, which is a joint project of the University of California, Los Angeles, and the Jet Propulsion Laboratory/California Institute of Technology, funded by the National Aeronautics and Space Administration. This research has made use of the JSDC Jean-Marie Mariotti Center database, available at http://www.jmmc.fr/jsdc.

\clearpage
\bibliographystyle{apj}            
\bibliography{apj-jour,paper}      


\newpage
\LongTables

\begin{deluxetable}{rccccc}
\tabletypesize{\tiny}
\tablewidth{0pt}
\tablecaption{Observation Log\label{tab:observations}}
\tablehead{
\colhead{\textbf{ }} &
\colhead{\textbf{UT}} &
\colhead{\textbf{ }} &
\colhead{\textbf{ }} &
\colhead{\textbf{\# of}} &
\colhead{\textbf{Calibrator}}  \\
\colhead{\textbf{Object}} &
\colhead{\textbf{Date}} &
\colhead{\textbf{Baseline}} &
\colhead{\textbf{Filter}} &
\colhead{\textbf{Brackets}} &
\colhead{\textbf{HD}}	
}

\startdata

HD 166	&	2010/09/17	&	E1/W1	&	$H$	&	4	&	HD~1404		\\
		&	2010/09/19	&	S1/E1	&	$H$	&	6	&	HD~2628$^{\dagger}$		\\
		&	2011/08/16	&	S1/E1	&	$H$	&	7	&	HD~112$^{\dagger}$, HD~2628$^{\dagger}$	\\ 
		&	2011/08/20	&	E1/W1	&	$H$	&	3	&	HD~2628$^{\dagger}$	\\ \\

HD 6210	&	2010/09/20	&	S1/E1	&	$H$	&	12	&	HD~3283, HD~9407 \\
		&	2012/11/12	&	E1/W1	&	$H$	&	3	&	HD~3283	\\	
		&	2012/11/14	&	E1/W1	&	$K^{\prime}$	&	5	&	HD~3283	\\
\enddata

\tablecomments{{\bf Stars marked with a dagger $^{\dagger}$ have estimated angular sizes $>0.45$~mas.} Refer to Section~\ref{sec:observations_chara} for details.}

\end{deluxetable}
\newpage
\begin{deluxetable}{lcccccc}
\tabletypesize{\tiny}
\tablewidth{0pt}
\tablecaption{New Angular Diameters\label{tab:diameters}}
\tablehead{
\colhead{\textbf{Star}} &
\colhead{\textbf{\# of}} &
\colhead{\textbf{Reduced}} &
\colhead{\textbf{$\theta_{\rm UD} \pm \sigma$}} &
\colhead{\textbf{ }} &
\colhead{\textbf{$\theta_{\rm LD} \pm \sigma$}} &
\colhead{\textbf{$\theta_{\rm LD}$}} \\
\colhead{\textbf{Name}} &
\colhead{\textbf{Obs.}}	&
\colhead{\textbf{$\chi^{2}$}}	&
\colhead{\textbf{(mas)}} &
\colhead{\textbf{$\mu_{\lambda}$}}	&
\colhead{\textbf{(mas)}}	&
\colhead{\textbf{\% err}}
}
\startdata

HD~166	&	20	&	0.27	&	$0.604\pm0.009$	&	0.386	&	$0.624\pm0.009$	&	1.5	\\
HD~6210	&	20	&	0.14	&	$0.508\pm0.006$	&	0.307	&	$0.520\pm0.006$	&	1.2	\\
HD~10476 &     12 &       0.65 & $0.963\pm0.004$ &      0.410 & $1.000\pm0.004$ &        0.4 \\    	
HD~10697 &     10 &       0.29 & $0.531\pm0.012$ &      0.362 & $0.547\pm0.013$ &        2.3 \\     
HD~11964 &     12 &       0.05 & $0.589\pm0.015$ &      0.362 & $0.607\pm0.015$ &        2.5 \\    
HD~16765	&	14	&	0.06	&	$0.486\pm0.007$	&	0.307	&	$0.497\pm0.007$	&	1.4	\\
HD~21019	&	13	&	0.23	&	$0.588\pm0.015$	&	0.382	&	$0.606\pm0.015$	&	2.5	\\
HD~38858	&	22	&	0.50	&	$0.556\pm0.009$	&	0.349	&	$0.572\pm0.009$	&	1.6	\\
\enddata

\tablecomments{Refer to Section~\ref{sec:observations_chara} for details.}

\end{deluxetable}
\newpage
\clearpage
\LongTables

\begin{landscape}
\begin{deluxetable}{lccccccccc}
\tabletypesize{\tiny}
\tablewidth{0pt}
\tablecaption{Angular Diameter Anthology\label{tab:diameters_literature}}
\tablehead{
\colhead{\textbf{Star}} &
\colhead{\textbf{Spectral}} &
\colhead{\textbf{Metallicity}} &
\colhead{\textbf{Radius}} &
\colhead{\textbf{Radius}} &
\colhead{\textbf{$F_{\rm BOL}$}} &
\colhead{\textbf{$L$}} &
\colhead{\textbf{$T_{\rm eff}$}} 	&
\colhead{\textbf{Age}} &
\colhead{\textbf{Mass}}  \\
\colhead{\textbf{HD}} &
\colhead{\textbf{Type}}	&
\colhead{\textbf{[Fe/H]}}	&
\colhead{\textbf{(R$_{\odot}$)}} &
\colhead{\textbf{Reference}}	&
\colhead{\textbf{(1e$-8$~erg~s$^{-1}$~cm$^{-2}$)}}	&
\colhead{\textbf{(L$_{\odot}$)}} &
\colhead{\textbf{(K)}} &
\colhead{\textbf{(Gyr)\tablenotemark{b}}} &
\colhead{\textbf{(M$_{\odot}$)\tablenotemark{b}}}
}
\startdata

166     &       G8V     &         $0.08$\phn    &       $0.9172\pm0.0090$       &       this work       &       $10.4400\pm0.0600$   \phn       &       $0.6078\pm0.0099$       &       $5327\pm39$\phn\phn     &       9.6             &       0.889           \\
3651    &       K0V     &         $0.15$\phn    &       $0.9470\pm0.0320$       &       1       &       $13.4700\pm0.0600$\phn     &       $0.5131\pm0.0043$       &       $5046\pm86$\phn\phn     &       14.9\phn                &       0.839           \\
4614    &       F9V     &        $-0.28$\phn\phs        &       $1.0386\pm0.0038$       &       2       &       $111.6000\pm0.1940$  \phn\phn   &       $1.2321\pm0.0074$       &       $5973\pm8$\phn\phn      &       5.9             &       0.967           \\
5015    &       F8V     &         $0.04$\phn    &       $1.7426\pm0.0233$       &       2       &       $31.5400\pm0.0590$   \phn       &       $3.4521\pm0.0432$       &       $5965\pm35$\phn\phn     &       5.3             &       1.194           \\
\enddata

\tablenotetext{}{{\bf References.} (1) \citet{bai08}, (2) \citet{boy12a}, (3) \citet{lig12}, (4) \citet{dif04}, (5) \citet{van09}, (6) \citet{the05}, (7) \citet{dav11}, (8) \citet{han74}, (9) \citet{dav86}, (10) \citet{ker03a}, (11) \citet{moz03}, (12) \citet{big11}, (13) \citet{chi12}, (14) \citet{nor01}, (15) \citet{ker04a}, (16) \citet{von11b}, (17) \citet{creevey12}, (18) \citet{ker03b}, (19) \citet{big06}, (20) \citet{baz11}, (21) \citet{hub12}, (22) \citet{nor99}, (23) \citet{cre12}, (24) \citet{bai12}}

\tablenotetext{a}{Bolometric flux from \citet{hub12}.}
\tablenotetext{b}{Stellar mass and age determined by interpolating the $Y^2$ isochrones to match the measured stellar radii, effective temperature, and metallicity.}
\tablenotetext{c}{Spectral type from Simbad.}
\tablenotetext{d}{The measurements and associated errors are incommensurate for the two stars HD~146233 and HD~185395, likely caused from calibration errors.  No measurement averages are taken due to this.}

\tablecomments{All measurements of stellar radii found in the literature, with precision of better than 5\%. Stars with multiple measurements are marked with a $^{\dagger}$. Metallicities are from \citet{and11}, and parallaxes are from \citet{van07}.  The bottom portion of the table lists the stars with multiple measurements, and the weighted mean for their radii and temperatures (all other parameters remain unaffected when combining the multiple sources for measured radii). All bolometric flux, luminosity, and temperature values are computed/measured in this work.  Refer to Sections~\ref{sec:observations_literature_averages}, \ref{sec:stellar_params}, and \ref{sec:evolution} for details.}

\end{deluxetable}

\clearpage
 \end{landscape}

\newpage
\clearpage
\LongTables

\begin{deluxetable}{lccccl}
\tablecolumns{6}
\tabletypesize{\scriptsize}
\tablewidth{0pc}
\tablecaption{Object photometry used in SED fits \label{tab:monsterPhot}}
\tablehead{
Star & System /  & Bandpass /  &  &  &  \\
 ID &  Wavelength &  Bandwidth & Value & Error & Reference
}
\startdata

HD166 & DDO  & m35 & 8.43 & 0.05 & \citet{1976AJ.....81..182M}\\
HD166 & WBVR  & W & 7.00 & 0.05 & \citet{1991TrSht..63....4K}\\
HD166 & Johnson  & U & 7.22 & 0.05 & \citet{1957ApJ...126..113J}\\
HD166 & Johnson  & U & 7.22 & 0.05 & \citet{1966CoLPL...4...99J}\\
HD166 & Johnson  & U & 7.22 & 0.05 & \citet{1966MNRAS.133..475A}\\
HD166 & Johnson  & U & 7.15 & 0.05 & \citet{1986EgUBV........0M}\\
HD166 & DDO  & m38 & 7.54 & 0.05 & \citet{1976AJ.....81..182M}\\
HD166 & DDO  & m41 & 8.20 & 0.05 & \citet{1976AJ.....81..182M}\\
\enddata
\tablecomments{The collections of photometry used in the SED fitting routine for all objects. Refer to Section~\ref{sec:stellar_params} for details.}

\end{deluxetable}



\newpage
\clearpage
\LongTables

\begin{landscape}
\begin{deluxetable}{lcccccccc|cccc}
\tabletypesize{\tiny}
\tablewidth{0pt}
\tablecaption{Table of Spectral Types and Bolometric Fluxes \label{tab:fbols_table}}
\tablehead{
\multicolumn{5}{c}{ }	&
\multicolumn{4}{c|}{\textbf{Photometry}}	&
\multicolumn{4}{c}{\textbf{Spectrophotometry}}	\\
	\cline{6-9} 
	\cline{10-13}
\colhead{\textbf{Star}} &	
\colhead{\textbf{Sp.Ty.}} &	
\colhead{\textbf{Sp.Ty.}} &	
\colhead{\textbf{Sp.Ty.}} &		
\colhead{\textbf{Sp.Ty.}} &	
\colhead{\textbf{ }} &		
\colhead{\textbf{ }} &		
\colhead{\textbf{Best Fit}} &	
\colhead{\textbf{$F_{\rm BOL} \pm \sigma$}} &	
\colhead{\textbf{ }} &		
\colhead{\textbf{Best Fit}} &	
\colhead{\textbf{$F_{\rm BOL} \pm \sigma$}} &	
\colhead{\textbf{Spectro.}}	\\	
\colhead{\textbf{HD}} &				
\colhead{\textbf{First}}	&		
\colhead{\textbf{Second}}	&		
\colhead{\textbf{Ref.}} &		
\colhead{\textbf{Other}} &			
\colhead{\textbf{DF}}	&			
\colhead{\textbf{$\chi^2$/DF}}	&		
\colhead{\textbf{Sp.Ty.}} &		
\colhead{\textbf{(1e$-8$~erg~s$^{-1}$~cm$^{-2}$)}} 	& 
\colhead{\textbf{$\chi^2$/DF}}	&		
\colhead{\textbf{Sp.Ty.}} &		
\colhead{\textbf{(1e$-8$~erg~s$^{-1}$~cm$^{-2}$)}} 	& 
\colhead{\textbf{Ref.}}		
}

166 & G8V &  \nodata  & 24 &  \nodata  & 53 & 0.48 & G8V & $10.44 \pm 0.06$ &  \nodata  &  \nodata  &  \nodata  &  \\
3651 & K0.5V & K0V & 21, 24 & K1V & 110 & 2.19, 4.16, 1.91 & K1V & $13.47 \pm 0.06$ &  \nodata  &  \nodata  &  \nodata  &  \\
4614 & F9V & F9V & 21, 23 &  \nodata  & 80 & 1.19, 1.19 & F9V & $112.60 \pm 0.59$ & 3.66 & F8V & $111.60 \pm 0.19$ & 3 \\
5015 & F9V & F8V & 28, 23 &  \nodata  & 76 & 0.68, 0.76 & F9V & $31.30 \pm 0.18$ & 5.06 & F9V & $31.54 \pm 0.06$ & 5 \\
6210 & F6V & F7IV & 9, 12 & G0V & 21 & 17.45, 13.93, 1.15 & G0V & $12.40 \pm 0.10$ & 5.72 & G0V & $12.38 \pm 0.02$ & 3 \\
9826 & F9V & F8V & 28, 23 &  \nodata  & 136 & 0.71, 0.64 & F8V & $60.07 \pm 0.25$ & 1.87 & F8V & $60.15 \pm 0.13$ & 3 \\
\enddata

\tablecaption{Bolometric fluxes of target stars based upon \citet{pic98} spectral templates fit to literature spectral types with photometry in the literature (in Table~\ref{tab:monsterPhot}), and extended by spectrophotometry when available.  References for the first and second spectral types are in the $4^{th}$ column, as found from the index by \citet{ski13}, with an additional spectral type if necessary for adequate SED fitting (column~5).  For the fits, degrees-of-freedom (DF) and $\chi^2$-per-DF metrics are given, along with spectral type of the template for the best fit, and its corresponding $F_{\rm BOL}$ value.}

\tablenotetext{}{{\bf Spectral type references:} (1) \citet{1952ApJ...116..592M}, (2) \citet{1967AJ.....72.1334C}, (3) \citet{1967PDDO....2..479H}, (4) \citet{1970ApJS...19..281B}, (5) \citet{1971ApJ...163...75S}, (6) \citet{1973ARA_A..11...29M}, (7) \citet{1974AJ.....79..682H}, (8) \citet{1975mcts.book.....H}, (9) \citet{1976PASP...88...95C}, (10) \citet{1978PASP...90..429L}, (11) \citet{1979PASP...91...83C}, (12) \citet{1981A_AS...45..455J}, (13) \citet{1981ApJS...45..437A}, (14) \citet{1981PASP...93...45B}, (15) \citet{1985ApJS...59...95A}, (16) \citet{1986ApJ...309..260A}, (17) \citet{1989AJ.....98.1049G}, (18) \citet{1989ApJS...69..301G}, (19) \citet{1989ApJS...70..623G}, (20) \citet{1989ApJS...71..245K}, (21) \citet{1995ApJS...99..135A}, (22) \citet{2001AJ....121.2148G}, (23) \citet{2003AJ....126.2048G}, (24) \citet{2006AJ....132..161G}, (25) \citet{2008ApJS..176..216A}, (26) \citet{2009A_A...501..297Z}, (27) \citet{2009ApJS..180..117A}.}

\tablenotetext{}{{\bf Spectrophotometry references:}
(1)	\citet{Burnashev1985BCrAO..66..152B},
(2)	\citet{Alekseeva1996BaltA...5..603A,Alekseeva1997BaltA...6..481A},
(3)	\citet{Kharitonov1988scsb.book.....K},
(4)	\citet{Glushneva1998yCat.3207....0G},
(5)	\citet{Glushneva1998yCat.3208....0G}.}

\end{deluxetable}

\clearpage
 \end{landscape}

\newpage
\clearpage
\LongTables

\begin{deluxetable}{lcccc}
\tabletypesize{\tiny}
\tablewidth{0pt}
\tablecaption{Spectral Type Lookup Table\label{tab:spectral_types}}
\tablehead{
\colhead{\textbf{Spectral}} &
\colhead{\textbf{ }} &
\colhead{\textbf{$B-V$}} &
\colhead{\textbf{$V-K$}} &
\colhead{\textbf{$T_{\rm eff} \pm \sigma$}} \\
\colhead{\textbf{Type}} &
\colhead{\textbf{$n$}}	&
\colhead{\textbf{(mag)}} &
\colhead{\textbf{(mag)}} &
\colhead{\textbf{(K)}}	
}

\startdata

    A0 \dotfill  &      2 &   0.00 &  -0.04 &   9394 $ \pm $     44 \\ 
    A1 \dotfill  &      2 &  -0.01 &  -0.03 &   9121 $ \pm $     71 \\ 
    A2 \dotfill  &      3 &   0.12 &   0.27 &   7965 $ \pm $     24 \\ 
    A3 \dotfill  &      2 &   0.10 &   0.10 &   8176 $ \pm $     12 \\ 
    A4 \dotfill  &      1 &   0.09 &   0.19 &   8459 $ \pm $     44 \\ 
    A5 \dotfill  &      1 &   0.12 &   0.29 &   7889 $ \pm $     60 \\ 
    A7 \dotfill  &      1 &   0.21 &   0.51 &   7734 $ \pm $     80 \\ 
    F0 \dotfill  &      3 &   0.30 &   0.79 &   6850 $ \pm $     28 \\
\enddata

\tablecomments{The value of the parameter given is the average value of all stars within the spectral type bin, and the $\sigma$ is the standard deviation of the parameter uncertainties for each spectral type bin. The spectral types with only one measurement ($n=1$) simply lists the individual value and the measured error of that measurement. Refer to Section~\ref{sec:stellar_params} for details.}

\end{deluxetable}

\clearpage
\newpage

\clearpage
\LongTables

\begin{landscape}
\begin{deluxetable}{lccccccccccccccccc}
\tabletypesize{\tiny}
\tablewidth{0pt}
\tablecaption{Photometry used in Color$-$Temperature Relations \label{tab:phot_forfits}}
\tablehead{
\colhead{\textbf{Star}} &
\colhead{\textbf{$B$}} &
\colhead{\textbf{$V$}} &
\colhead{\textbf{$R_J$}} &
\colhead{\textbf{$I_J$}} &
\colhead{\textbf{$J$}} &
\colhead{\textbf{$H$}} &
\colhead{\textbf{$K$}} &
\colhead{\textbf{$R_C$}}	&
\colhead{\textbf{$I_C$}}	&
\colhead{\textbf{$R_K$}} &
\colhead{\textbf{$I_K$}}	&
\colhead{\textbf{$g$\tablenotemark{a}}}	&
\colhead{\textbf{$r$\tablenotemark{a}}} &
\colhead{\textbf{$i$\tablenotemark{a}}} &
\colhead{\textbf{$z$\tablenotemark{a}}} &
\colhead{\textbf{$W3$\tablenotemark{b}}}	&
\colhead{\textbf{$W4$\tablenotemark{b}}} 
}

\startdata
166 \dotfill & 6.89	&	6.14	&	\nodata	&	\nodata	&	4.65\tablenotemark{c}	&	4.60\tablenotemark{c}	&	4.28\tablenotemark{c}	&	\nodata	&	\nodata	&	\nodata	&	\nodata	&	6.42	&	5.86	&	5.67	&	5.62	&	4.33	&	4.12	\\
3651 \dotfill & 6.71	&	5.86	&	5.21	&	4.82	&	4.48 	&	4.03 	&	3.97 	&	\nodata	&	\nodata	&	5.52	&	5.25	&	6.27	&	5.66	&	5.44	&	5.35	&	3.93	&	3.91	\\
4614 \dotfill & 4.02	&	3.44	&	2.94	&	2.58	&	2.35 	&	2.02 	&	1.96 	&	\nodata	&	\nodata	&	3.30	&	3.08	&	\nodata	&	\nodata	&	\nodata	&	\nodata	&	\nodata	&	\nodata	\\
5015 \dotfill & 5.35	&	4.82	&	4.34	&	4.04	&	3.85 	&	3.56 	&	3.54 	&	\nodata	&	\nodata	&	\nodata	&	\nodata	&	\nodata	&	\nodata	&	\nodata	&	\nodata	&	3.50	&	3.46	\\
6210 \dotfill & 6.38	&	5.84	&	\nodata	&	\nodata	&	4.67\tablenotemark{c}	&	4.76\tablenotemark{c}	&	4.41\tablenotemark{c}	&	\nodata	&	\nodata	&	\nodata	&	\nodata	&	6.08	&	5.71	&	5.58	&	5.55	&	4.46	&	4.43	\\
9826 \dotfill & 4.64	&	4.10	&	3.64	&	3.35	&	3.17 	&	2.99 	&	2.85 	&	\nodata	&	\nodata	&	\nodata	&	\nodata	&	\nodata	&	\nodata	&	\nodata	&	\nodata	&	2.88	&	2.84	\\
10476 \dotfill & 6.08	&	5.24	&	4.55	&	4.12	&	3.85 	&	3.44 	&	3.21 	&	\nodata	&	\nodata	&	4.87	&	4.58	&	5.68	&	5.04	&	4.84	&	4.75	&	3.09	&	3.31	\\
10697 \dotfill & 6.99	&	6.26	&	\nodata	&	\nodata	&	4.98 	&	4.66 	&	4.58 	&	\nodata	&	\nodata	&	\nodata	&	\nodata	&	6.57	&	6.07	&	5.93	&	5.88	&	4.64	&	4.58	\\
\enddata

\tablenotetext{a}{Average from \citet{ofe08} and \citet{pic10}.}
\tablenotetext{b}{The WISE W3 and W4 magnitudes have been filtered to only allow values that have not reached saturation limits ($W3<3.8$~mags and $W4<-0.4$~mags).}
\tablenotetext{c}{$2MASS$ magnitudes converted to the Johnson system. See Section~\ref{sec:discussion} for details. } 

\tablecomments{Photometry sources include: \citet{1966CoLPL...4...99J,
1968ApJ...152..465J,
1972RGOB..176..127E,
1974MNSSA..33...53G,
1975MNRAS.171P..19G,
1977AJ.....82..598G,
1979MNRAS.188..847B,
1981PASJ...33..373N,
1986AJ.....91.1140S,
1989A&A...215..305A,
1990A&A...232..396B,
1991ApJ...368..264A,
1994A_AS..107..365A,
1996MNRAS.279..915S,
1997yCat.2168....0M,
2002yCat.2237....0D,
cou80, 
kro57,
wri10}.  Refer to Section~\ref{sec:discussion} for details.}

\end{deluxetable}

\clearpage
 \end{landscape}

\newpage
\clearpage
\LongTables

\begin{deluxetable}{lccccccc}
\tabletypesize{\tiny}
\tablewidth{0pt}
\tablecaption{Solutions to Temperature Relations\label{tab:poly3_coeffs}}
\tablehead{
\colhead{\textbf{Color}} &
\colhead{\textbf{\# of}} &
\colhead{\textbf{Range}} &
\colhead{\textbf{ }} &
\colhead{\textbf{ }} &
\colhead{\textbf{ }} &
\colhead{\textbf{ }} &
\colhead{\textbf{$\sigma$}} 	\\
\colhead{\textbf{Index}} &
\colhead{\textbf{Points}}	&
\colhead{\textbf{(mag)}}	&
\colhead{\textbf{$a_0 \pm \sigma$}} &
\colhead{\textbf{$a_1 \pm \sigma$}}	&
\colhead{\textbf{$a_2 \pm \sigma$}}	&
\colhead{\textbf{$a_3 \pm \sigma$}}	&
\colhead{\textbf{(\%)}}
}

\startdata

        $(B-V)$\tablenotemark{a} &   124 &  [$-0.02-1.73$]  & $    9084 \pm   15$ & $   -7736 \pm   57$ & $    4781 \pm   69$ & $ -1342.9 \pm 25.1$ &      3.3 \\ 
 		$(B-V)$\tablenotemark{a,b} &    111 &  [$ 0.32-1.73$]  & $    7722 \pm   38$ & $   -3144 \pm  129$ & $     112 \pm  136$ & $   114.2 \pm  44.3$ &      3.0 \\ \\

         $(V-J)$ &    122 &  [$-0.12-4.24$]  & $    9052 \pm   13$ & $   -3972 \pm   20$ & $    1039 \pm    9$ & $  -101.0 \pm  1.5$ &      3.7 \\
		 $(V-J)$\tablenotemark{c} &  109 &  [$ 0.60-4.24$]  & $    9041 \pm   25$ & $   -3950 \pm   39$ & $    1028 \pm   17$ & $   -99.4 \pm   2.3$ &      3.7 \\
         $(V-J)$\tablenotemark{d} &   95 &  [$-0.12-4.24$]  & $    9127 \pm   13$ & $   -4084 \pm   21$ & $    1082 \pm   10$ & $  -106.0 \pm  1.5$ &      2.6 \\
        $(V-J)$\tablenotemark{c,d} &    85 &  [$ 0.60-4.24$]  & $    9183 \pm   26$ & $   -4163 \pm   40$ & $    1114 \pm   17$ & $  -109.9 \pm   2.3$ &      2.5 \\ \\
\enddata

\tablenotetext{a}{The ($B-V$) relation expressed as a $3^{rd}$ order polynomial is insufficient for the full range of AFGKM stars.  See Section~\ref{sec:discussion_bmv} for discussion and alternate solutions.  }
\tablenotetext{b}{For a metallicity dependent solution, refer to Equation~\ref{eq:poly2cm} in Section~\ref{sec:discussion_bmv}.}  
\tablenotetext{c}{Solutions derived omitting hot stars ($T_{\rm eff} > 6750$~K).  See Section~\ref{sec:discussion_astars_evolution}.} 
\tablenotetext{d}{Solutions derived omitting all stars that only have $2MASS$ magnitudes. See Section~\ref{sec:discussion_2mass}.}

\tablecomments{Refer to Section~\ref{sec:discussion},  Section~\ref{sec:discussion_bmv}, Section~\ref{sec:discussion_astars_evolution}, Section~\ref{sec:discussion_2mass}, and Equation~\ref{eq:poly3c} for details.}

\end{deluxetable}

\clearpage
\newpage
\clearpage
\LongTables

\begin{deluxetable}{lc}
\tabletypesize{\tiny}
\tablewidth{0pt}
\tablecaption{Maximum Difference in Temperature Solution when Omitting Hot Stars\label{tab:max_diff_omit_hot}}
\tablehead{
\colhead{\textbf{Color}} &
\colhead{\textbf{Max.}} 	\\
\colhead{\textbf{Index}} &
\colhead{\textbf{\% Diff.}}
}

        		 $(B-V)$ &  $-$4.3 \\ 
                  $(V-J)$ &  $-$0.1 \\ 
                  $(V-H)$ &  $-$1.1 \\ 
                  $(V-K)$ &  $-$1.0 \\ 
          $(V-R_{\rm J})$ &  $-$1.7 \\ 
          $(V-I_{\rm J})$ &  $-$2.6 \\ 
          $(V-R_{\rm C})$ &  $-$0.7 \\ 
          $(V-I_{\rm C})$ &  $-$0.4 \\ 
          $(V-R_{\rm K})$ & $-$11.5\phn \\ 
          $(V-I_{\rm K})$ &  $-$0.7 \\ 

 $(R_{\rm J}-J)$ &  $-$7.5 \\ 
  $(R_{\rm J}-H)$ & $-$11.8\phn \\ 
  $(R_{\rm J}-K)$ & $-$14.8\phn \\ 
  $(R_{\rm C}-J)$ &  $-$1.7 \\ 
  $(R_{\rm C}-H)$ &  $-$0.2 \\ 
  $(R_{\rm C}-K)$ &  $-$0.2 \\ 
  $(R_{\rm K}-J)$ &  $-$0.8 \\ 
  $(R_{\rm K}-H)$ &  $-$4.2 \\ 
  $(R_{\rm K}-K)$ &  $-$4.6 \\ 

		         $(g-z)$ &  $-$0.5 \\ 
                  $(g-i)$ &  $-$0.5 \\ 
                  $(g-r)$ &  $-$1.2 \\ 
                  $(g-J)$ &  $-$0.2 \\ 
                  $(g-H)$ &  0.0\phn \\ 
                  $(g-K)$ &  $-$0.3 \\ 

        $(V-W3)$ &  $-$0.1 \\ 
                $(V-W4)$ &  $-$0.4 \\ 
        $(R_{\rm J}-W4)$ &  $-$4.5 \\ 
        $(I_{\rm J}-W4)$ &  $-$4.2 \\ 
        $(R_{\rm C}-W4)$ &  $-$0.4 \\ 
        $(I_{\rm C}-W4)$ &  $-$1.0 \\ 
        $(R_{\rm K}-W4)$ &  $-$1.1 \\ 
        $(I_{\rm K}-W4)$ &  $-$2.0 \\

\enddata

\tablecomments{For each color index, we show the maximum difference in predicted temperature using the full AFGKM sample to the sample that omits the early-type stars.  Refer to Section~\ref{sec:discussion_astars_evolution} for details.}

\end{deluxetable}

\clearpage

\newpage

\newpage

\newpage

\newpage

\newpage

\clearpage
\newpage

\clearpage
\newpage

\clearpage
\newpage

\clearpage
\newpage

\clearpage
\newpage

\clearpage
\newpage

\clearpage
\newpage

\clearpage

\newpage

\newpage

\newpage

\newpage

\newpage

\clearpage
\newpage

\clearpage
\newpage

\clearpage
\newpage

\clearpage
\newpage


\end{document}